\documentclass{aa}  

\usepackage{graphicx}

\usepackage{txfonts}
\newcommand{\ebv}{$E(B-V)$}

\newcommand{\hb}{H$\beta$}
\newcommand{\ha}{H$\alpha$}

\newcommand{\hei}{He\,\textsc{i}}
\newcommand{\siii}{\textsc{[S\,iii]}}

\newcommand{\oii}{[O\,\textsc{ii}]}

\newcommand{\sii}{\textsc{[S\,ii]}}

\newcommand{\cliii}{[Cl\,\textsc{iii}]}
\newcommand{\oi}{[O\,\textsc{i}]}

\newcommand{\ariv}{[Ar\,\textsc{iv}]}
\newcommand{\niA}{\textsc{[N\,i]}}
\newcommand{\nii}{\textsc{[N\,ii]}}
\newcommand{\ariii}{[Ar\,\textsc{\,iii]}}
\newcommand{\oiii}{\textsc{[O\,iii]}}
\newcommand{\heii}{He\,\textsc{ii}}

\usepackage{xcolor}

\begin{document}

   \title{The MUSE view  of the planetary nebula \object{NGC\,3132}}
   \author{Ana Monreal-Ibero
          \inst{1,2}
          \and
          Jeremy R. Walsh
          \inst{3}
             \fnmsep \thanks{Based on observations collected at the European Organisation for Astronomical Research in the Southern Hemisphere, Chile (ESO Programme 60.A-9100)}
          }
    \institute{
    Instituto de Astrof\'{\i}sica de Canarias (IAC), E-38205 La Laguna, Tenerife, Spain
    \and
    Universidad de La Laguna, Dpto.\ Astrof\'{\i}sica, E-38206 La Laguna, Tenerife, Spain\\
    \email{amonreal@iac.es}
    \and
    European Southern Observatory. Karl-Schwarzschild Strasse 2, 85748 Garching, Germany 
    }

   \date{Received October XX, 2019; accepted XXXX XX, XXXX}

 
  \abstract
   {}
   {2D spectroscopic data for the whole extent of the NGC\,3132 planetary nebula have been obtained. We deliver a reduced data-cube and high-quality maps on a spaxel-by-spaxel basis for the many emission lines falling within the MUSE spectral coverage over a range in surface brightness $>$1000. Physical diagnostics derived from the emission line images, opening up a variety of scientific applications, are discussed.}
   {Data were obtained during MUSE commissioning on the ESO Very Large Telescope and reduced with the standard ESO pipeline. Emission lines were fitted by Gaussian profiles. The dust extinction, electron densities and temperatures of the ionised gas and abundances were determined using Python and \texttt{PyNeb} routines.}
   {The delivered datacube has a spatial size of $\sim63^{\prime\prime}\times123^{\prime\prime}$, corresponding to $\sim0.26\times0.51$~pc$^2$ for the adopted distance,  and a contiguous wavelength coverage of 4750-9300~\AA\, at a spectral sampling of 1.25~\AA\,pix~$^{-1}$. 
   The nebula presents a complex reddening structure with high values (c(\hb)$\sim0.4$) at the rim.
   Density maps are compatible with an inner high-ionisation plasma at moderate high density ($\sim$1000~cm$^{-3}$) while the low-ionisation plasma presents a structure in density peaking at the rim with values $\sim$700~cm$^{-3}$.
   Median $T_e$ using different diagnostics decreases according to the sequence \nii,\sii $\rightarrow$ \siii  $\rightarrow$ \oi $\rightarrow$ \hei $\rightarrow$ Paschen Jump. Likewise the range of temperatures covered by recombination lines is much larger than those obtained from collisionally excited lines (CELs), with large spatial variations within the nebula.
   If these differences were due to the existence of high density clumps, as previously suggested, these spatial variations suggest changes in the properties and/or distribution of the clumps within the nebula.
   We determined a median helium abundance He / H = 0.124, with slightly higher values at the rim and outer shell. 
   The range of measured ionic abundances for light elements are compatible with literature values.
   Our kinematic analysis nicely illustrates the power of 2D kinematic information in many emission lines to shed light on the intrinsic structure of the nebula.
   Specifically,
   our derived velocity maps support a geometry for the nebula similar to the diabolo-like model proposed by \citet{Monteiro00}, but oriented with its major axis roughly at P.A.$\sim-22^\circ$. 
   We identified two low-surface brightness arc-like structures towards the northern and southern tips of the nebula, with high extinction, high helium abundance, and strong low-ionisation emission lines. They are spatially coincident with some extended low-surface brightness mid-IR emission. The characteristics of the features are compatible with being the consequence of precessing jets caused by the binary star system. 
   A simple 1D Cloudy model is able to reproduce the strong lines in the integrated spectrum of the whole nebula with an accuracy of $\sim$15\% .
   }
   {Together with the work on \object{NGC\,7009} presented by \citet{Walsh18}, the present study illustrates the enormous potential of wide field  integral field spectrographs for the study of Galactic PNe.}


   \maketitle
%

\section{Introduction}

Planetary Nebulae (PNe) are ionised nebulae resulting from the evolution and death of low-mass ($\sim0.8-8$~M$_\odot$)  stars. They present a large variety of physical and morphological structures, covering zones from high to low ionisation conditions, and ranging from compact knots at high density to low-density extended haloes. In that sense, PNe constitute ideal laboratories to study the interstellar medium \citep[ISM, see][for a review]{Kwok94,Kwitter14}.
Likewise, they may present a mix of chemical abundances, a consequence of the complex interplay between the recent ejection history and conditions of the evolved star as well as those of the surrounding ISM. A long-standing issue in this regard is that the measured abundances are higher when using optical recombination lines than from collisionally excited lines \citep[CELs, e.g.][]{Liu00,Corradi15,Wesson18}.

In view of their diversity and proximity, Galactic PNe constitute a corner-stone for studies of ISM in general, including diffuse ionised gas, \textsc{H\,ii} regions, starburst galaxies and active galactic nuclei. 
However, this complexity can typically be displayed within a single object and Galactic PNe have a relatively large angular size. In that sense, observations with a small aperture or slit are clearly insufficient to grasp the full range of conditions that one may encounter within a given object.
Instead, having access to full 2D spectroscopic coverage, mapping the whole nebula would be much more desirable.

In the past, this need for spectral mapping was partially fulfilled by multiple slit observations at key positions in the nebula \citep[e.g.][]{Meaburn81,Cuesta93,Akras16,GarciaDiaz12,Lago16}.
This is an expensive observational strategy requiring a substantial amount of observing time. Moreover, the effect of the atmospheric differential refraction can be difficult to quantify and the data, as a whole, are not necessarily homogeneous since they can be affected by changes in the observing conditions. Besides, the spatial coverage is not continuous but restricted to representative portions of the nebula.

The use of Fabry-P\'erot interferometers, with a relatively large field of view, can overcome this lack of continuous spatial coverage \citep{Pismis89,Lame96}. Moreover, they provide relatively high spectral resolution, thus being an excellent option for kinematic studies. However the spectral range covered by these intruments is typically small, and thus, studies using Fabry-P\'erot interferometers focus on the information that can be extracted from one or a few lines.

The advent of Integral Field Spectroscopy (IFS), able to record simultaneously spectra of a relatively large area in the sky, provides an opportunity to obtain at once 2D information of many emission lines, allowing for an accurate determination of the physical and chemical nebular parameters. \citet{MonrealIbero05,MonrealIbero06} were pioneers in applying this technique to PN research. They made use of the VIMOS IFU, the one with the largest available field-of-view at the time, to 
study the physical properties of the faint halo of \object{NGC\,3242}.
In the following years, the technique gained popularity in slow but steady mode. Thus, soon after, \citet{Tsamis08} used the FLAMES instrument in Argus mode to map a considerable portion of \object{NGC\,7009}, \object{NGC\,5882}, and \object{NGC\,6153}. Likewise, \citet{Sandin08} used PMAS to characterise the faint halos of several PNe. Later, \citet{Monteiro13} derived spatially resolved maps of the electron densities, temperatures, and chemical abundances for
\object{NGC 3242}. In recent years, several southern hemisphere PNe have been studied using data obtained with the Wide Field Spectrograph (WiFeS) on the 2.3-m ANU telescope \citep[e.g.][and references therein]{Ali16}. All these works nicely illustrate how IFS is an excellent approach to study Galactic extended PNe. Still, IFS-based existing instrumentation up to now covered a small to moderate field-of-view (f.o.v.), and thus one must choose between 
fully mapping relatively far (and thus small angular size) PNe, or studying (previously identified) key portions of the nebula.
The Multi-Unit Spectroscopic Explorer \citep[MUSE, ][]{Bacon10} in its Wide Field Mode can map at once an area of $\sim60^{\prime\prime}\times60^{\prime\prime}$ at exquisite spatial sampling. These dimensions nicely suit the size of many nearby Galactic PNe. \cite{Walsh18,Walsh16} demonstrated its potential with a detailed study on \object{NGC\,7009}. The team obtained spatially resolved maps for electron densities, and temperatures, as traced by several ($>$3) diagnostics, as well as chemical abundances for oxygen and helium and ionic abundances for several other species. The wealth of derived information might well sound like opening Pandora's box, but is essential to ultimately obtain a fully self-consistent 3D picture of the physical and chemical properties of the individual nebula.

With the same esprit, we present here our work on the MUSE data for \object{NGC\,3132}. The aim of the contribution is two-fold.
On the one hand, we provide the community with the fully reduced MUSE datacube, as well as the derived emission line maps on a spaxel-by-spaxel basis for public use.
On the other hand, we illustrate the potential of the data by addressing some scientific cases. Specifically, i) we will use the emission line maps created on spaxel-by-spaxel basis to explore the physical (extinction, electron density and temperature) and chemical (ionic and total abundances) properties in the nebula; ii) we will characterise two newly identified structures at the northern and southern extremities of the nebula; iii) we will discuss what can be learned for the kinematics of the nebula at the MUSE spectral resolution; iv) we will provide a simple 1D photoionisation model to evaluate how the 2D discrepancies affect our conclusions about the structure of the nebula.
However these topics by no means exhaust the full exploitation of the data. Further analysis could be done by e.g., a smart binning of the datacube to increase the signal-to-noise of the spectra at the expense of some loss of spatial resolution, or by using (3D) ionisation codes to reproduce the mapped quantities. Even if these and other examples are out of the scope of this contribution, by making  the reduced datacube publicly available, it is our expectation that these and other studies can be addressed in the future.

\object{NGC\,3132} is a PN with a relatively low abundance discrepancy factor \citep[adf=2.4,][]{Tsamis04}.
With an angular size of $\sim58^{\prime\prime}\times85^{\prime\prime}$ \citep{Mata16}, this PN is too big to fit in one single MUSE pointing. Still, areas much larger than MUSE f.o.v. can successfully be mapped by mosaicing, as has been proven for the Orion Nebula \citep{Weilbacher15,Weilbacher15b}. 
\object{NGC\,3132}  presents an elliptical ring inner structure and an outer irregular ellipse of lower surface brightness \citep{Juguet88}. This would suggest an intrinsic ellipsoidal geometry. However, an elliptical model is not able to reproduce 
all the observational features for the nebula. Instead, a diabolo-like model seems more adequate to represent its structure \citep{Monteiro00}.
\citet{Evans68} determined that the spectral type of the bright central star was about A3V and thus it could not be the ionising star.
This was discovered  by \citet{Kohoutek77}, who confirmed that there was actually a binary system at the centre of the nebula.
Our current understanding of the central object is that it is a wide visual binary 
with a most-likely A0 central-star companion \citep{Ciardullo99} and an ionising star with a luminosity of $\log$(L/L$_\sun$) = 2.19 and a temperature of $T_{\rm eff}$=100\,000~K \citep{Frew08}.
Distances reported for \object{NGC~3132} range between 540 and 1240 pc \citep{Gathier86,Schoenberner18,Frew08,Monteiro00}.
In particular, \cite{Kimeswenger18}, using Gaia DR2, report a distance of $\sim$824-904~pc. Here, we will adopt the mean, 864~pc, as distance to the nebula.

   \begin{figure}
   \centering
   \includegraphics[angle=0,width=0.49\textwidth, clip=, viewport=105 0 510 550,]{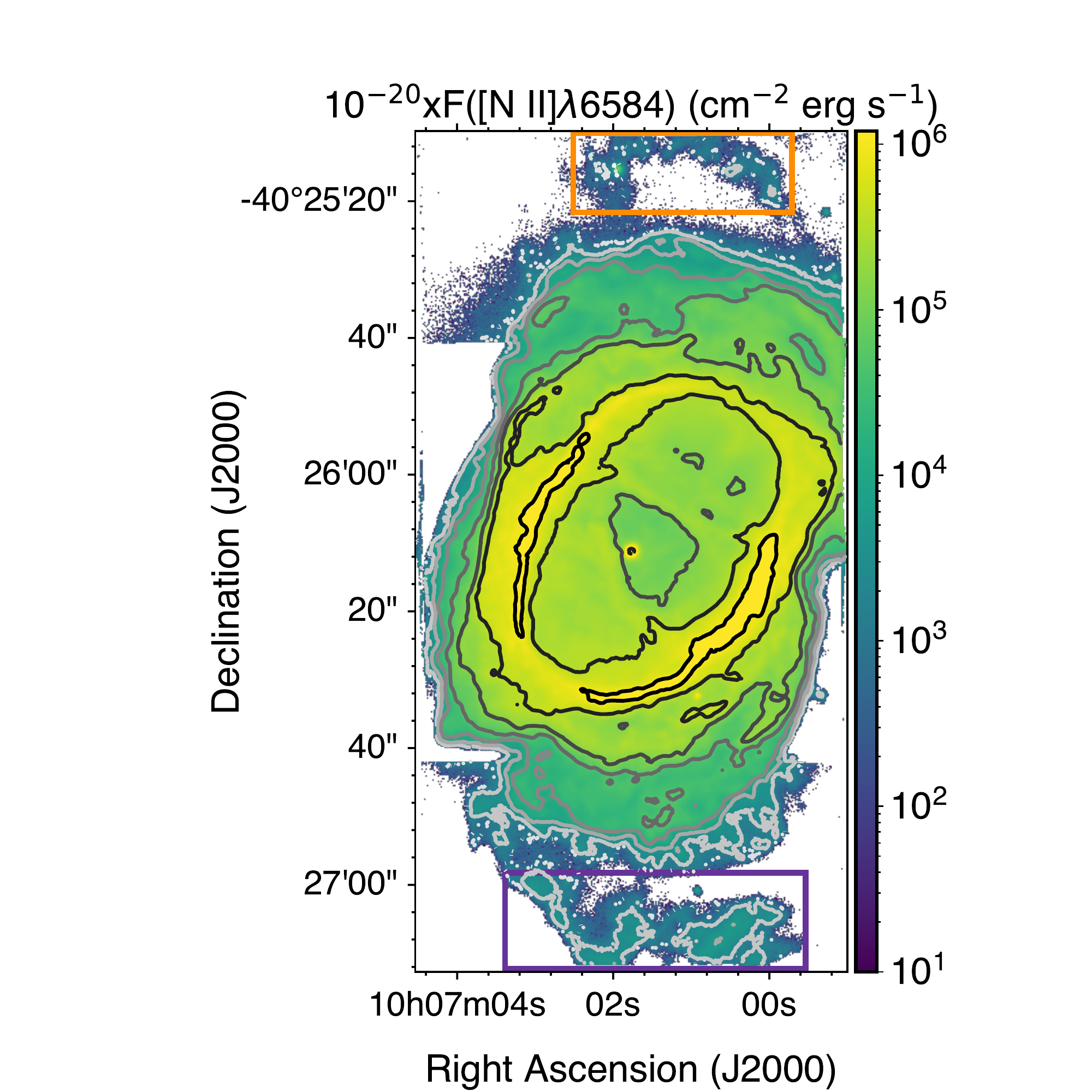}
   \caption{Reconstructed image of \object{NGC\,3132} in the [\ion{N}{ii}]$\lambda$6584 emission line, made by simulating the action of a narrow filter (see main text for details). The orange and violet rectangles mark the area used to extract the spectra of the newly detected northern and southern arcs (see Sections \ref{seclinempas} and \ref{secarms}).
   Here, and in forthcoming maps, this image will be presented for reference with ten evenly spaced contours (in logarithmic scale), ranging from 1$\times$10$^{-18}$ erg~cm$^{-2}$~s$^{-1}$ (white) to 1$\times$10$^{-14}$
~erg~cm$^{-2}$~s$^{-1}$ (black).
}
   \label{apuntado}
    \end{figure}

The characteristics of the data utilised in this contribution, as well as the methodology to extract the line information, are presented in Sect. \ref{datared}. Then, an overview of the morphological appearance of the nebula according to flux maps in a range of emission lines is presented in Sect. \ref{seclinempas}. Following, Sect. \ref{secextin} discusses the derived extinction structure.
Maps of the physical properties are presented and discussed in Sect. \ref{sectene}. The ionic and element abundances are  derived in sections  \ref{secionic} and \ref{secelement}, respectively.
Although the spectral resolution of MUSE is low, some kinematic information can be derived: in Sect. \ref{seckinema}, this is compared to previous work and proposed models.
Sect. \ref{secarms} discusses the properties and nature of the two newly identified structures.
Sect. \ref{sec1Dmodel} contains the 1D photoionisation model based on the integrated fluxes of the strongest emission lines.
Finally, Sect. \ref{secconclu} summarises the main conclusions of this work and outlines perspectives for further studies. 

\section{The data \label{datared}}

\subsection{Observations and data reduction}

The Planetary Nebula \object{NGC\,3132} (\object{PN G272.1+12.3}) was observed as part of
commissioning run 2a (Bacon et al. 2014) of the Multi Unit Spectroscopic Explorer
(MUSE) instrument at the VLT on the night of 19 February 2014 and released by ESO to the community afterwards.
The standard Wide Field Mode was used, thus covering an area of $\sim60^{\prime\prime}\times60^{\prime\prime}$ per pointing with a spatial sampling of 0\farcs2, a wavelength coverage of 4750-9300 \AA\, with spectral sampling of 1.25~\AA\, and a typical spectral resolution of $\sim$2500.
Exposures were taken with t$_{exp}$=60~s at three pointings, with offsets between consecutive pointings of $\sim30^{\prime\prime}$ ($\sim$half of the MUSE field of view) in declination.
For each of the two pointings at the northern and southern edges, three exposures were performed at positions angles (PA) 0$^\circ$, 180$^\circ$ and 270$^\circ$. The central pointing was observed with a total of seven exposures (two at PA=0$^\circ$, three at PA=180$^\circ$, and two at PA=270$^\circ$).
With this strategy, an area of $\sim1^\prime\times2^\prime$ was mapped with a total integration time of either 180~s (at the edges) or 600~s (in the central $\sim60^{\prime\prime}\times60^{\prime\prime}$ area).
Observing conditions were clear and the seeing, as measured as the FWHM of the central star and a field star in the final reduced cube (see below), ranged from $\sim0\farcs7$ at 9\,100~\AA\, to $0\farcs8$ at 4\,800~\AA.

   \begin{figure*}
   \centering
   \includegraphics[angle=0,width=0.30\textwidth, clip=, viewport=60 130 300 520,]{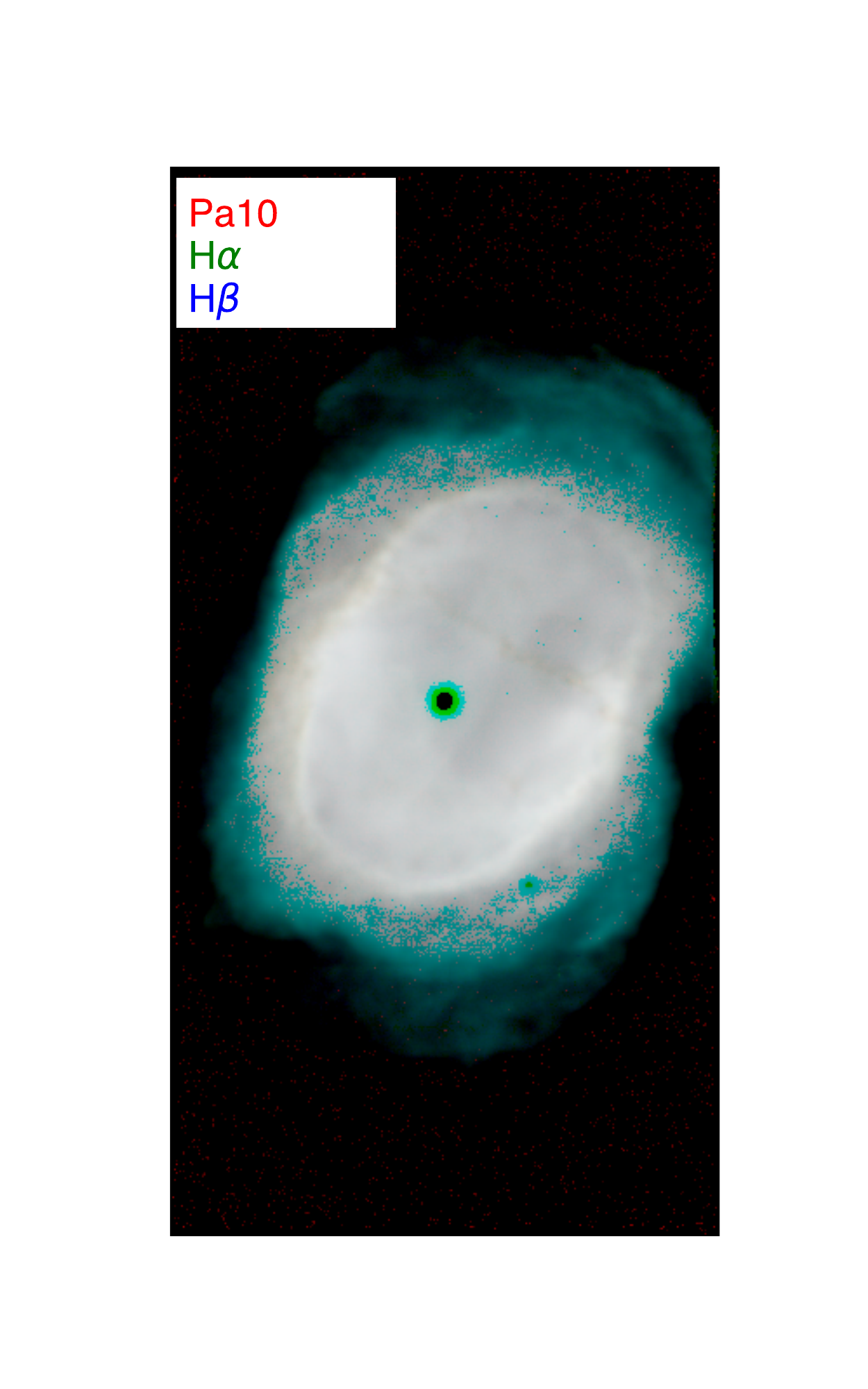}
   \includegraphics[angle=0,width=0.30\textwidth, clip=, viewport=60 130 300 520,]{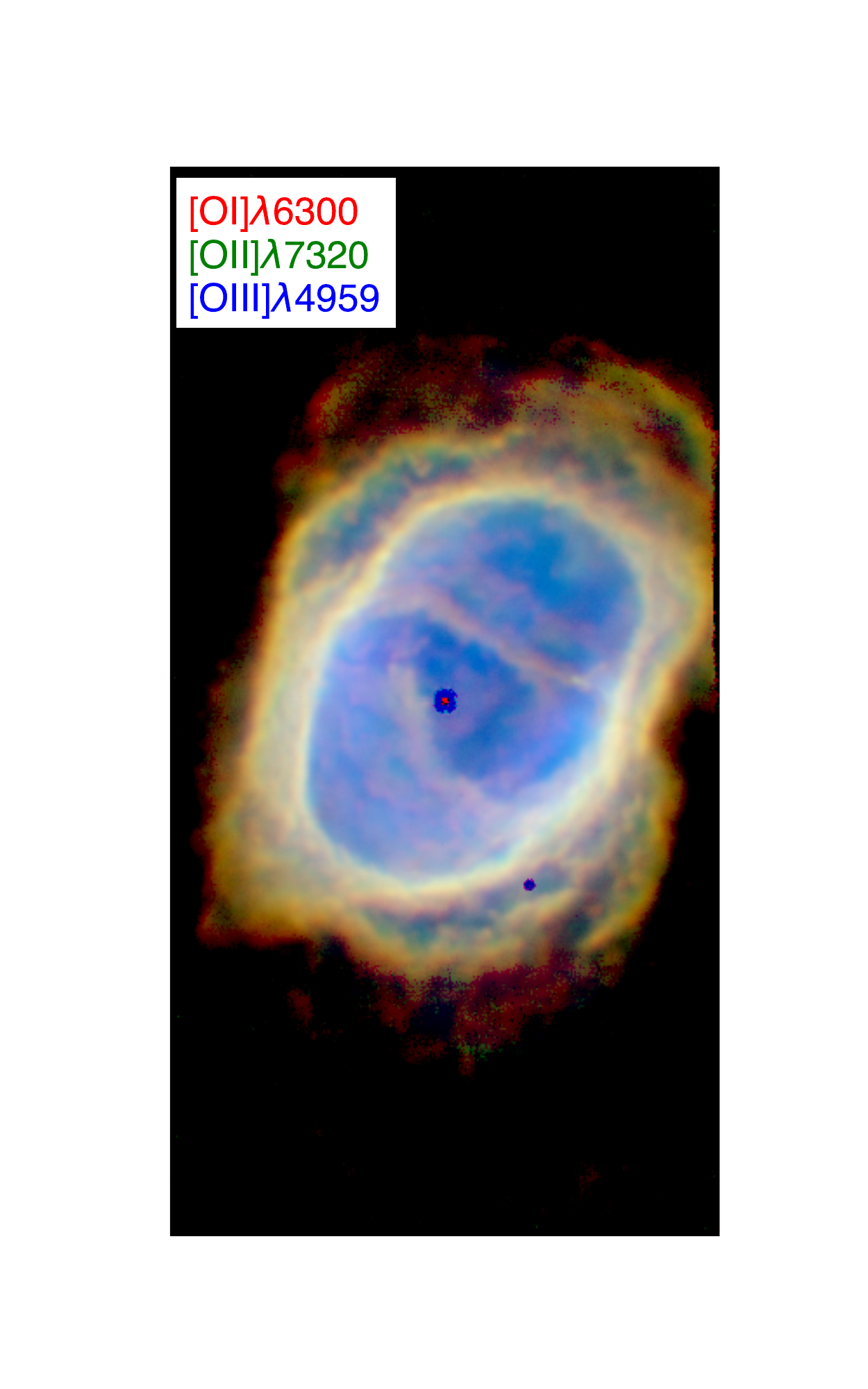}
   \includegraphics[angle=0,width=0.30\textwidth, clip=, viewport=60 130 300 520,]{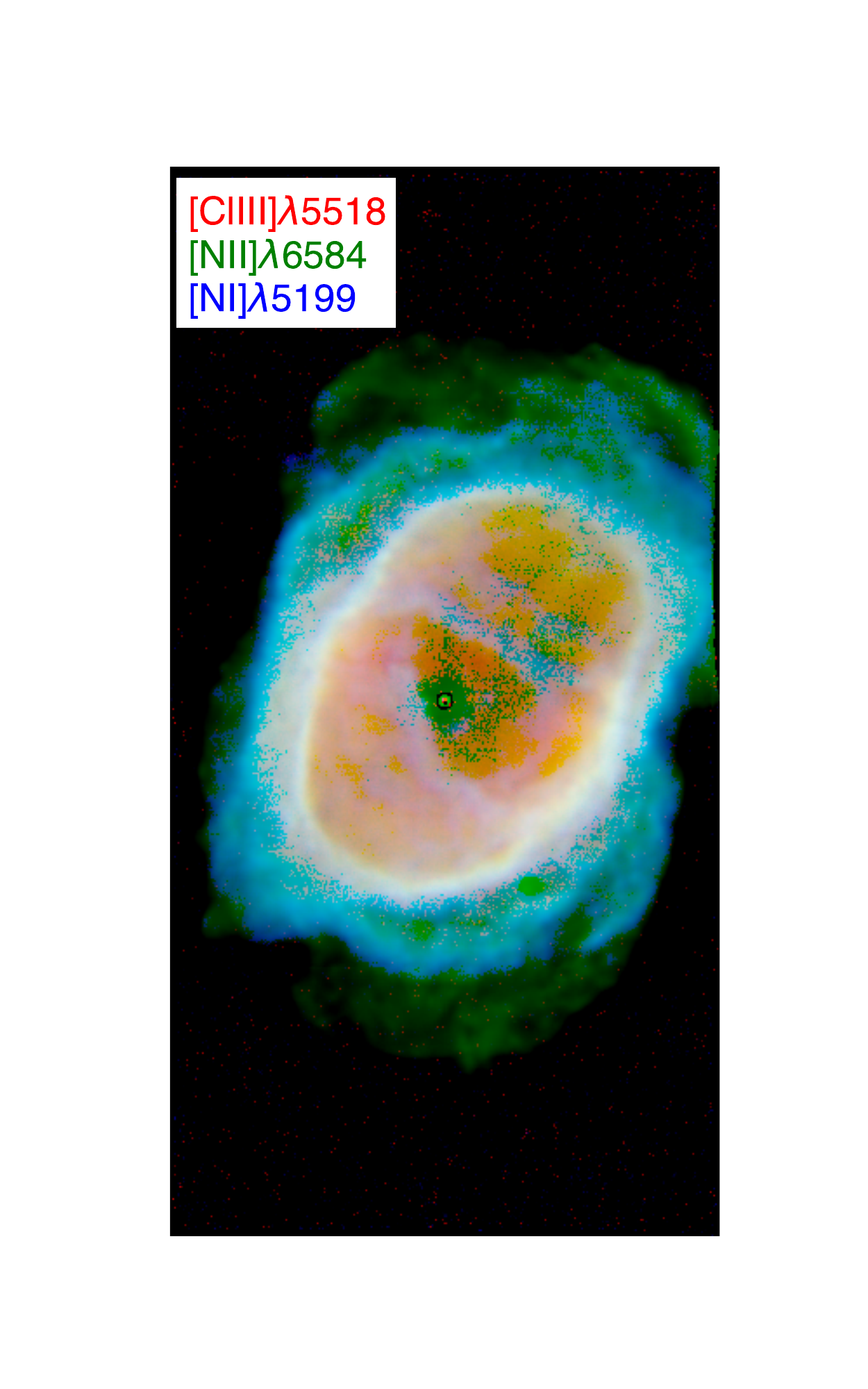}
   \includegraphics[angle=0,width=0.30\textwidth, clip=, viewport=60 130 300 520,]{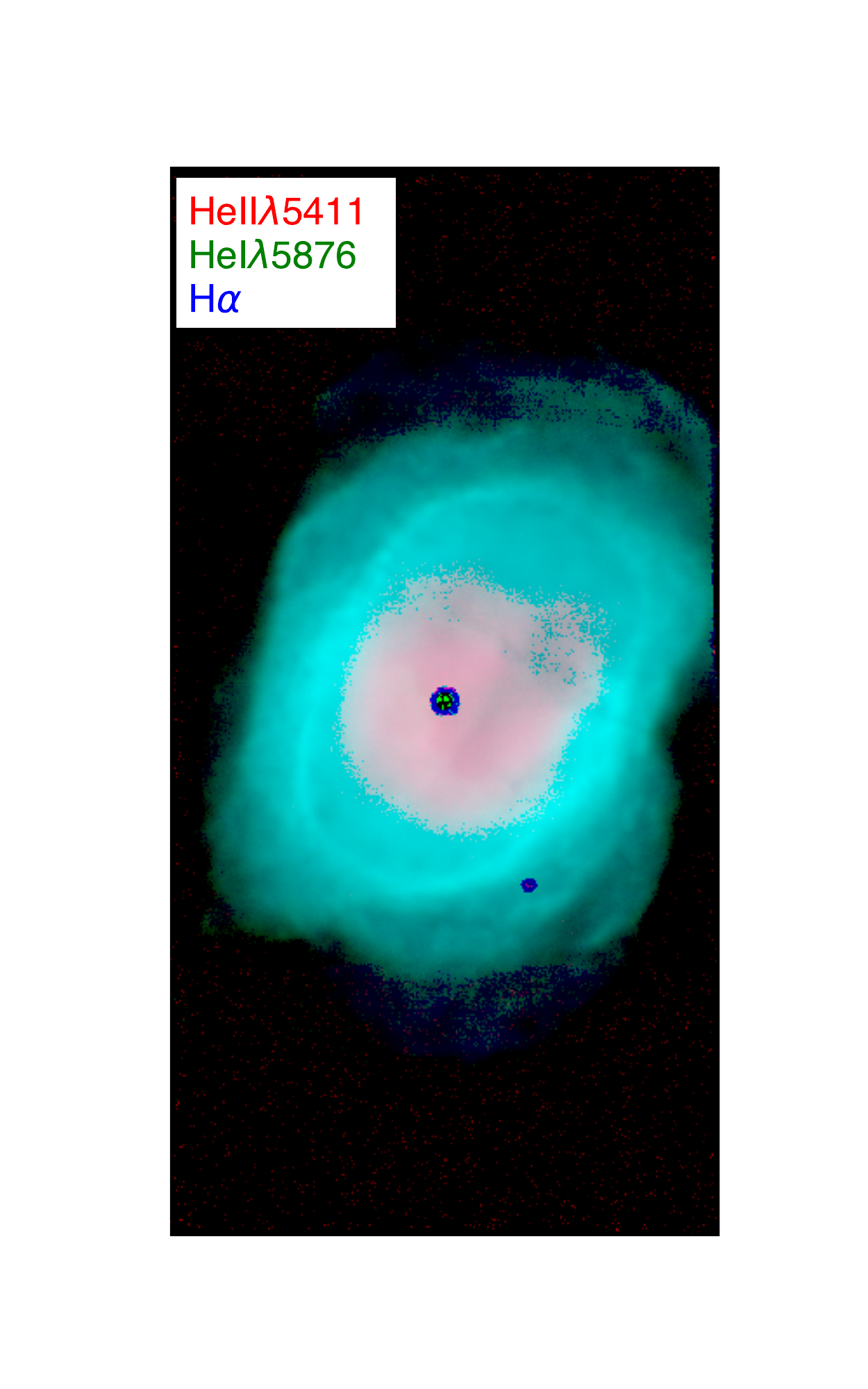}
   \includegraphics[angle=0,width=0.30\textwidth, clip=, viewport=60 130 300 520,]{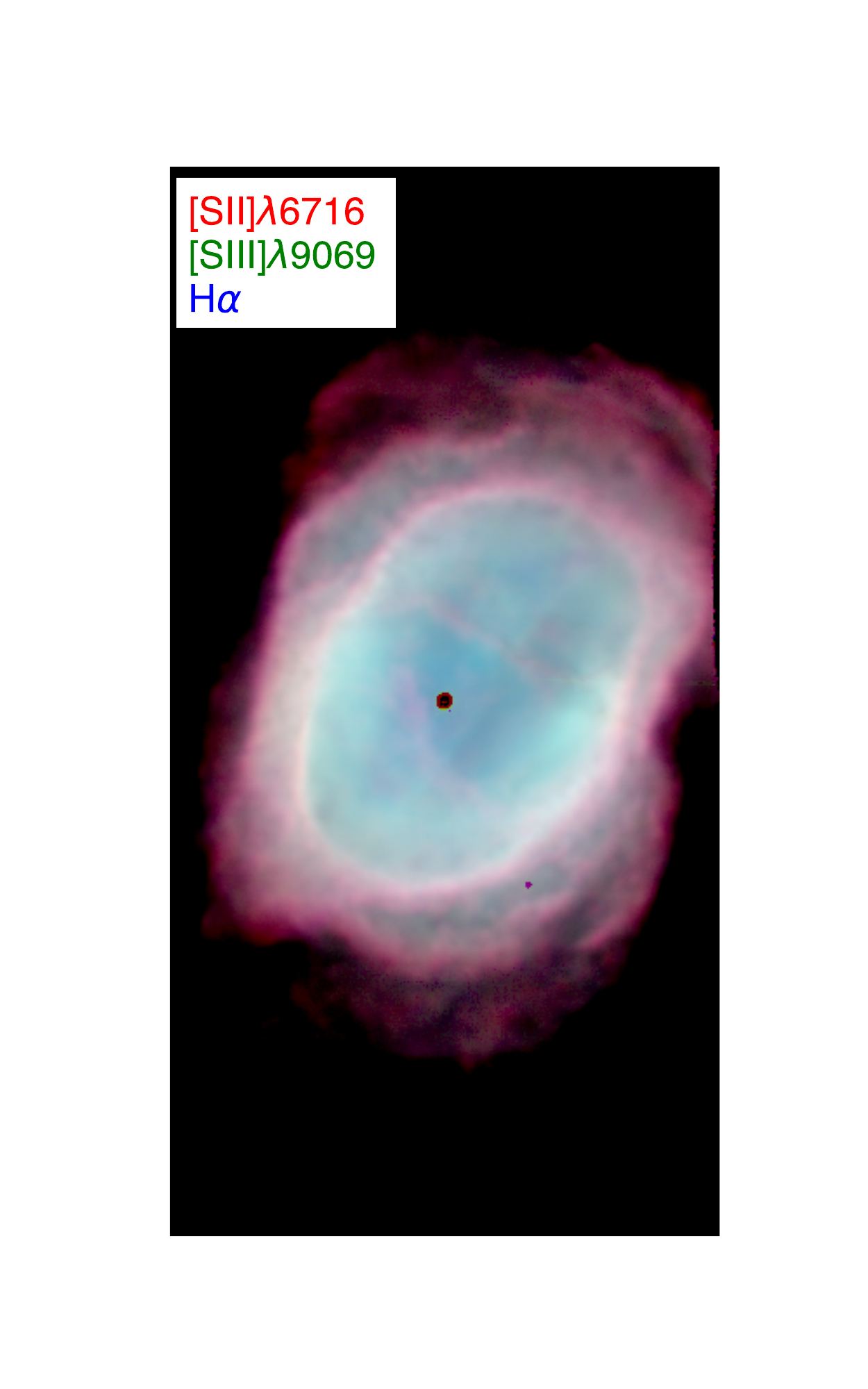}
   \includegraphics[angle=0,width=0.30\textwidth, clip=, viewport=60 130 300 520,]{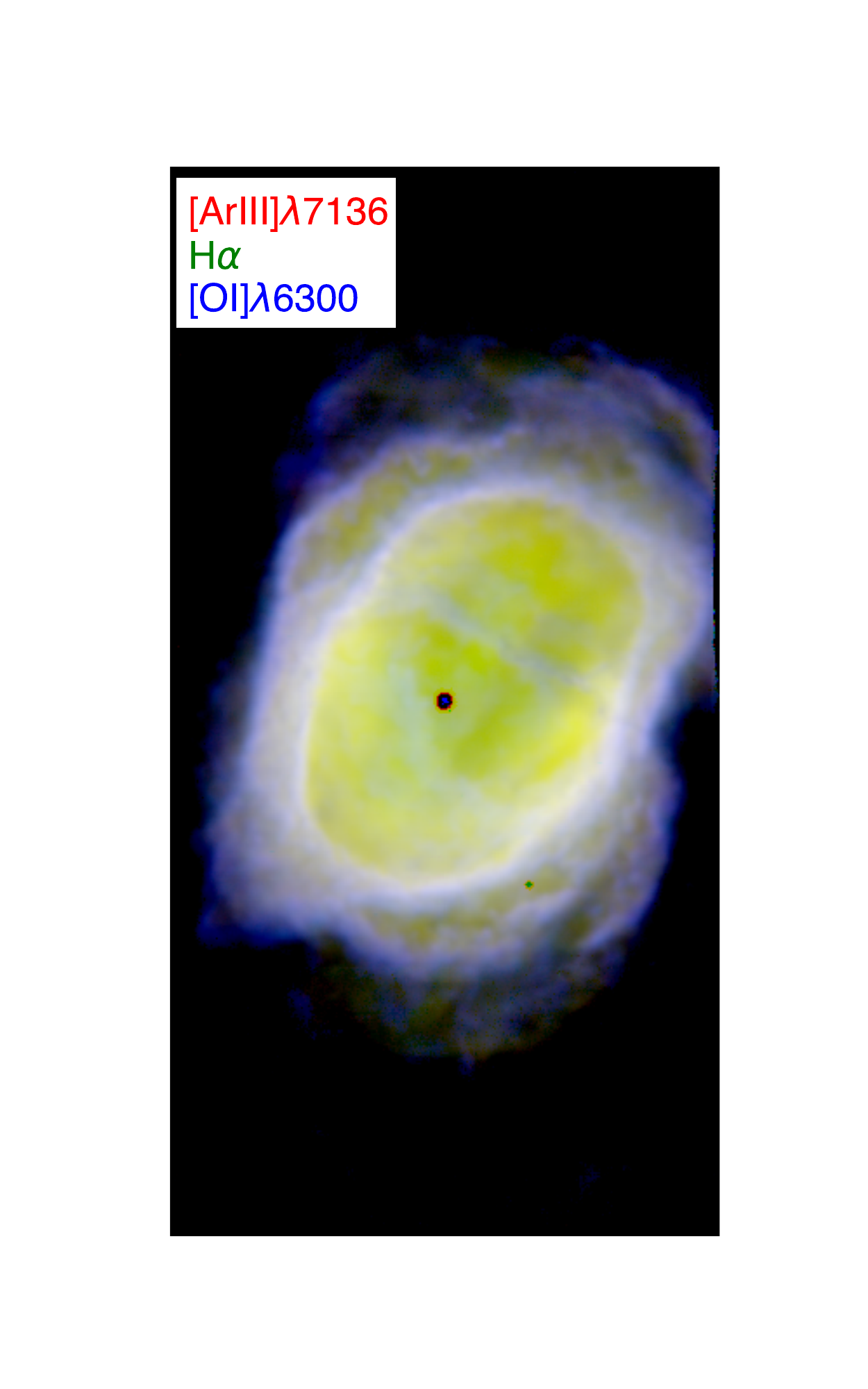}
   \caption{Set of colour composite images using fluxes of three emission lines.  The panels summarise in a synthetic manner the richness in ionisation structure. The emission lines used to create the individual images are listed in the upper left corner of each panel, coloured according to the corresponding RGB channel. North is up and east towards the left. 
   See text for a detailed description of the images. \label{panel3colores}}
    \end{figure*}

We reduced the data with the public MUSE Instrument Pipeline v1.6.1 \citep{Weilbacher14}  and EsoRex, version 3.12.3, using the delivered calibration frames (bias, flat field, arc lamp, and twilight exposures), the pipeline astrometric, and geometric files  for commissioning data, the vignetting mask and the corresponding bad pixels, atmospheric extinction and sky line tables.
The reduction for each individual frame includes bias subtraction, flat-fielding and slice tracing, wavelength calibration, twilight sky correction,  sky subtraction, and correction to barycentric velocities.
In the last step, individual exposures were flux calibrated - using data for the spectrophotometric standard HD\,49798 -, and corrected for differential atmospheric diffraction.
The MUSE pipeline also reconstructs a variance cube (\texttt{STAT}) and stores it in an extension in the same file.
All exposures for the PN were combined in a single file (i.e. extensions \texttt{DATA} and \texttt{STAT}) of 5.4~Gib. The final output data cube has $316\times615\times2681$ voxels. To have an idea of the mapped area, a reconstructed image in [\ion{N}{ii}]$\lambda$6584, created by integrating from $\lambda$6578 to $\lambda$6589 from the MUSE datacube, is displayed in Fig.~\ref{apuntado}.

\subsection{Line emission measurements}

Information  was derived on a spaxel-by-spaxel basis. We fitted the different spectral features needed for our analysis by Gaussians using the Python package \texttt{LMFIT}\footnote{https://lmfit.github.io/lmfit-py/}. Line width, $\sigma$, was always bounded to be between 1.05\,\AA\, and 1.40\,\AA. 
For strong and isolated lines, we created images simulating the action of a narrow filter, which were used as initial conditions for the flux (see Fig \ref{apuntado} for the corresponding image for [\ion{N}{ii}]$\lambda$6584). For fainter or blended lines, a scaled version of one these images (e.g., one tracing the same ion, or an ion with similar ionisation potential) was used instead. The initial condition for the H$\beta$ central wavelength was set to its rest-frame value, reasonable to ensure fit convergence for the expected velocities in Galactic objects. For the other lines, results for a stronger line of the same ion, when available, or for H$\beta$, otherwise, was used. Most of the lines were independently fitted. The exceptions were: i) H$\alpha$, that was fitted together with  the [\ion{N}{ii}]$\lambda\lambda$6548,6584 doublet, and assuming the same width for all the lines; ii) the [\ion{S}{ii}]$\lambda\lambda$6717,6731 and [\ion{Cl}{iii}]$\lambda\lambda$5518,5538 doublets, each of them fitted using the same width for both lines; iii) the [\ion{O}{i}]$\lambda\lambda$6300,6364 doublet, fitted together with the [\ion{S}{iii}]$\lambda$6312 line. The statistical errors for each voxel delivered by the MUSE pipeline were used to weight the data to be fitted.  In all the fits we modeled the local underlying nebular continuum as a one-degree polynomial. The tables with the fitted line fluxes, widths and central wavelengths (and its translation to velocities) as well as the corresponding errors were then reformatted to 2D arrays that were stored as FITS files. These will be referred to throughout the manuscript as both images and maps.

Additionally, maps for the continuum in three spectral windows (from  blue to red: 4880-4920~\AA, 6800-6840~ \AA, and 9180-9220~\AA) were also created. These were used to estimate the effective spatial resolution quoted above. Finally, maps of the corresponding standard deviation within those same windows were also made.  These were used to establish a threshold for reliable emission line measurement. In order to minimize the inclusion of bad fits results, line flux images were cleaned by masking those spaxels satisfying at least one of the following criteria: i) an error for the centroid of the line  $\ge$0.2~\AA; ii) a value for flux error in the continuum equal or larger than half the value for the line flux; iii) a velocity smaller than $-60$~km~s$^{-1}$ or larger than $+60$~km~s$^{-1}$; iv) a flux smaller than 10$^{-18}$~cm$^{-2}$~erg~s$^{-1}$.

\section{Presentation of the emission line maps \label{seclinempas}}

In Fig. \ref{apuntado}, we present the map for the \nii$\lambda$6584 emission line, the richest one in terms of structure. The image displays an inner bright elliptical rim, and several secondary inner structures (e.g., two lanes of enhanced surface brightness). %
The basic geometrical parameters for the rim can be recovered by selecting the brightest spaxels and  performing a principal component analysis, since this is basically a translation and  rotation of the coordinate system. We tried several thresholds in \nii$\lambda$6584 flux to delineate the rim. The position angle (P.A.) determined in this manner was always around $\sim-22^{\circ}$ (ranging between $-20^{\circ}$ and $-27^{\circ}$). The derived ellipticity was $b/a \sim 0.70$, in good agreement with previous characterisations of  the morphology of the nebula \citep[e.g.][]{Juguet88} .
On its side,
the derived centre of the nebula was always towards the SW of the bright primary star (HD 87892), in the direction of the ionising white dwarf.
The specific position depended on the selected threshold in flux, ranging typically between $\sim$1\farcs3 and  $\sim$3\farcs1.
As a reference, for f(\nii$\lambda$6584) $>2\times10^{-14}$~cm$^{-2}$~erg~s$^{-1}$, the centre was at $\sim$1\farcs6, roughly coincident with the position of the ionising star.

Fig. \ref{apuntado}  also shows that the nebula extends well beyond the bright rim, presenting an external doily-like shell and two low-surface brightness arcs at the northern and southern tips of the nebula.
The dynamical range covered in this image is extremely high with about three orders of magnitude in surface brightness from the faintest  (i.e., the arcs) to the brightest (i.e., the rim) structures. 
To our knowledge, the presence of these two faint structures has not been reported before.
On a spaxel-by-spaxel basis, these are actually only detected in \nii\, and, to a much lower extent, \ha\, and \sii\, (not shown). These detections suggest that they are some kind of low-ionisation structures \citep[LIS, see][for a review]{Gonsalves04}.
In Sect. \ref{secarms}, we will characterise  them by analysing their integrated spectra.

   \begin{figure}
   \centering
   \includegraphics[angle=0,width=0.36\textwidth, clip=, viewport=30 0 540 550,]{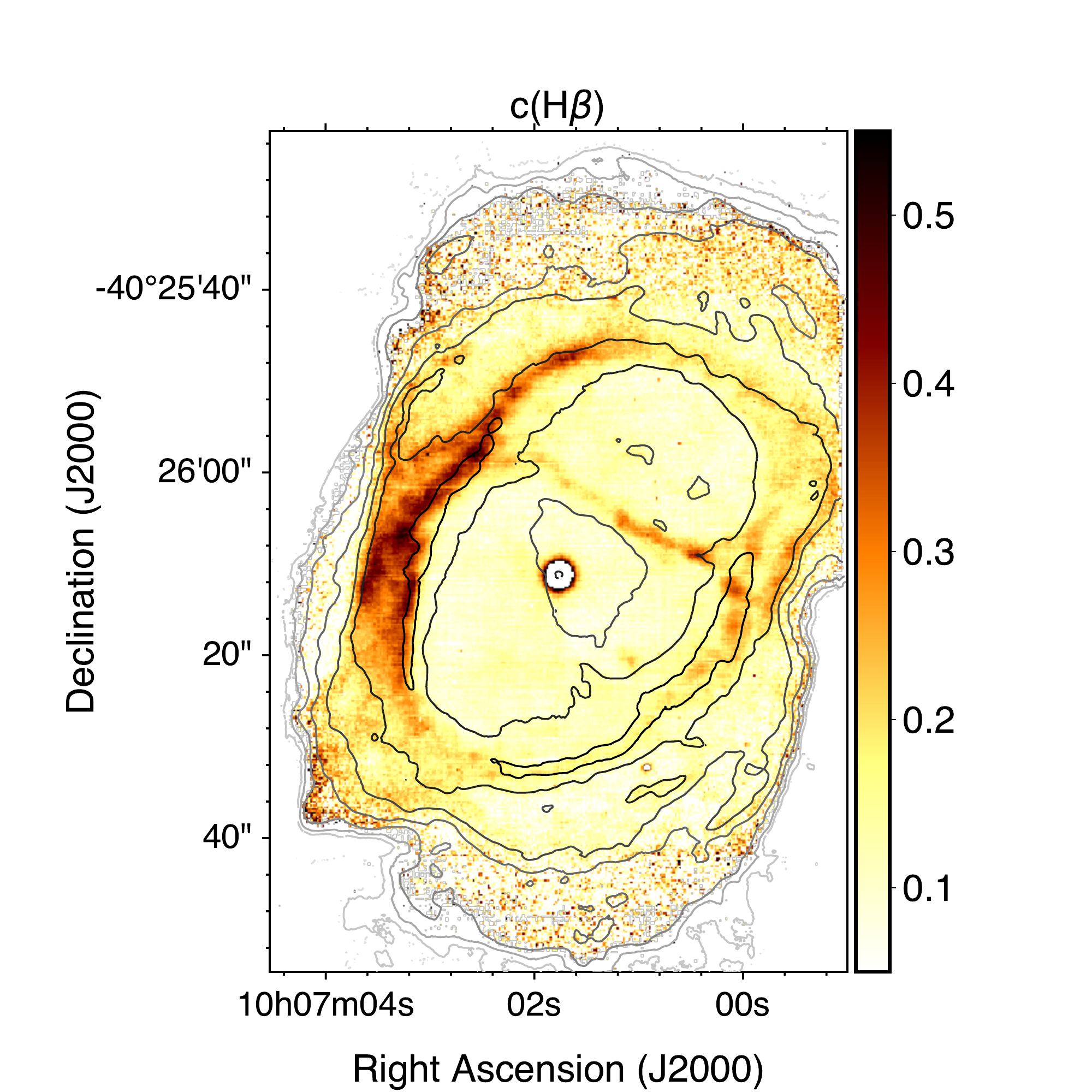}
   \caption{Log extinction at \hb, c(\hb), for \object{NGC\,3132} using \ha\, and \hb\, recombination lines, assuming \ha/\hb=2.89, and the \citet{Cardelli89} extinction law. The estimated median uncertainty was $\sim$0.04, with 90\% of the uncertainties ranging between 0.02 and 0.18.}
   \label{mapchb}
    \end{figure}

Even if  Fig. \ref{apuntado} is rich in structure, it does not transmit by itself the degree of complexity inherent to a PN: given the complex ionisation stratification in PNe, different structures are better traced by different emission lines. 
Fig.  \ref{panel3colores} illustrates this in a synthetic way. This panel contains several colour composite images using flux maps for different emission lines. To better emphasise the structure in the main body of the nebula, the colour tables have been selected such that the low-surface brightness northern and southern arcs, seen in the \nii\, image in Fig. \ref{apuntado}, are not visible (and in any subsequent map in this contribution).

The image in the upper left corner of Fig. \ref{panel3colores} was made using three hydrogen recombination lines, with the R/G/B channel allocated for the reddest/central/bluest line, in such a way that areas with higher extinction would appear reddish. Both the rim and a lane crossing the northern side of the nebula are slightly reddish, indicating some extinction. This will be quantified in Sect. \ref{secextin}.
The other two images in the upper row contains maps of ions with very different ionisation degree. The one in the centre was made using lines -- from R to B channel -- of neutral, singly ionised and doubly ionised oxygen, while the one on the right contains the neutral (B) and singly ionised (G) nitrogen, and doubly ionised chlorine (R). Both images trace a similar structure. Even though the ionisation potentials for [\ion{Cl}{iii}] and   [\ion{O}{iii}] are comparable, the area of high ionisation as traced by [\ion{O}{iii}] extends well beyond that traced by [\ion{Cl}{iii}], since the $\lambda$5518 emission line is much weaker than the $\lambda$4959 emission line.
Images in the lower row always contain a map in \ha\, for reference. The one on the left summarises the emissivity distribution in the main recombination lines. Since \ha\, is found all over the nebula, those locations with an overabundance of neutral helium, should this exist at all, would have been seen in dark blue.
This colour is absent in this panel.
Instead most of the nebula is coloured in different shades of cyan (=B+G) indicating that singly ionised hydrogen and helium occupy a similar extension. Still, an inner circular area of about $32^{\prime\prime}$ in diameter is displayed in magenta (=B+R), indicating the presence of double ionised helium. The sharp transition between the magenta and cyan areas indicates that the nebula is ionisation bounded in \heii.
The other two figures display a structure similar to that for the three oxygen lines but with softer contrast.

%
\begin{table*}[th!]
\caption{Extinction, electron temperature and density statistics. Number of spaxels, mean and standard deviation were calculated for the data between the 5 and 95 percentiles. \label{TabTeNe}} 
\label{tab_TeNe}      
\centering                          
\begin{tabular}{lcccccccc}        
\hline\hline                 
 & N spaxels & 5\% & Q1 & median & Q3 & 95\% & mean & $\sigma$\\
\hline 
c(\hb) (\ha/\hb) &  89663 &     0.05 &     0.10 &     0.14 &     0.19 &     0.33 &     0.16 &     0.08 \\
Uncert. &   &      0.02 &     0.03 &     0.04 &     0.07 &     0.18 &     0.06 &     0.05 \\
c(\hb) (\ha/Pa9)   &  53264 &     0.06 &     0.14 &     0.20 &     0.27 &     0.41 &     0.23 &     0.28 \\
Uncert. &   &      0.06 &     0.07 &     0.09 &     0.12 &     0.19 &     0.10 &     0.04 \\
c(\hb) (\ha/Pa10) &  50225 &     0.06 &     0.15 &     0.22 &     0.29 &     0.43 &     0.25 &     0.32 \\
Uncert. &   &      0.06 &     0.08 &     0.10 &     0.14 &     0.21 &     0.12 &     0.05 \\
c(\hb) (\ha/Pa11)  &  39250 &     0.02 &     0.08 &     0.14 &     0.22 &     0.39 &     0.19 &     0.39 \\
Uncert. &   &      0.07 &     0.09 &     0.12 &     0.16 &     0.23 &     0.13 &     0.05 \\
c(\hb) (\ha/Pa12)  &  28862 &     0.02 &     0.07 &     0.13 &     0.20 &     0.35 &     0.17 &     0.41 \\
Uncert. &   &      0.08 &     0.10 &     0.13 &     0.17 &     0.23 &     0.14 &     0.05 \\
\hline
$N_e$(\nii,\sii)  (cm$^{-3}$)       &  77908 &      151 &      256 &      339 &      447 &      666 &      364 &      154 \\
Uncert. &   &         8 &       20 &       34 &       52 &       96 &       40 &       30 \\
$N_e$(\cliii)  (cm$^{-3}$)  &  32384 &      124 &      544 &     1048 &     1724 &     3410 &     1309 &     1146 \\
Uncert. &   &      136 &      292 &      429 &      627 &     1144 &      510 &      362 \\
\hline
$T_e$(\nii,\sii) (K)     &  77915 &     9198 &     9528 &     9758 &     9981 &    10527 &     9790 &      412 \\
Uncert. &   &    23 &    57 &     94 &    148 &    309 &    121 &    114 \\
$T_e$(\siii)  (K)    &  57738 &     8897 &     9210 &     9528 &     9909 &    10640 &     9613 &      555 \\
Uncert.  &   &    76 &   108 &    143 &    195 &    314 &    162 &     84 \\
$T_e$(\oi)  (K)  &   2418 &     8314 &     8664 &     8925 &     9210 &     9717 &     8949 &      413 \\
Uncert.  &   & 78 &   141 &    182 &    226 &    302 &   187 &     70\\
$T_e$(\hei) (K)  &  42596 &     4476 &     6218 &     7890 &    10395 &    12953 &     8322 &     2660 \\
Uncert. &  &   266 &   375 &    504 &    771 &   2973 &    821 &    881 \\
$T_e$(PJ) (K)     &  48940 &     3460 &     5373 &     6632 &     8508 &    12092 &     7094 &     2536 \\
Uncert.          &   &      818 &     1318 &     2005 &     2652 &     3328 &     2008 &      817 \\
\hline                                   
\end{tabular}
\end{table*}

\section{Mapping the extinction structure \label{secextin}}

 
Extinction was derived using the \texttt{RedCorr()} method in \texttt{PyNeb}\footnote{http://www.iac.es/proyecto/PyNeb/}, assuming an intrinsic Balmer emission
line ratio of \ha/\hb = 2.89 \citep{Osterbrock06}, which is the mean expected value for $T_e$'s between 6\,000 and 12\,000~K and $N_e$'s between 200 and 1200~cm$^{-3}$. Variations of this ratio within this range are always $\lesssim$1\%.
The assumed extinction curve was that provided by \cite{Cardelli89} with $R_V=3.1$. 
The obtained c(\hb) map is presented in Fig. \ref{mapchb}.
Errors were computed using a Monte Carlo approach with 50 trials and assuming that the errors on both lines followed a normal distribution.

Additionally, similar maps were created by comparing \ha\, with other hydrogen recombination lines (Paschen 9\ldots12). The obtained reddending maps were comparable to that shown in Fig. \ref{mapchb}, but with somewhat greater uncertainties.

About 90\% of the spaxels have c(\hb)$<$0.32. In particular, most of the surface of nebula presents relatively low reddening with median value of c(\hb)=0.14, corresponding to  \ebv=0.09. This suggests that this extinction is caused by foreground material, and not intrinsic to the nebula.
For the assumed distance here of $\sim$864~pc  \citep{Kimeswenger18}, \citet{Capitanio17} reports an \ebv$\sim$0.05$\pm$0.03 along this line of sight.
Our estimated foreground extinction is slightly larger, but consistent with this value, allowing for the existence of plausible inhomogeneities in the foreground ISM, smaller than the resolution of the 3D reddening maps.

Even if most of the nebula has a uniform extinction distribution, Fig. \ref{mapchb} shows that there is actually considerable structure, with non-negligible areas of much higher extinction (e.g., over the inner northern lane). The rim displays a particularly high extinction reaching values of up to c(\hb)$\sim$0.50, just beyond its eastern side.
Reddening is expected to be higher in regions with higher content in molecular gas. In that sense, it would be interesting to compare the map presented in Fig. \ref{mapchb} with mid-IR images tracing H$_2$. Mid-IR spectroscopy with
Spitzer/IRS has revealed both H$_{2}$ line emission and Unidentified Infrared Emission  features or bands (UIEs or UIBs), believed to be caused by PAHs and related species \citep{Draine07}\footnote{UIBs or UIEs are often directly referred in the literature as the PAH bands or PAH emission, as in the refered work by \citet{Mata16}. Here, we will use the acronym UIBs to refer to the spectroscopic features and reserve the acronym PAHs to refer to the molecules
themselves, to avoid ambiguities.}, in the main body of the nebula \citep{Mata16}.
\citet{Hora04} presented images with IRAC on board of \emph{Spitzer} in its four available bands for \object{NGC\,3132}. They are sensitive to different H$_2$ line emissions (IRAC 4.5 $\mu$m, 5.8 $\mu$m, and 8 $\mu$m bands), and several UIBs (IRAC 3.6 $\mu$m, 4.5 $\mu$m, 5.8 $\mu$m bands). In spite of the differing spatial resolution (higher in the optical image), a comparison of Fig. \ref{mapchb} with their figures 1 and 2 show how the regions within the nebula with high extinction also display high surface brightness in the IRAC bands, as expected.

2D maps of extinction in PNe from optical recombination lines are rare. A similar but less extreme result was also found recently for \object{NGC\,7009} \citep{Walsh16}. Actually, the level of structure, as well as the covered range in reddening in \object{NGC\,3132}, is much higher than that found for  \object{NGC\,7009}.
Large small-scale variations in reddening in other PNe determined using other IFUs (e.g. \object{NGC\,5882}, \citet{Tsamis08}), or other techniques (e.g. \object{NGC\,7027}, \citet{Woodward92};  \object{NGC\,40}, \citet{LealFerreira11}) have also been reported.
In particular, evidence for peripheral enhancements of extinction have been reported in e.g., \object{NGC\,2346} \citep{Phillips00}.
In view of these examples, it is tempting to speculate that many PNe possess internal dust and hence display reddening inhomogeneities. Further detailed 2D extinction maps in other PNe would be desirable to address this issue. 

\begin{figure*}
   \centering
   \includegraphics[angle=0,width=0.32\textwidth, clip=, viewport=30 0 540 550,]{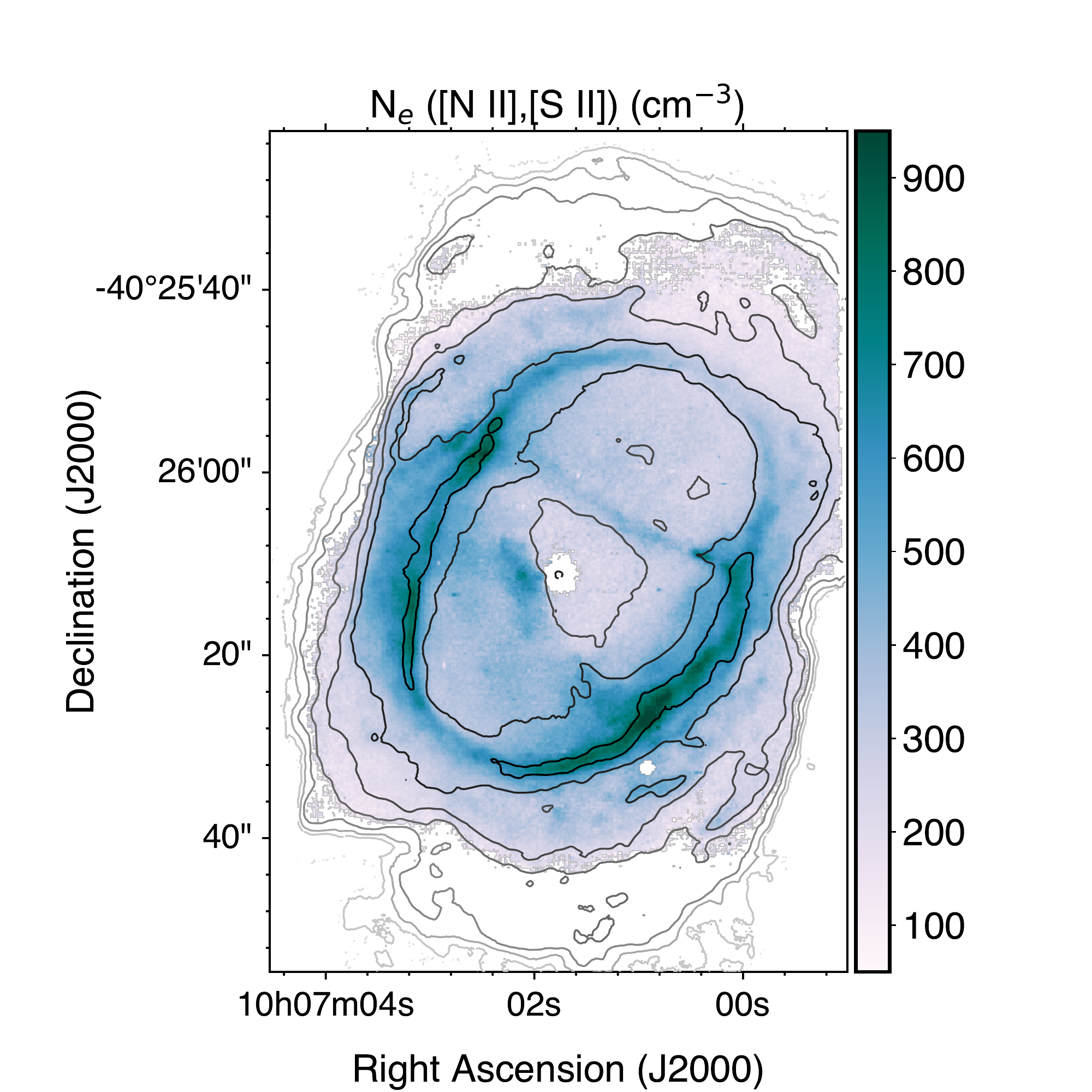}
   \includegraphics[angle=0,width=0.32\textwidth, clip=, viewport=30 0 540 550,]{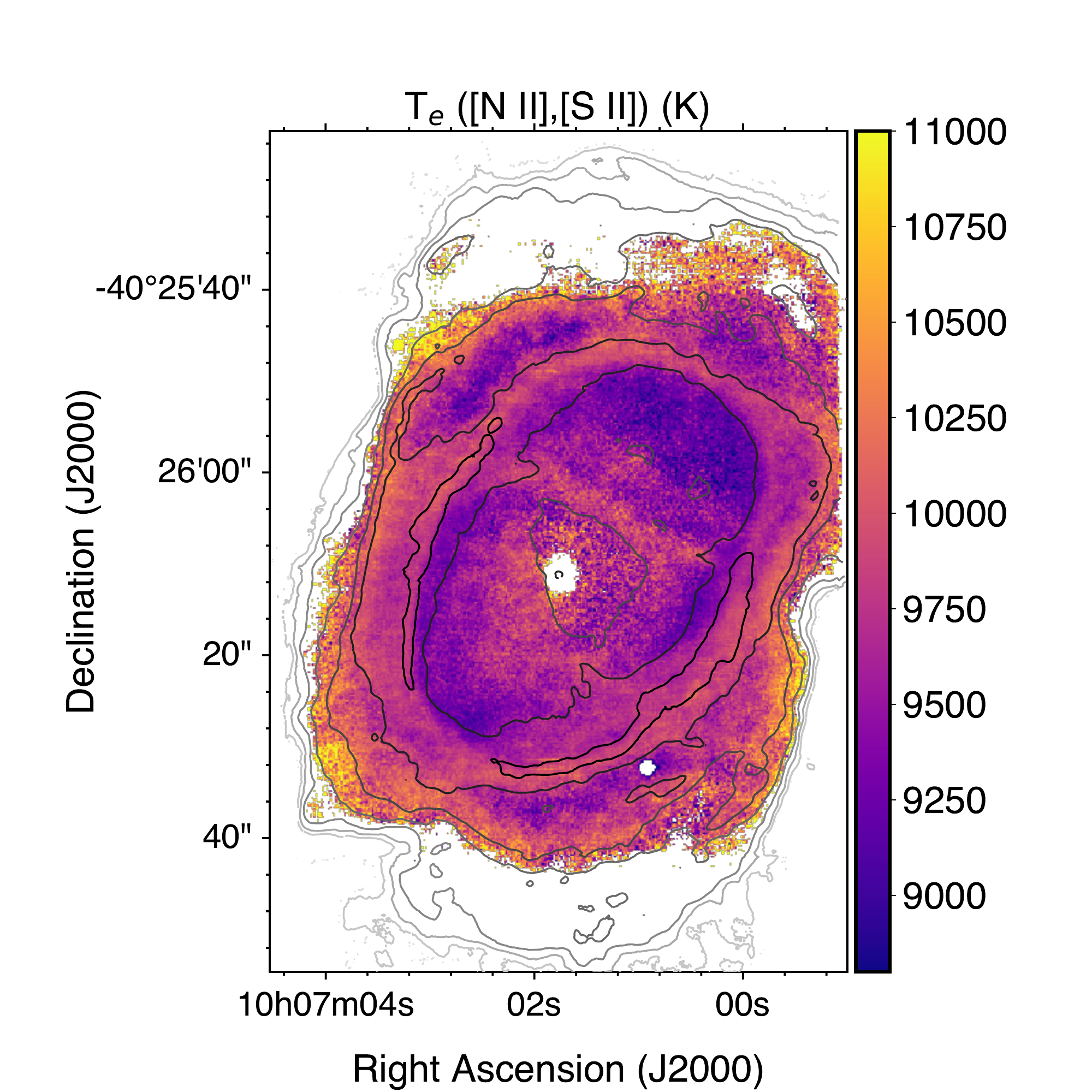}
 \includegraphics[angle=0,width=0.32\textwidth, clip=, viewport=30 0 540 550,]{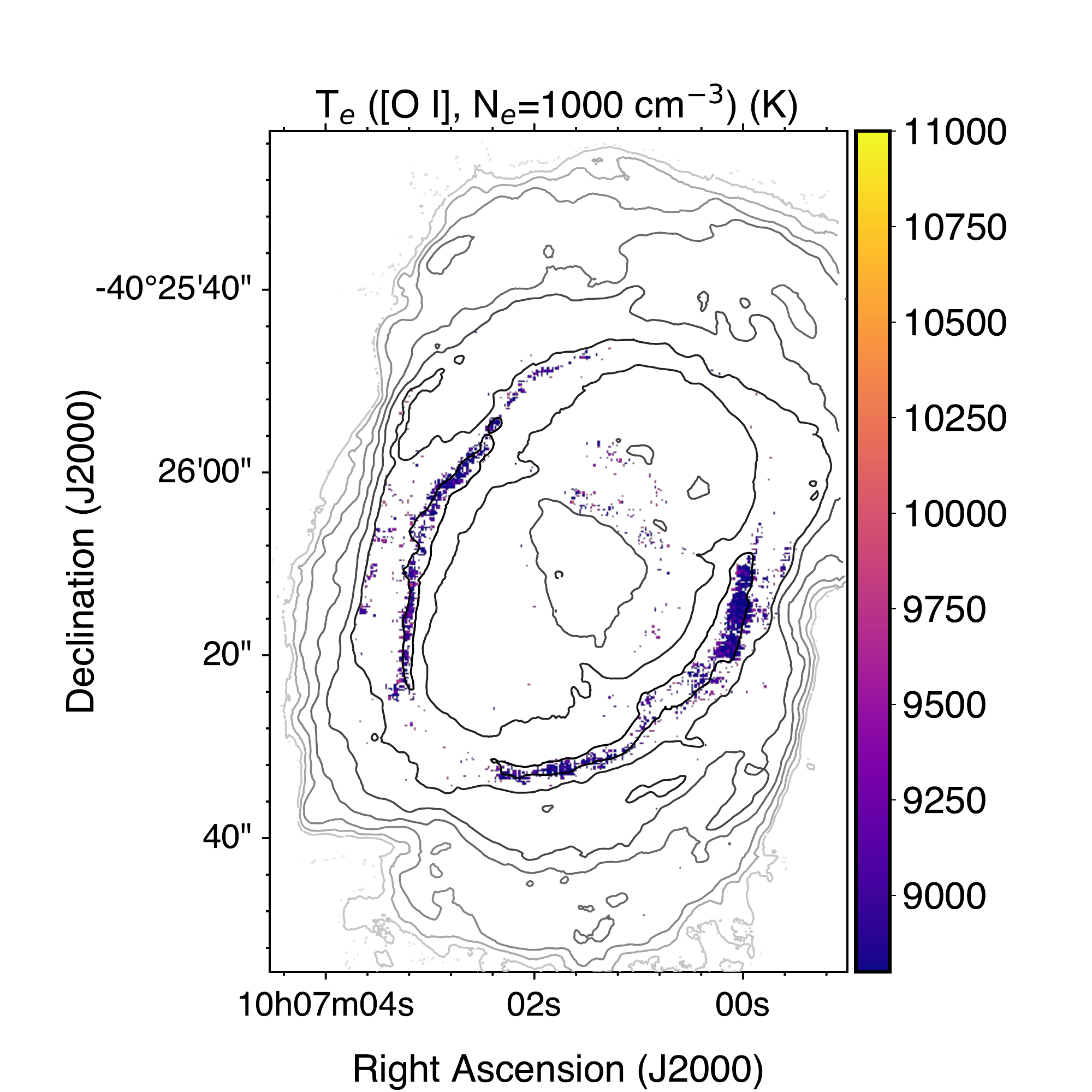}
   \includegraphics[angle=0,width=0.32\textwidth, clip=, viewport=30 0 540 550,]{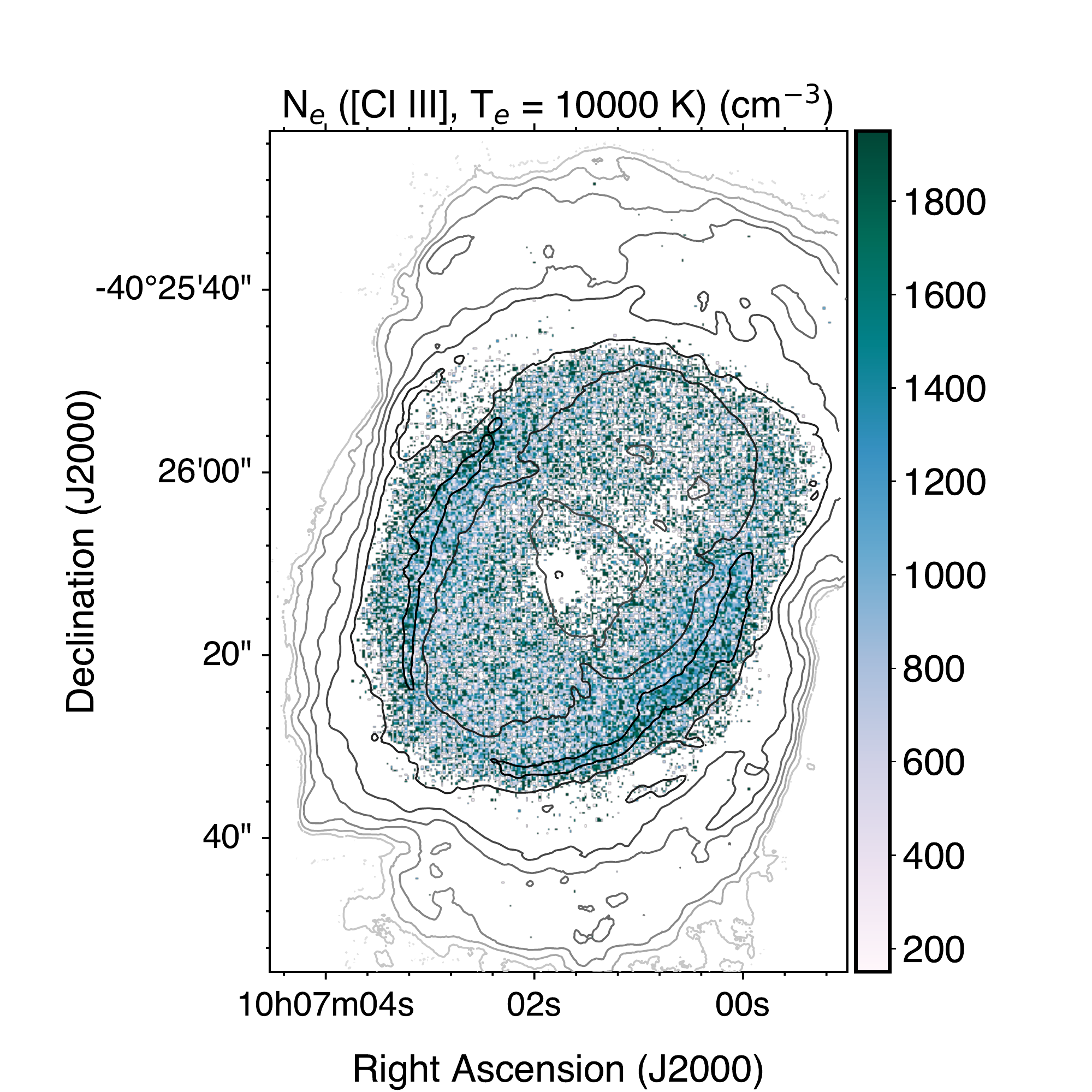}
       \includegraphics[angle=0,width=0.32\textwidth, clip=, viewport=30 0 540 550,]{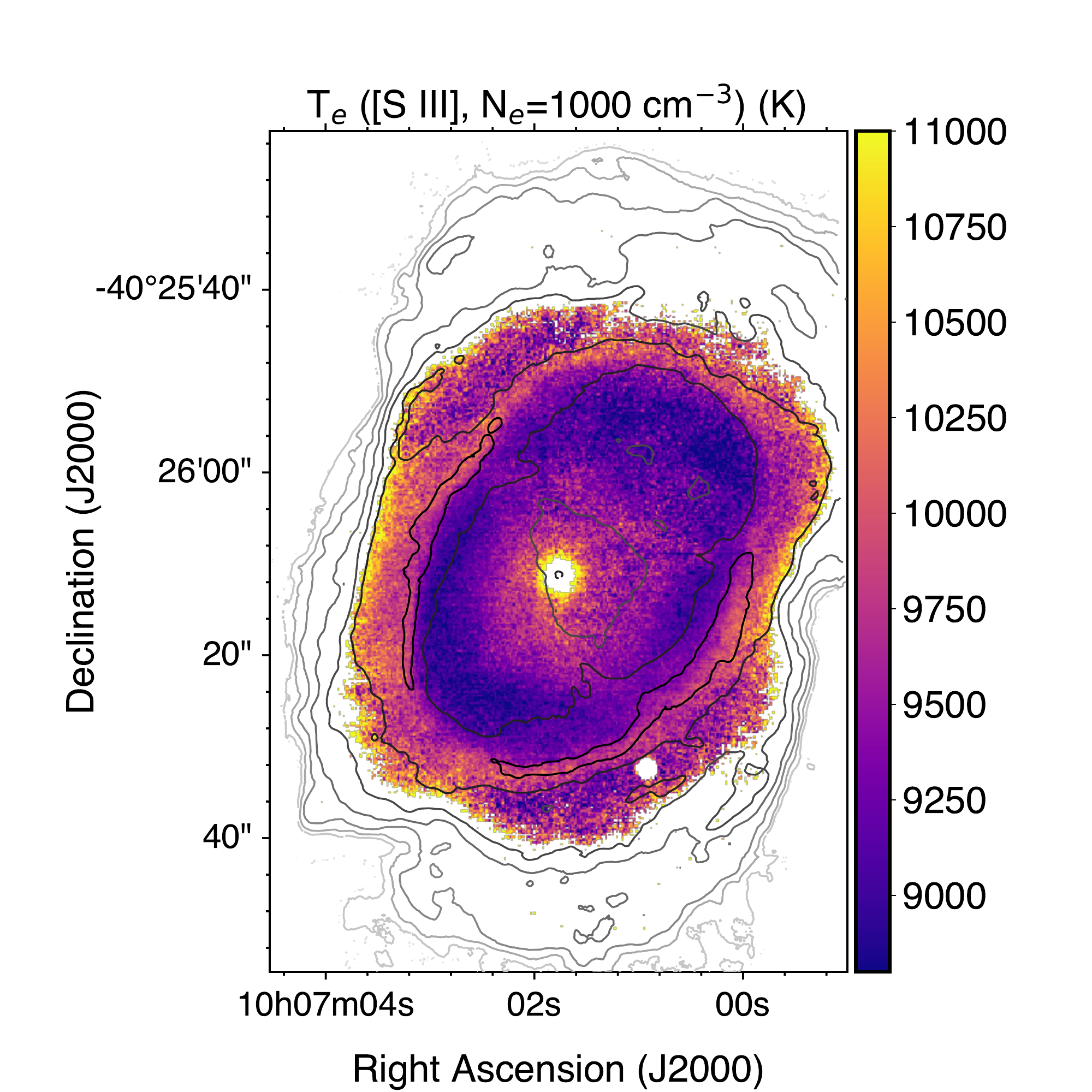}
    \includegraphics[angle=0,width=0.32\textwidth, clip=, viewport=30 0 540 550,]{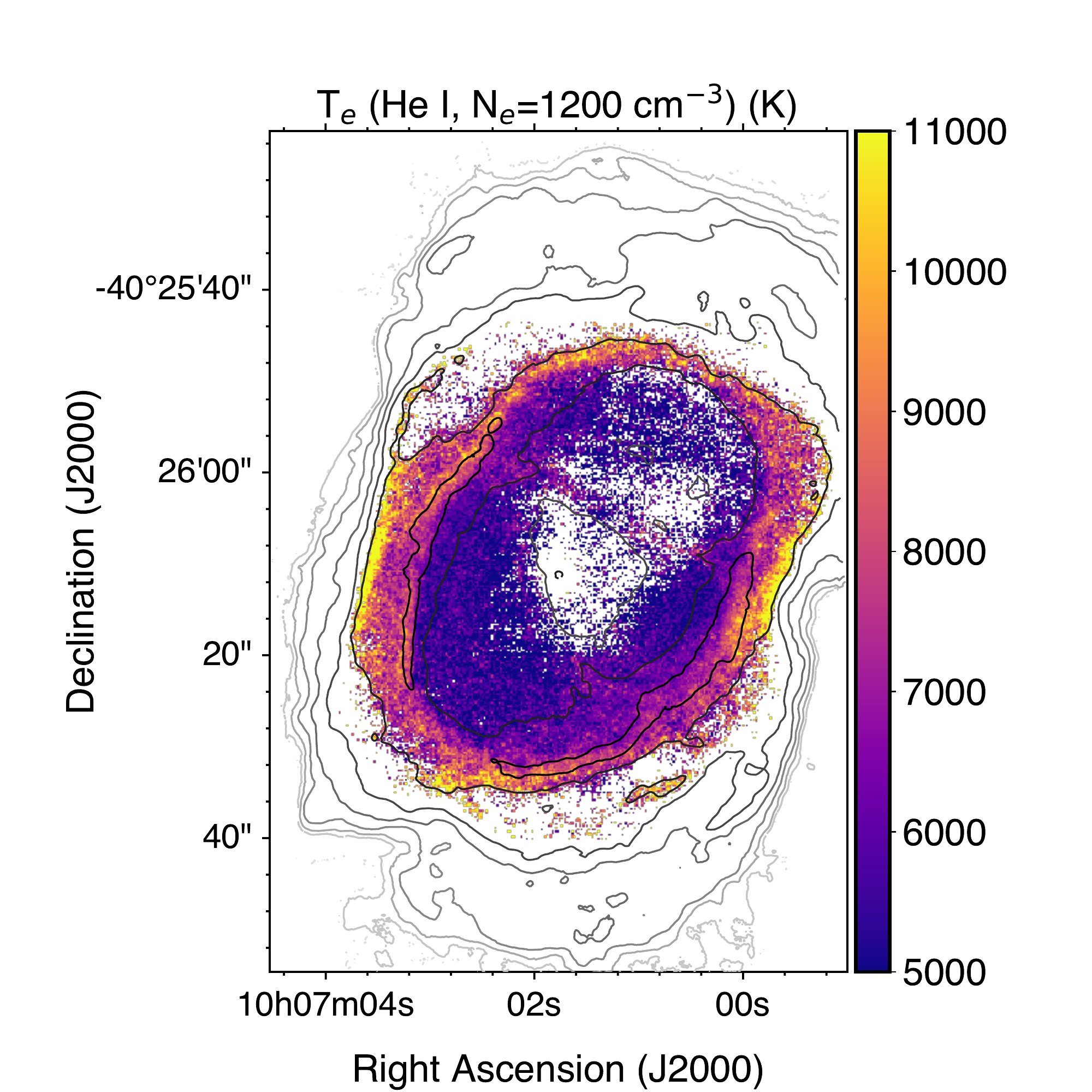}
   \caption{\emph{Left column:} Maps of $N_e$ determined from \sii\, and \nii\, lines (\emph{upper}) and \cliii\, and a $T_e$=10\,000 ~ K (\emph{lower}).
    The electron densities for collisional de-excitation of the \sii\, and \cliii\, $^2D_{3/2}$ levels for these diagnostics are 3.1$\times$10$^3$ and 2.4$\times$10$^4$ cm$^{-3}$ at 10\,000 K respectively.
    \emph{Centre column:} The maps of $T_e$ determined from \sii\, and \nii\, lines (\emph{upper}) and \siii\, and a $N_e$=1\,000 cm$^{-3}$  (\emph{lower}).
    \emph{Right column:} Maps of $T_e$ for the \oi\, lines (\emph{upper}) and the \hei\, recombination lines (\emph{lower}).
    }
   \label{mapNeTe}
    \end{figure*}

   \begin{figure*}
   \centering
   \includegraphics[angle=0,width=0.32\textwidth, clip=, viewport=30 0 540 550,]{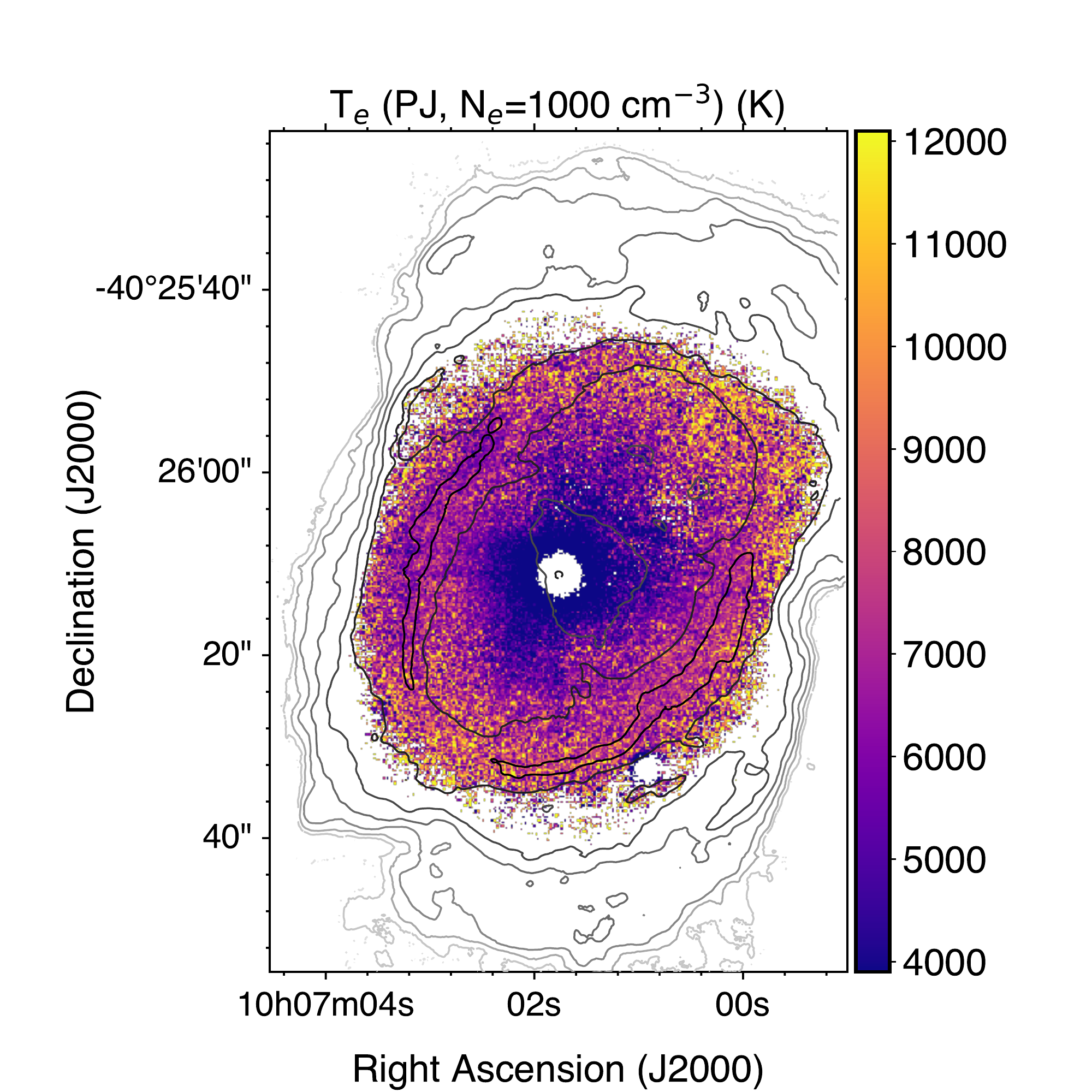}
      \includegraphics[angle=0,width=0.32\textwidth, clip=, viewport=30 0 540 550,]{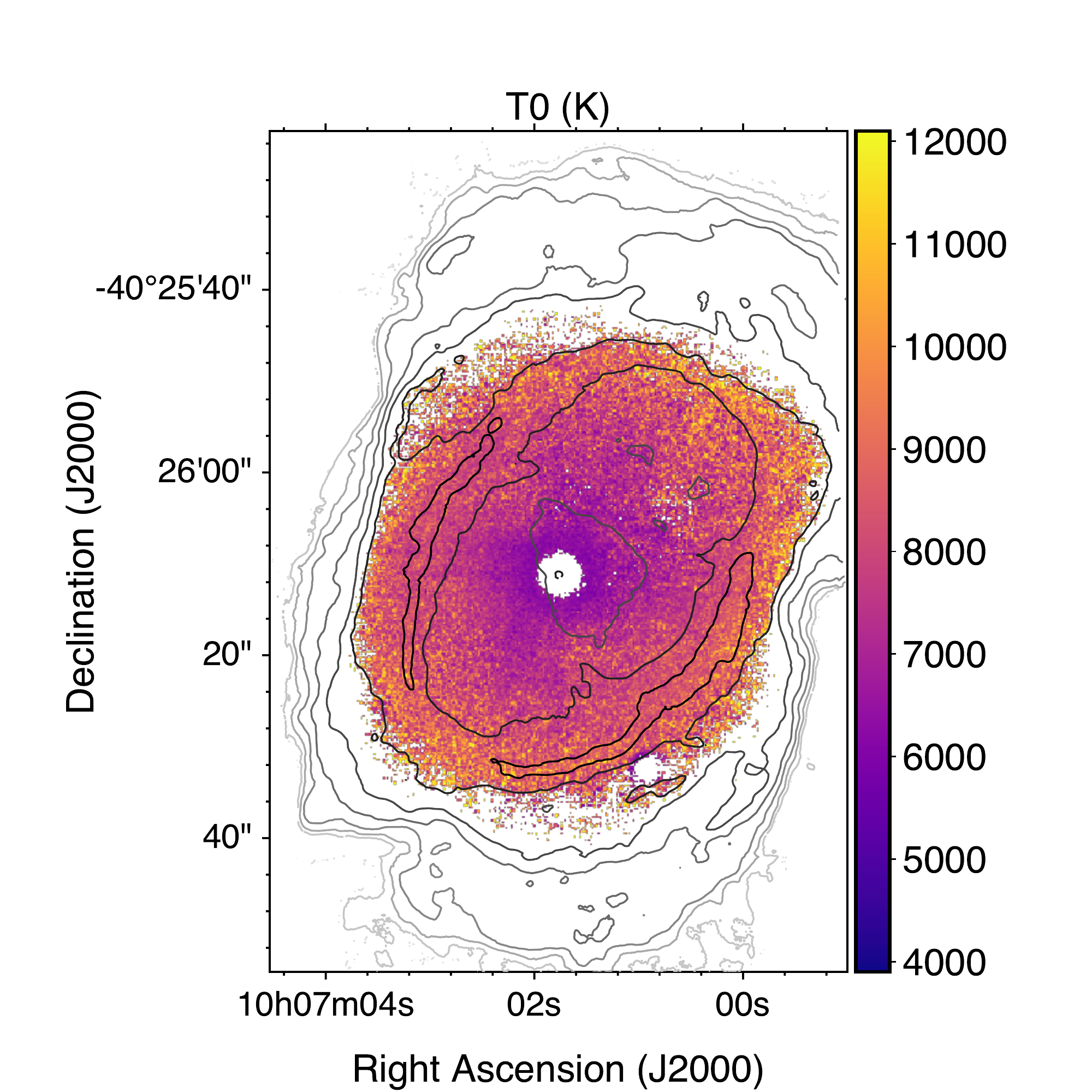}
   \includegraphics[angle=0,width=0.32\textwidth, clip=, viewport=30 0 540 550,]{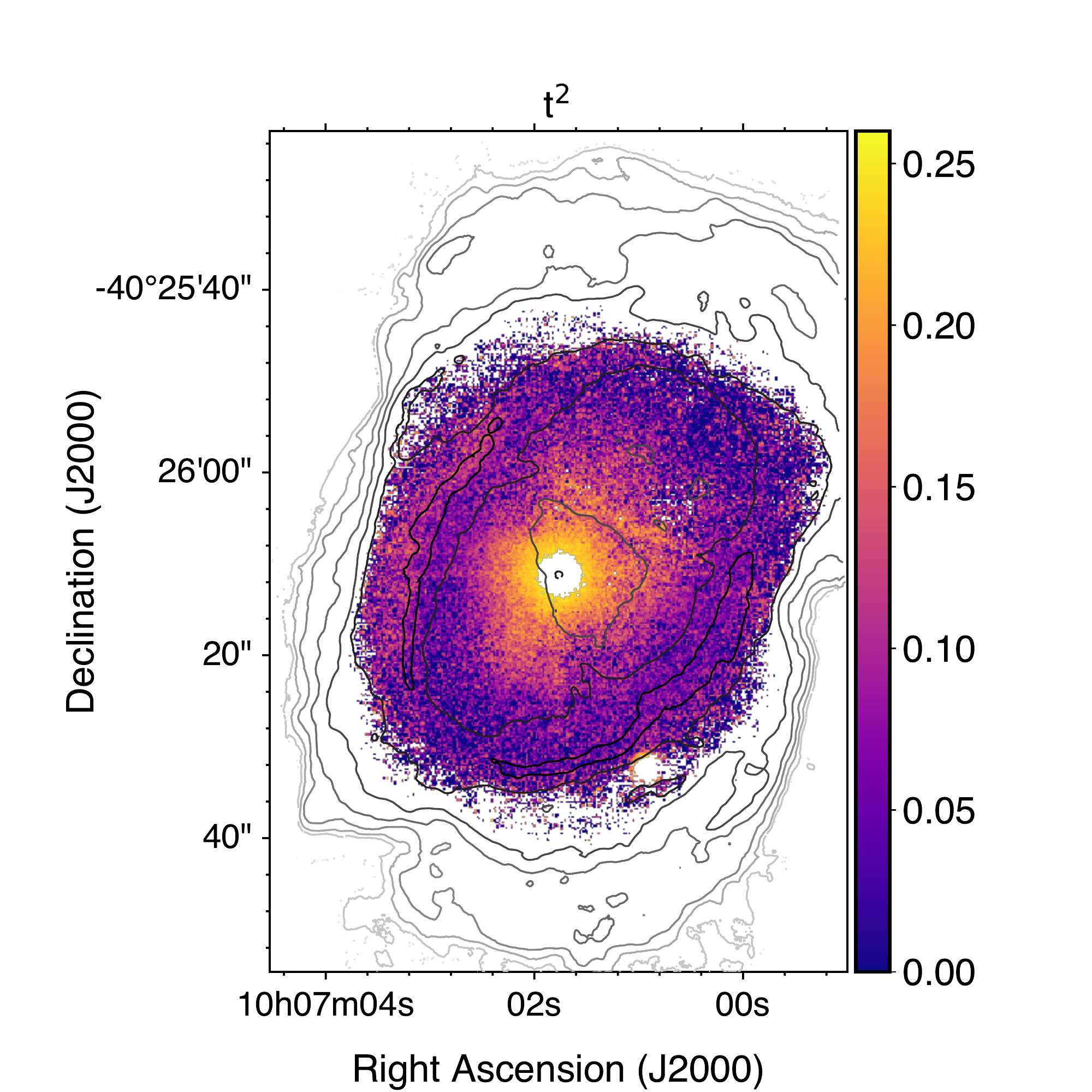}
   \caption{\emph{Left:} 
   Map of $T_{\rm e}$ derived from the 
magnitude of the Paschen continuum jump at 8250\,\AA\ ratioed by the 
dereddened \ion{H}{i} Paschen 11 (8862.8\,\AA) emission line strength. 
Initial estimates of $T_{\rm e}$ and $N_{\rm e}$ from the [\ion{S}{iii}] 
and [\ion{Cl}{iii}] ratio maps were applied; see text for details. Low  
$T_{\rm e}$ values over the two bright stars in the field (central star and field star to south-west) are produced by the stellar continua mimicking a low $T_{\rm e}$.
\emph{Centre:} Map of the average temperature, $T0$. \emph{Right:} Map of the temperature fluctuation parameter, $t^{2}$. Both were determined from the map presented in the left column and the $T_e$(\siii) map shown in Fig. \ref{mapNeTe}.
    }
   \label{mapTePJ}
    \end{figure*}

\section{Mapping electron temperature and density \label{sectene}}

There are several spectral features within the MUSE wavelength coverage that can be used to determine the electron temperature ($T_e$) and density ($N_e$). In the following, we discuss the maps derived from these tracers on a spaxel-by-spaxel basis. Likewise, to get a synthesized overview of the measured quantities, we compile a summary describing the derived Gaussian parameters, as well as the estimated uncertainties, in Table \ref{TabTeNe}. The uncertainties were estimated by means of Monte Carlo simulations with 50 trials in each case and assuming that the flux errors followed a normal distribution. 

\subsection{$T_e$ and $N_e$ from collisionally excited species \label{subsec_tenecol}}

We produced $N_e$ and $T_e$ maps from the extinction corrected maps of the CEL ratios \nii\,5755/6584, \oi\,5577/6300, \sii\,6731/6716, (lower ionisation plasma), \siii\,6312/9069, and \cliii\,5538/5518 (higher ionisation plasma).
As a first step, we used the \texttt{Atom.getTemDen} method in the  \texttt{PyNeb} package  \citep{Luridiana15}, version 1.1.4 with the default set of atomic data (transition probabilities and collision cross sections).
The method  evaluates either $T_e$ or $N_e$ given the other variable for a selected line ratio of known intensity and it is relatively fast.
The  \nii\,5755/6584, \oi\,5577/6300, and  \siii\,6312/9069 line ratios are almost independent of  $N_e$ while the  \sii\,6731/6716 and  \cliii 5538/5518 line ratios depend only marginally on $T_e$. Thus, this is a good starting point to have a first idea of the covered range in $T_e$ and $N_e$.

We derived $T_e$ maps assuming three different values of  $N_e$: 10$^{2}$, 10$^{3}$, and 10$^{4}$ cm$^{-3}$.
For a given ion, maps derived using $N_e$=10$^{2}$~cm$^{-3}$ and $N_e$=10$^{3}$~cm$^{-3}$ where comparable. Assuming $N_e$=10$^{4}$~cm$^{-3}$ resulted in slightly lower temperatures (differences $<1$\% for the  \oi\, and \siii\, ions, and $<4$\% for \nii).

Likewise, we derived $N_e$ maps assuming a uniform $T_e$ with three values: 8\,000, 10\,000, and 12\,000~K. In this case, maps using the \cliii\, emission lines were comparable independent of the assumed $T_e$ for the estimated uncertainties, while differences for the maps using the \sii\, emission lines could be as large as 10\%.

Then we used the \texttt{diags.getCrossTemDen} method, that cross-converges the $T_e$ and $N_e$  derived from two sensitive line ratios, by inputing the quantity 
derived from one line ratio into the other and then iterating. After exploring the outcome of \texttt{Atom.getTemDen}, we assumed $T_e$=9\,000~K as initial condition. We derived two pairs of maps: one for the \nii\,5755/6584, and \sii\,6731/6716 line ratios, representative of lower ionisation plasma, and another one for the \siii\,6312/9069, and \cliii\,5538/5518 line ratios, for higher ionisation.
However since, i) densities derived with the \texttt{Atom.getTemDen} method were $<$2\,000 cm$^{-3}$, ii) the derived temperature is almost independent of $N_e$ in this range, iii) the number of spaxels with information for the \cliii\, lines is much smaller than for the \siii\, lines, the $T_e$ map discussed hereafter, will be the one assuming a constant  $N_e$ = 10$^{3}$  cm$^{-3}$. For similar reasons, the $N_e$(\cliii) map assuming  $T_e$=10\,000~K will be used for the higher ionisation plasma.
Results are summarised in Fig. \ref{mapNeTe}.

The derived $N_e(\nii,\sii)$ map is consistent with the available  $N_e$(\sii) profiles for the E-W and N-S directions  \citep{Krabbe05,Juguet88}, and extend those to the whole surface of the nebula:  $N_e(\nii,\sii)$ is highest at the rim. There are also other locations with high $N_e(\nii,\sii)$, coincident with the dust lanes. 

The $N_e$(\cliii) map, made on a spaxel-by-spaxel basis is subject to much larger uncertainties and does not reveal any clear structure, other than a possible increase of $N_e$(\cliii) at the rim.
\citet{Tsamis03} report $N_e$(\cliii)=720\,cm$^{-3}$ for an integrated spectrum of the nebula. This is somewhat smaller than the typical derived values from the MUSE data, but still in agreement within the estimated uncertainties.

The $T_e(\nii,\sii)$ and $T_e(\siii)$ maps are pretty similar in terms of covered range of temperatures and structure: both are typically between 8\,800 and 10\,700 ~K and both display higher values at the rim. Still, variations of the $T_e(\siii)$ are slightly more extreme, with the region inside the rim presenting a higher contrast between the innermost and highest ionisation zone (i.e., that with \heii\, detection, see Fig. \ref{panel3colores}), and that closer to the rim, with lower  $T_e$.
Overall, the derived $T_e(\nii,\sii)$  values are compatible to those reported in the literature ($T_e(\nii)$=9350 K, for the integrated spectrum \citep{Tsamis03}, mean value of $T_e(\nii)$=10\,163 K, along a N-S slit \citep{Krabbe05}). The \nii$\lambda$5755 line can be affected by recombination, leading to an apparent enhancement of $T_e$. However, this effect has been reported  to be negligible in \object{NGC\,3132}  \citep{Krabbe05}; a finding confirmed by the photoionization model
presented in Sect. \ref{sec1Dmodel}.

Finally, in a somewhat smaller fraction of spaxels, we could also derive $T_e$(\oi), although with large uncertainties.
We used the \texttt{Atom.getTemDen} method and assumed $N_e$=1\,000 cm$^{-3}$.
In general, $T_e$(\oi) was always lower than both $T_e(\nii,\sii)$ and $T_e$(\siii). 
The map with the difference between $T_e(\nii,\sii)$ and $T_e$(\oi) display values around $\sim$800~K without any relevant structure.
\object{NGC\,3132} is known to have strong near-infrared (vibrationally excited) molecular hydrogen \citep{Storey84} and CO \citep{Sahai90} emission, both peaked on the 
shell where \oi\, is strong. In addition, mid-infrared rotational H$_{2}$  lines have been detected \citep{Mata16} as well as CO (J=3-2) emission \citep{GuzmanRamirez18}. These neutral gas indicators suggest that there may be a photodissociation region (PDR) at the outer shell of the nebula. If so, then a component of the \oi\, emission can arise in the PDR, c.f. 
\citet{Stoerzer00}, from thermal excitation of O$^{0}$ or photodissociation of OH. However the measured electron densities in \object{NGC\,3132} 
(Figs. \ref{mapNeTe} and \ref{2DhistoNe}) are only moderate and there is no indicator of densities in the regime important for PDR production of \oi\, ($10^{4} \leq n \leq 10^{7}$ cm$^{-3}$). Additionally the observed 
\oi\,6300/5577 ratio is too large for the range expected for PDR emission \citep{Stoerzer00} at the low densities in \object{NGC\,3132}. 
Although a minor contribution of PDR emission to the measured \oi\, line strengths cannot be ruled out, it seems unlikely to be important for 
\oi\, as a $T_{\rm e}$ and O$^{0}$ abundance indicator.

\subsection{$T_e$  from \ion{He}{i} \label{subsec_tenehei}}

The $T_e$ and $N_e$  diagnostics presented in previous section are based on collisionally excited species. In order to prove the physical properties of the optical recombination lines (ORL) emitting regions, one can use line ratios of certain recombination lines.
In particular, \citet{Zhang05} showed that the ratio of the \hei\, $\lambda$7281 and $\lambda$6678 recombination lines constitutes a suitable diagnostic for the ORL temperature.
 We used the extinction corrected \hei\,7281/6678 line ratio map and minimized the analytic fits presented by \citet{Benjamin99}, but with the most recent emissivities for these lines \citep{Porter13}. According to \citet{Zhang05}, the \hei\,7281/6678 ratio is not very dependent on the electron density. So the value was fixed in the minimization process. We tried three options: uniform low (170 cm$^3$) and high (1200 cm$^3$) $N_e$, and that derived from the \nii\, and \sii\, line ratios (Fig. \ref{mapNeTe}, upper left panel). The resulting $T_e$(\hei) maps were equivalent. 
Likewise, an assumption for the initial $T_e$ was needed. Again, we tried three options:  uniform low (8\,000 K) and high (12\,000 K) $T_e$, and the derived $T_e$ map from the \nii\, and \sii\, line ratios. Again, the resulting map was stable against the assumed initial condition.
The lower right panel in Fig. \ref{mapNeTe} contains the resulting $T_e$(\hei) map. The mean and median values are consistent with the $T_e$(\hei) reported by \citet{Zhang05} while the range in temperatures is much more extreme than those found from the CELs. 

\subsection{$T_e$ from \ion{H}{i} Paschen Jump \label{subsec_tepj}}

There is another $T_e$ diagnostic available within the MUSE spectral range: the magnitude of the series continuum jump for bound-free transitions of \ion{H}{i}. 
The identical method to derive the flux difference across the Paschen Jump 
as in \object{NGC~7009} \citep{Walsh18} was applied to the equivalent data 
for \object{NGC~3132}. The mean continuum in carefully selected line-free windows 
on both sides of the Paschen Jump was measured in each reddening 
corrected spaxel spectrum. Then, the calibration of the difference of the 
blue and red continua with respect to the flux of the (dereddened) P11 line 
was applied as a function of $T_{\rm e}$ and $N_{\rm e}$ (Fig. \ref{mapNeTe})
and the fractions of He$^{+}$ and He$^{++}$ with respect to 
H$^{+}$ \citep[see Appendix A in][]{Walsh18}. The maps of He$^{+}$ 
and He$^{++}$ (Sect. \ref{secionic}) were employed. For the
$N_{\rm e}$ map, the closest match in terms of ionization is the one
calculated from [\ion{S}{iii}] and [\ion{Cl}{iii}]
but since there are a substantial number of spaxels with
low signal-to-noise in this map, we adopted a single mean value of 1000\,cm$^{-3}$ 
(for all spaxels with a detected Paschen Jump). An initial estimate of
$T_e$ from \siii\, was employed to determine $T_e$(PJ) followed by one iteration with the derived $T_e$(PJ) map (see \citet{Walsh18} for details). Since the  dependence of the Paschen Jump on $N_{\rm e}$ is very weak at such 
densities \citep[see Fig. A2 in ][]{Walsh18}, little loss of 
accuracy results from this strategy.    

As for NGC~7009, no correction for the presence of stellar continuum on 
the Paschen Jump was applied \citep[c.f.,][]{Zhang04}, so spaxels over the 
seeing disk of the central star, and the field star to the south-west of the central 
star, will produce erroneous estimates of $T_{\rm e}$(PJ). The left panel in Fig. 
\ref{mapTePJ} shows the resulting map of $T_{\rm e}$(PJ). The mean 
value is 7020 K with a root mean square (RMS) of 
2430 K (3 $\times$ 3 $\sigma$ clipped mean); the mean signal-to-error over 
the map, based on 100 Monte Carlo trials using the propagated errors on the 
measured PJ, is 3.6. Spaxels over the seeing disks of both stars have not 
been included in these statistics.

\subsection{Mapping the temperature fluctuation parameter $t^2$ \label{subsec_t2}}

Comparison of a map of $T_{\rm e}$ from the CEL line ratio for 
[\ion{S}{iii}] and the ORL Paschen Jump provides 
the temperature fluctuation parameter, $t^{2}$, introduced by 
\citet{Peimbert67}. An identical procedure was followed as for 
NGC\,7009 \citep[see Sect. 5.4. in][]{Walsh18} and the mean temperature,
$T0$, and the temperature fluctuation parameter, $t^{2}$, were determined
by minimizing the difference of the observed $T_{\rm e}$'s against the 
difference given by \citet[Eqn. 1 \& 2]{Walsh18}. The central and right panels in Fig. \ref{mapTePJ} 
show the $T0$ and $t^{2}$ maps: the (3-$\sigma$ clipped) mean values are 
8250 K and 0.111, respectively. Similar to NGC~7009, large values of $t^{2}$
are derived, much larger than expected from the small values encompassed
by the $t^{2}$ formulation \citep{Peimbert67}.

   \begin{figure}[th!]
   \centering
   \includegraphics[angle=0,width=0.40\textwidth, clip=, viewport=0 25 550 480]{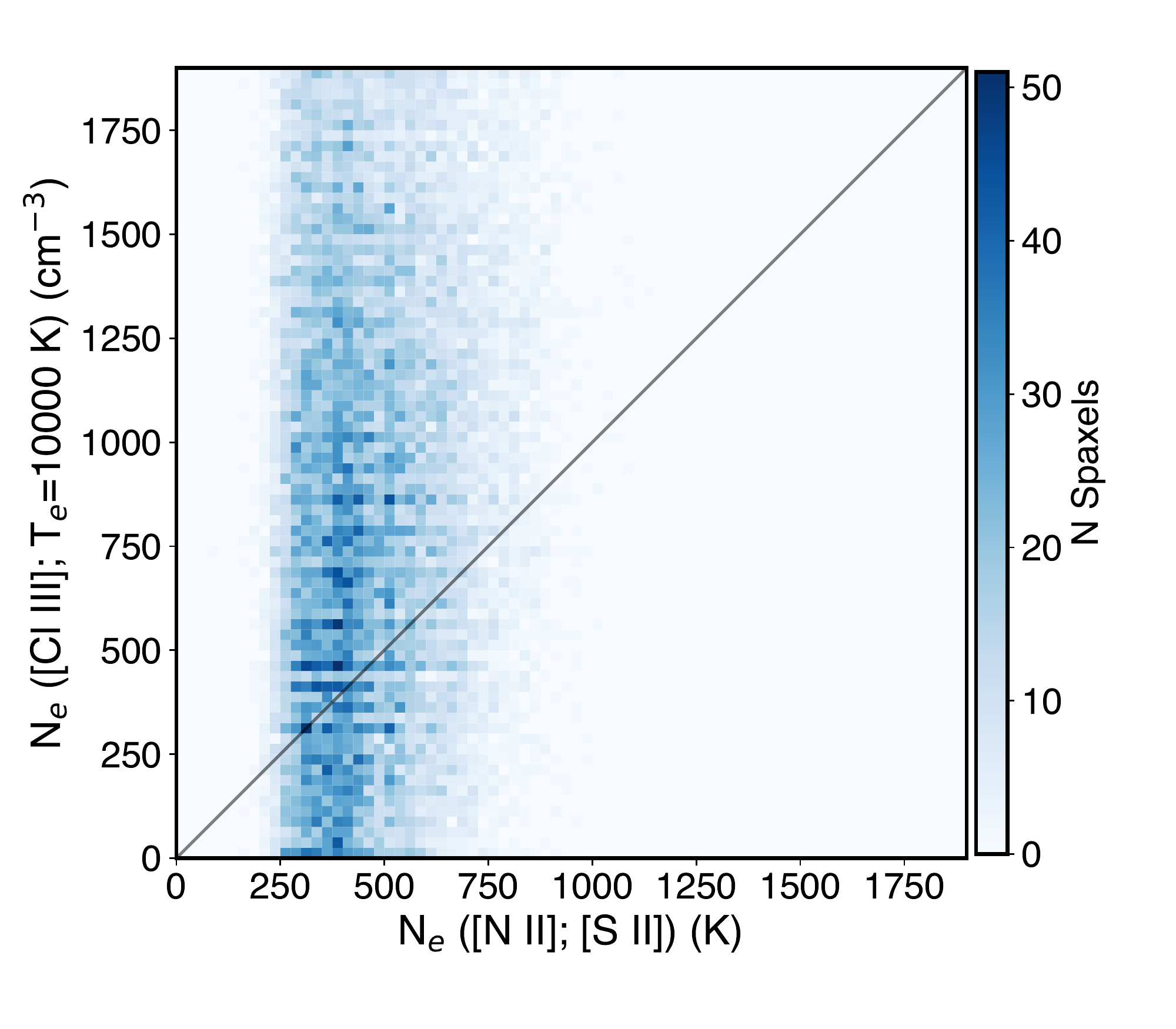}
   \caption{2D histogram to compare the different $N_e$ maps. The black diagonal signals the locus of equal densities.
}
   \label{2DhistoNe}
    \end{figure}
   
   \begin{figure*}[th!]
   \centering
   \includegraphics[angle=0,width=0.26\textwidth, clip=,viewport=0 25 550 480]{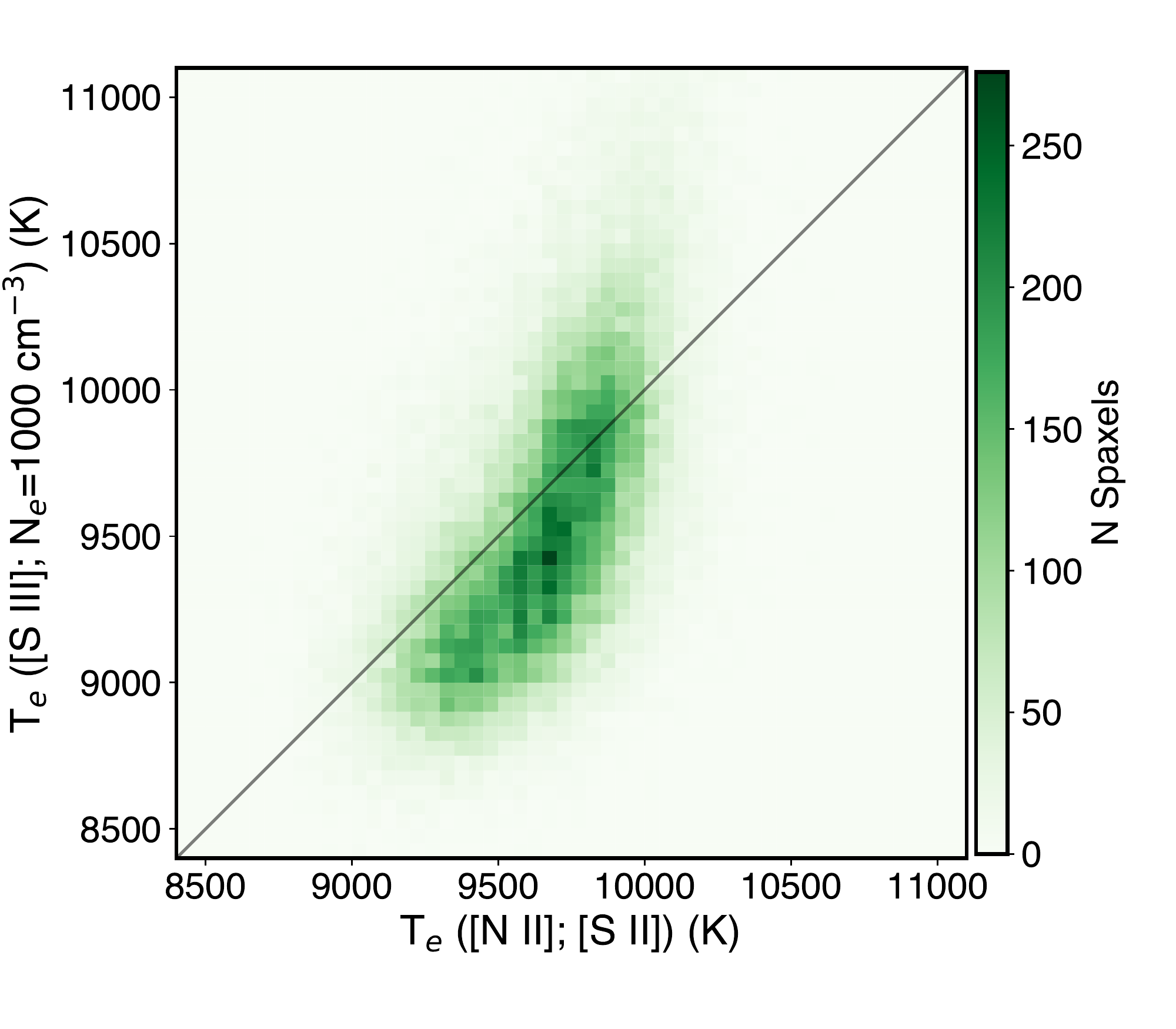}
     \includegraphics[angle=0,width=0.22\textwidth, clip=, viewport=30 0 540 550,]{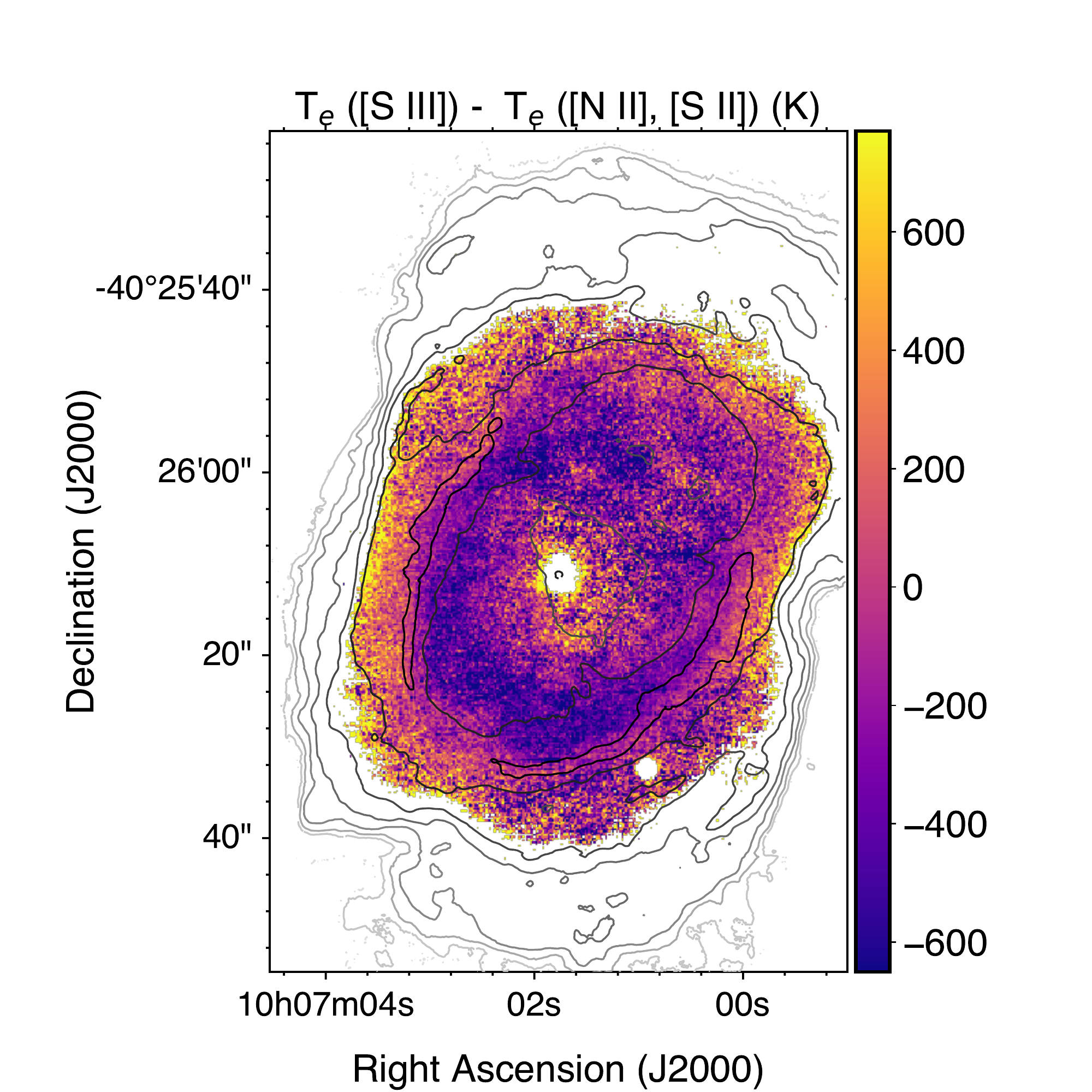}
    \includegraphics[angle=0,width=0.26\textwidth, clip=,viewport=0 25 550 480]{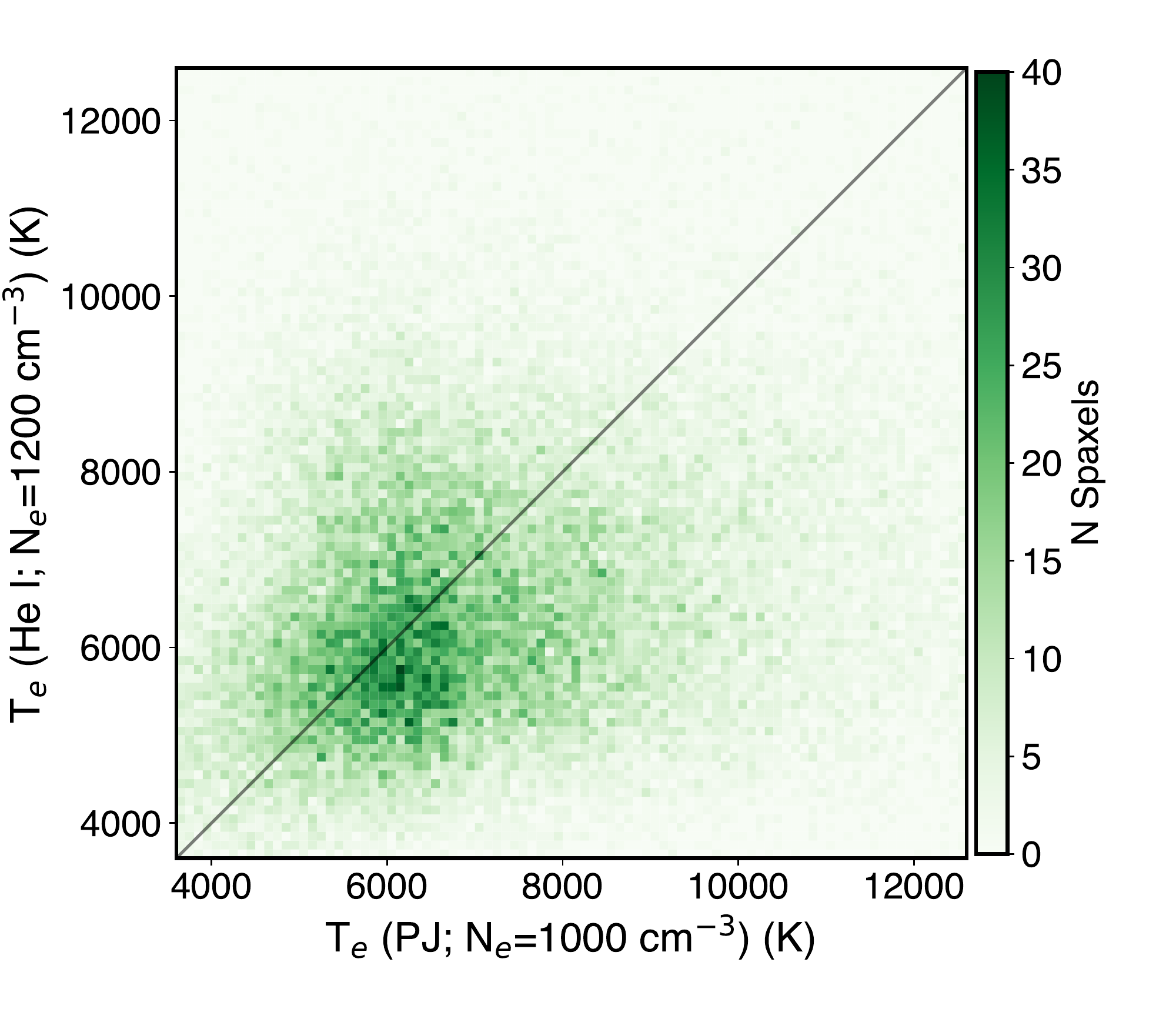}
     \includegraphics[angle=0,width=0.22\textwidth, clip=, viewport=30 0 540 550,]{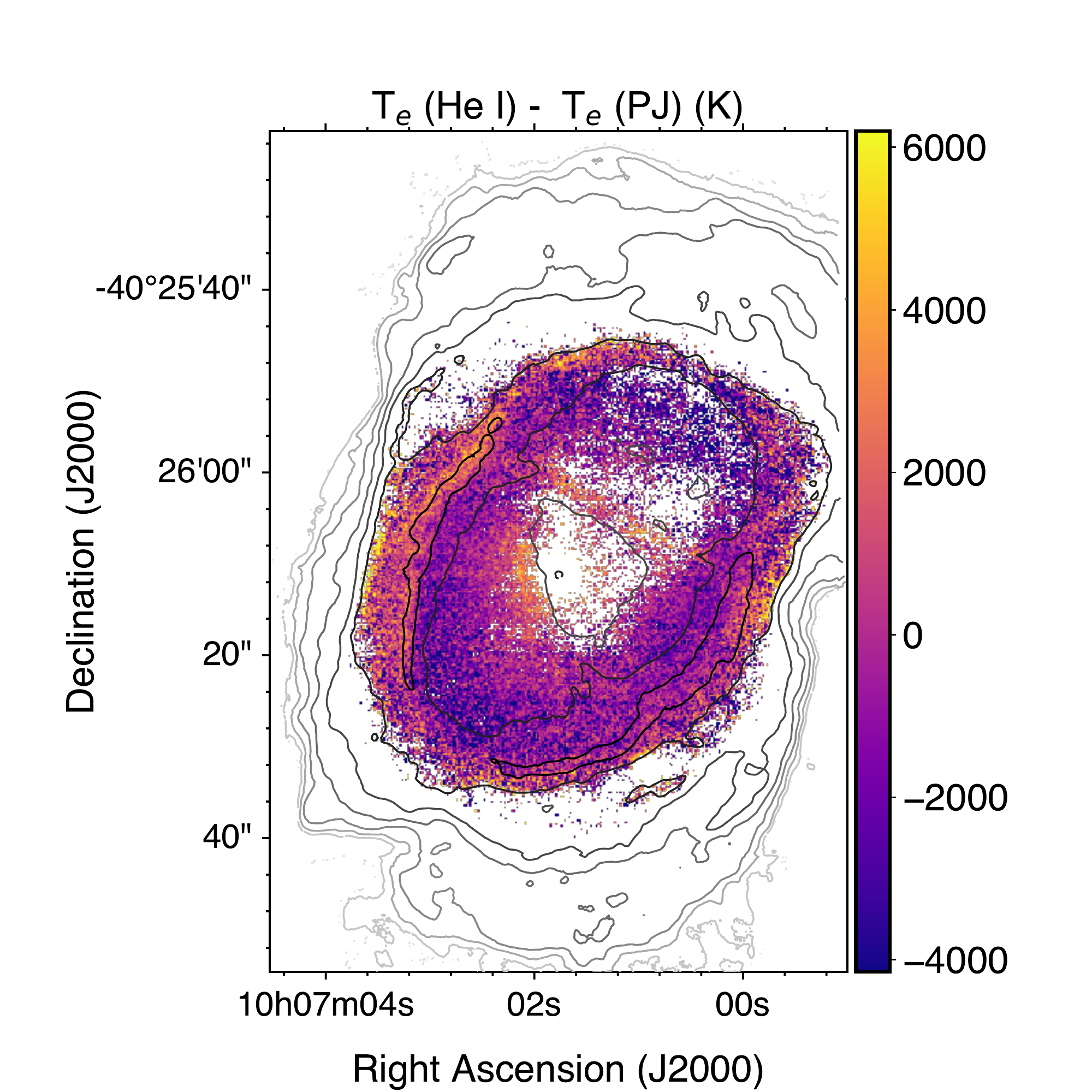}
   \includegraphics[angle=0,width=0.26\textwidth, clip=,viewport=0 25 550 480]{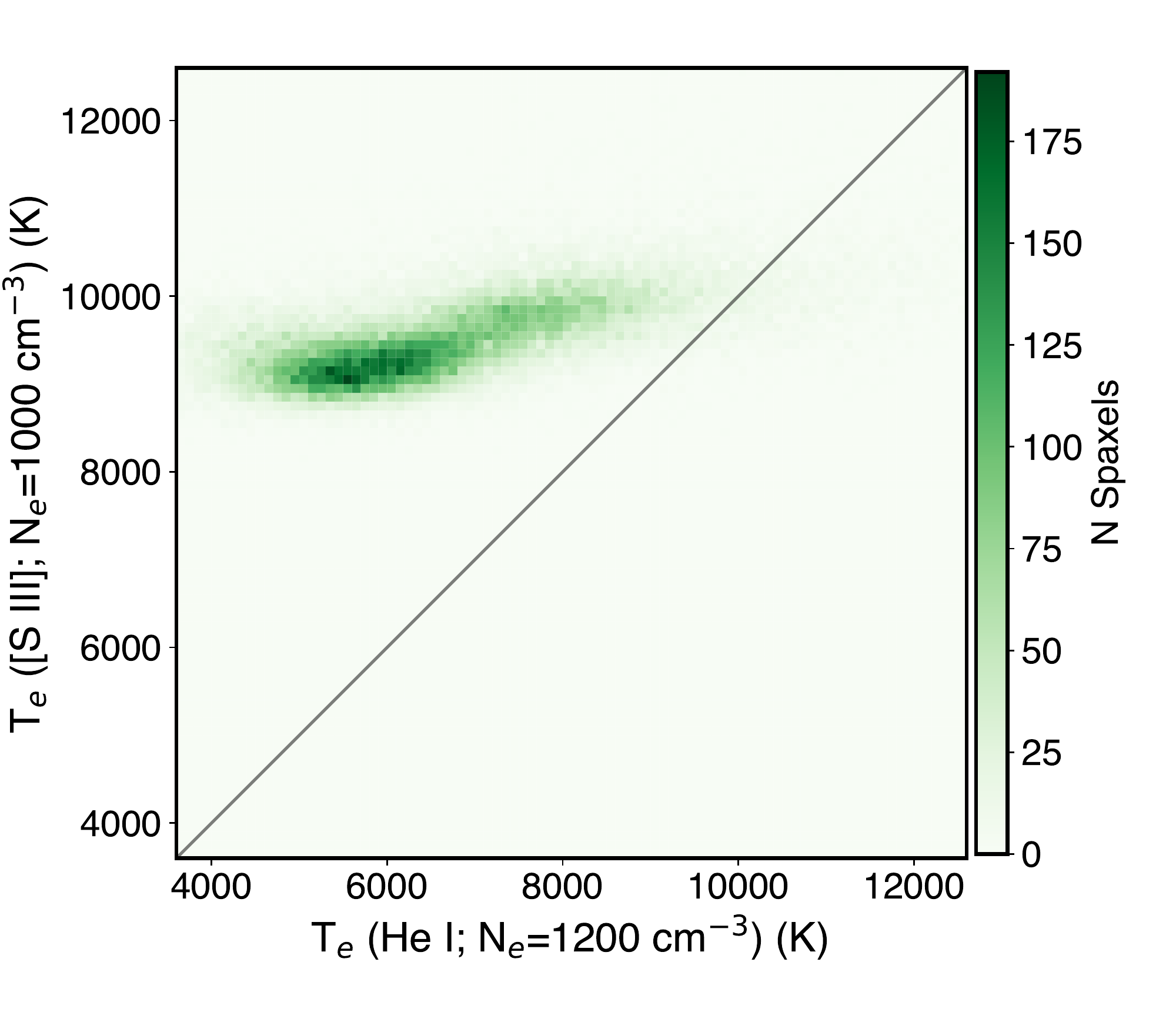}
     \includegraphics[angle=0,width=0.22\textwidth, clip=, viewport=30 0 540 550,]{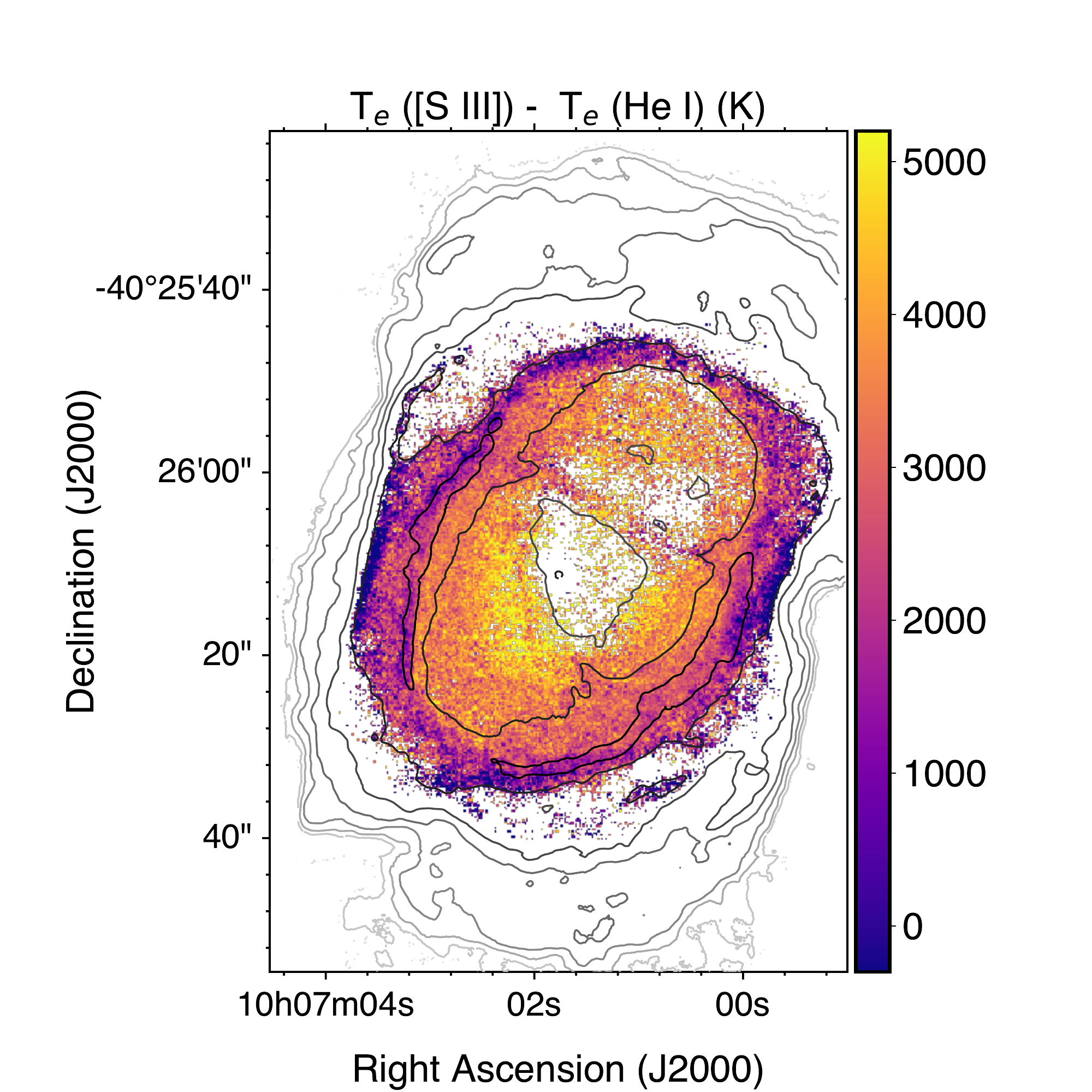}
   \includegraphics[angle=0,width=0.26\textwidth, clip=,viewport=0 25 550 480]{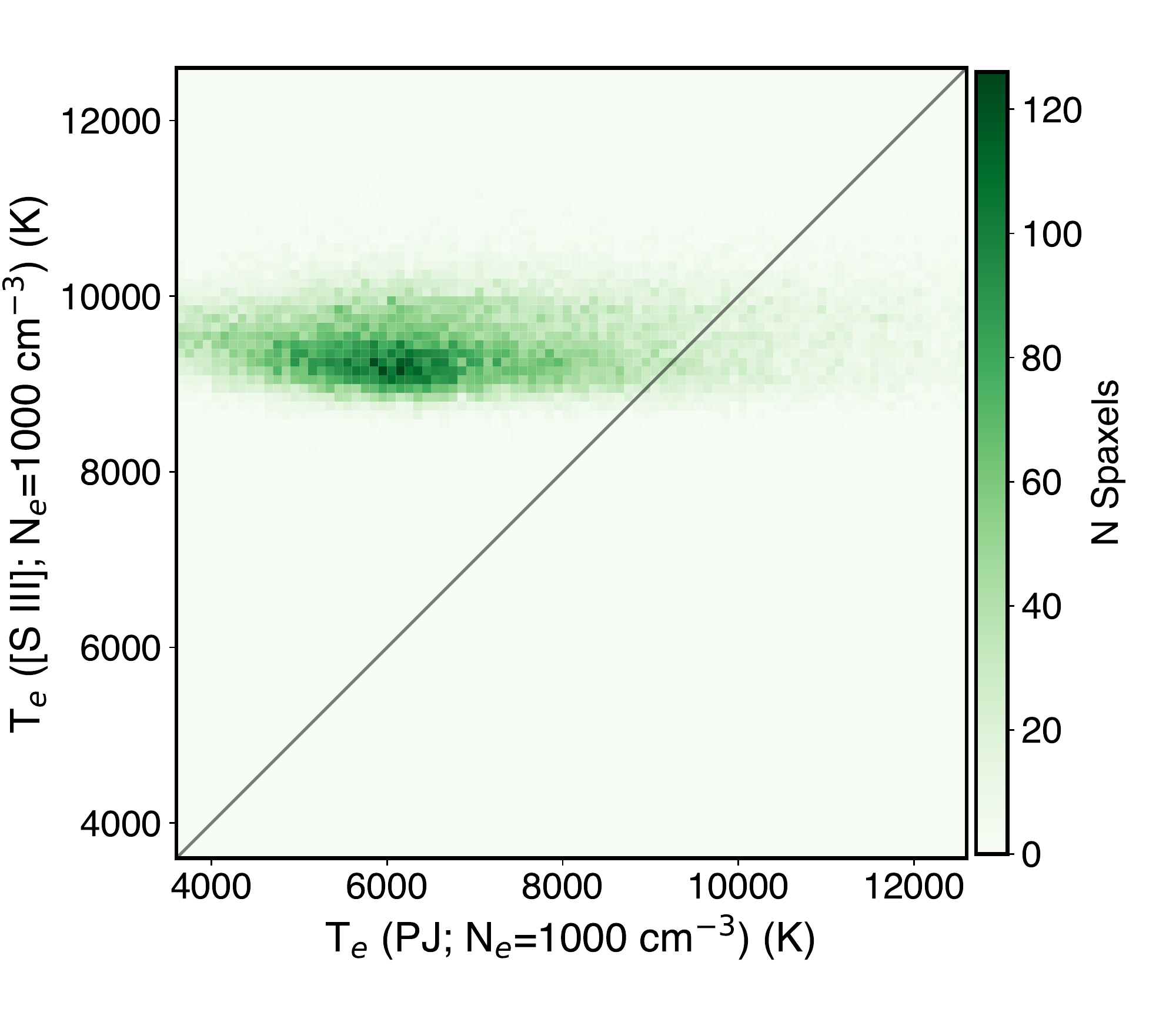}
     \includegraphics[angle=0,width=0.22\textwidth, clip=, viewport=30 0 540 550,]{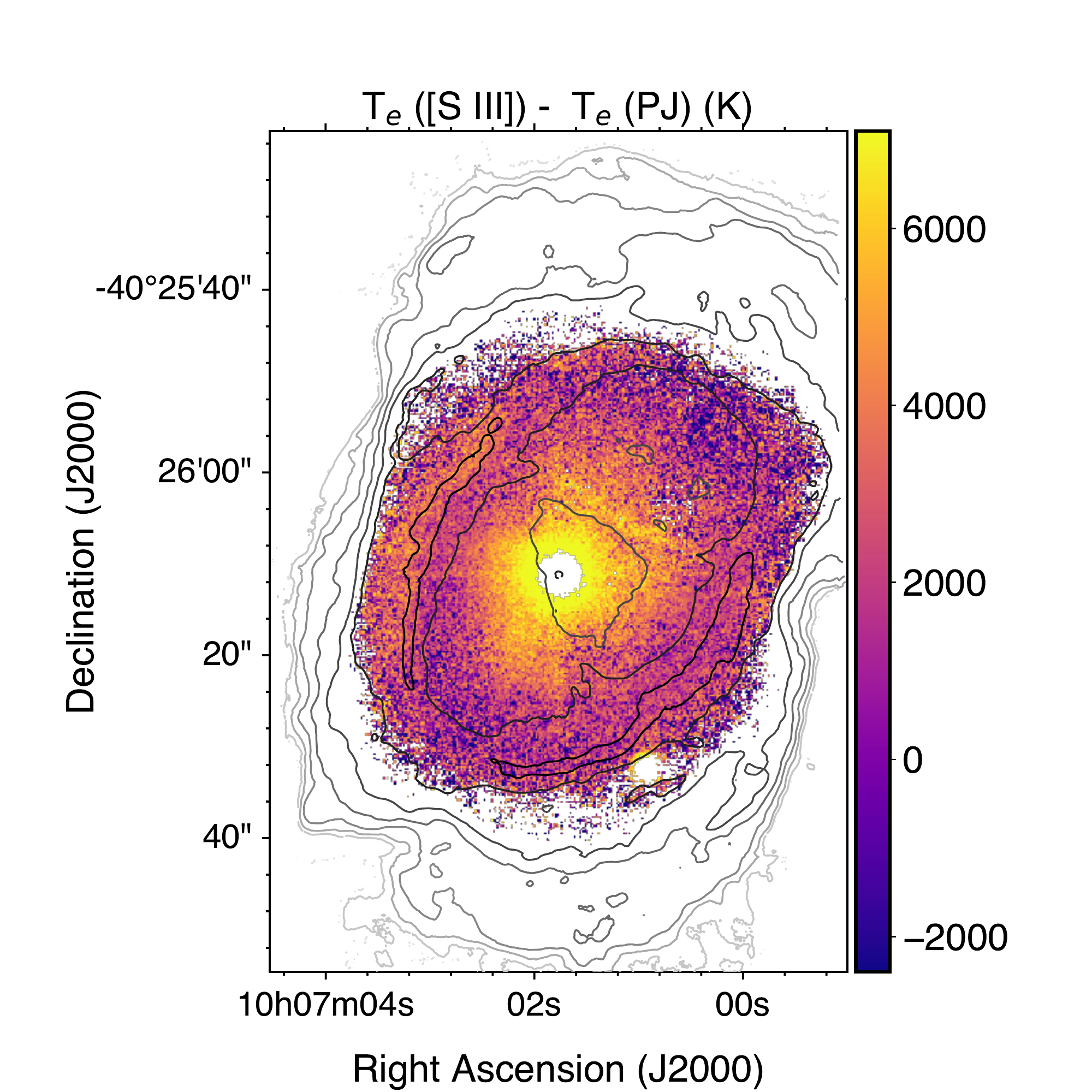}
   \caption{Comparison of the different $T_e$ maps. Each subfigure is made of two panels. On the left side, a 2D histogram to compare two different $T_e$ maps is presented. The black diagonal signals the locus of equal temperatures. On the right side, the map with difference between the ordinate and abscissa axes. The pairs of temperature considered are those involving \siii\, and \sii,\nii\, (\emph{top left}), \hei\, and Paschen Jump  (\emph{top right}), \siii\, and \hei\,  (\emph{bottom left}), and  \siii\, and Paschen Jump (\emph{bottom right}).
   $T_e$(\oi) was derived for a much smaller number of spaxels than the other $T_e$ maps and not included here.
}
   \label{2DhistoTe}
    \end{figure*}

\subsection{Discussion of physical properties\label{subsec_tenecompa}}

\subsubsection{Comparison of electron densities}

The \sii\, line ratio is a density diagnostic tracing the low-ionisation parts of the nebula, while that using the \cliii\, lines traces better the density in the high-ionisation parts, since the S$^+$ zone is basically separated from the Cl$^{+2}$ zone, assuming a simple ionisation stratification. The relation between these two tracers on a 
spaxel-by-spaxel basis is shown in Fig. \ref{2DhistoNe} and for most of the spaxels, $N_e$(\cliii) $> N_e$(\sii). With the reasonable assumption of more high-energy photons in 
the inner regions of the nebula (i.e., closer to the ionising star), this implies a decrease of density with increasing distance from the central star. 
Whether the assumed geometry of the PN is a diabolo or an elliptical shell, previous models of the nebula based only on $N_e$(\sii) profiles for the density suggest \object{NGC 3132} as a hollow structre, or at least filled by low density material. 
The measured $N_e$(\cliii) values add new information that complement any of these pictures, and point towards a more layer-like structure, with a higher-density and higher-ionisation inner layer filling the outer lower-ionisation layer.

\subsubsection{Comparison of electron temperatures}

The different electron temperatures are compared on a spaxel-by-spaxel basis in Fig. \ref{2DhistoTe} both as
maps and histograms.
Using measurements of $\sim$30 positions in \object{NGC\,3132}, \citet{Krabbe05} found that $T_e(\nii,\sii) \approx T_e(\oiii)$. Most of their data display $T_e(\nii,\sii) > T_e(\oiii)$, with differences of $\sim500-800$~K. It is not possible to derive $T_e(\oiii)$ with these MUSE data. Of the available tracers, $T_e(\siii)$ would be the most comparable one. The comparison shown in the upper left panel of Fig. \ref{2DhistoTe}, this time based on measurements on $>5\times10^5$ spatial elements, supports and delineates much better the finding hinted at \citet{Krabbe05}: there is clear correlation between $T_e(\nii,\sii)$ and  
$T_e$(\siii), in the sense that higher $T_e$(\siii) is linked to higher $T_e$(\nii,\sii).
However, a bi-modal behaviour is revealed: at $T_e(\nii,\sii)\lesssim9700$~K, $T_e(\nii,\sii)>T_e(\siii)$ by roughly a constant offset of $\sim200$~K; at higher temperatures, this behaviour is inverted with $T_e$(\siii)$>T_e$(\nii,\sii) and with increasing differences at increasing temperatures.
This second branch with the highest $T_e$'s corresponds to the spaxels mapping the rim and beyond, as well as the innermost region, with the highest ionisation.

The other 2D histograms in Fig. \ref{2DhistoTe} also reveal substructures. For example, the lower left panel shows a region of high $T_e$(\hei) and $T_e$(\siii) with lower temperature differences (i.e., closer to the diagonal line). An inspection of the accompanying $T_e$ difference map and those presented in Figs. \ref{mapNeTe} reveals that this substructure again delineates the rim. A similar behaviour, although much more subtle, is displayed in the two 2D histograms on the right side of Fig. \ref{2DhistoTe}: the rim is revealed as a substructure with high  $T_e$(\hei) and $T_e$(\siii), and  $T_e$(PJ)$\sim$8\,000~K.

Comparisons between $T_e$ (and $N_e$ too) traced by different ions at different positions within a given PNe exist in the literature \citep[e.g.][]{Krabbe05} but are rare. They are based typically on a relatively low number of data points, and thus they are not adequate for the construction of these kind of 2D histograms. It is the capacity of IFS-based instruments, like MUSE, to provide a large amount of high quality data points that allows for their construction. As hinted at the example of \object{NGC\,3132}, they may be well used to define regions that share ionisation conditions. Perspectives are even better for foreseen instruments such as HARMONI or BlueMUSE, since their higher spectral resolution would allow a much better definition of the regions in terms of radial velocity.
A similar discussion would apply to highly multiplexed  multi-object spectrographs sampling for example populations of unresolved PNe in other galaxies.
Should  the number of measured physical magnitudes become too large to be managed with a set of few 2D histograms, PNe (or regions within a given PN) with similar ionisation conditions could well be identified by means of unsupervised machine learning methods, where these would appear as members of the same cluster.

For  \object{NGC\,3132}, 
temperatures derived from the recombination lines ($T_e$(PJ) and $T_e$(\hei)) are lower than those from CELs: median $T_e$ decreases according to the following sequence: \nii,\sii $\rightarrow$ \siii  $\rightarrow$ \oi $\rightarrow$ \hei $\rightarrow$ PJ.
Temperatures derived from recombination lines also cover a much larger range than those derived from CELs ($T_e(\nii,\sii)$ and $T_e(\siii)$).
\citet{Zhang04} reported similar tendencies using the electron temperatures derived from the Balmer Jump and the \oiii\, forbidden lines for single measurements in a sample of PNe.
They proposed the existence of dense clumps within the PN as a possible explanation for this result (together with large $t^2$ for some of the nebula).
A similar situation may well be at play here. But, maps in Figs. \ref{mapTePJ} and \ref{2DhistoTe} show further how both the fractional mean-square temperature variations and temperature differences between different diagnostics change depending on the location within the nebula suggesting that the distribution and/or characteristics of the clumps, should these exist, vary across the nebula.

\begin{table*}[th!]
\caption{Ionic and total abundance statistics for helium. Number of spaxels, mean and standard deviation were calculated for the data between the 5 and 95 percentiles. \label{TabIonHe}} 
\label{tab_ion}      
\centering                          
\begin{tabular}{lcccccccc}        
\hline\hline                 
 & N spaxels & 5\% & Q1 & median & Q3 & 95\% & mean & $\sigma$\\
\hline                        
$\tau$($\lambda$3889) ($\omega$=3) & 55964 &   0.0082 &   0.1756 &   1.1731 &   3.1005 &   6.5061 &   1.9700 &   2.2118 \\
\hline
He$^+$/H$^+$ ($\lambda$5876) &  81356 &   0.0416 &   0.1129 &   0.1251 &   0.1369 &     0.1662 &      0.1221 &       0.0306 \\
Uncert.     &   &  0.0041 &   0.0061 &   0.0086 &   0.0136 &     0.0368 &    0.0126 &   0.0116 \\
He$^+$/H$^+$ ($\lambda$6678) &  71055 &   0.0904 &   0.1119 &   0.1218 &   0.1332 &     0.1632 &      0.1235 &       0.0209 \\
Uncert.     &  &   0.0040 &   0.0057 &   0.0077 &   0.0110 &     0.0225 &   0.0098 &   0.0076 \\
He$^+$/H$^+$ ($\lambda$7281) &  44027 &   0.0668 &   0.0875 &   0.1051 &   0.1332 &   0.1815 &   0.1128 &   0.0352 \\
Uncert.     & &   0.0045 &   0.0064 &   0.0085 &   0.0120 &   0.0209 &   0.0108 &   0.0103 \\
He$^{++}$/H$^+$  &   16614 &   0.0064 &   0.0136 &   0.0217 &   0.0322 &     0.0418 &      0.0232 &    0.0124 \\     
Uncert. &  &   0.0002 &   0.0005 &   0.0010 &   0.0017 &     0.0033 &      0.0014 &       0.0022 \\
He/H    &  71395 &   0.1123 &   0.1183 &   0.1239 &   0.1334 &     0.1631 &      0.1284 &       0.0166 \\
Uncert. &   &   0.0044 &   0.0062 &   0.0080 &   0.0112 &     0.0226 &      0.0101 &       0.0075 \\
\hline                                   
\end{tabular}
\end{table*}

\subsubsection{$T_{\rm e}$(PJ) and $t^{2}$ morphology}

Comparison of the morphology of the  $T_{\rm e}$(\siii) (Fig. \ref{mapNeTe}) map with that for the Paschen Jump (Fig. \ref{mapTePJ}) reveals a correspondence between the central region of elevated $T_{\rm e}$(\siii) and low $T_{\rm e}$(PJ), but no evidence for an association with the higher $T_{\rm e}$(\siii) values  around the prominent shell in the PJ map. In spatial coincidence to the chordal filament crossing the shell to the NW of the central star, the $T_{\rm e}$(PJ) map displays depressed values, although this feature is not apparent on the $T_{\rm e}$(\siii) image (but does display a diffuse signature in the $T_{\rm e}$(\nii,\sii) map -- Fig.~\ref{mapNeTe}). The morphology of the $T0$ image (central panel in Fig. \ref{mapTePJ}) is dominated by the $T_{\rm e}$(PJ) map with a mean value increased by $\sim$ 1200\,K and no evidence of the morphology of the outer shell so clearly apparent in the  $T_{\rm e}$(\siii) image. The $t^{2}$ map however shows closer correspondence to the $T_{\rm e}$(\siii)  image: a signature of the bright shell (as elevated $t^{2}$); the central region showing elevated $t^{2}$ in spatial association with the region of lowered $T_{\rm e}$(PJ). Values of $t^{2}$ are high around the central star with the highest values (up to 0.24) at smaller stellar offset radii.

\section{Ionisation maps \label{secionic}}

We derived maps of the ionic fraction with respect to H$^+$ by using 
those lines with sufficient signal-to-noise  on a spaxel-by-spaxel basis over a significant fraction of the nebula. 
Tables \ref{TabIonHe} and \ref{TabIon} contain summaries describing the derived values, as well as the estimated uncertainties.
When a given abundance could be derived from several lines (e.g. O$^{++}$), all of them where used, to identify possible discrepancies due to e.g. a saturated line in certain spaxels.
In all the cases, ionic abundances from CELs were determined using the method \texttt{getIonAbundance} of the \texttt{Atom} class in \texttt{PyNeb}.
Ionic abundances from recombination lines - here only for helium - were calculated using the method \texttt{getEmissivity} of the \texttt{RecAtom} class in \texttt{PyNeb}.
As with the physical properties, uncertainties were estimated using Monte Carlo simulations with 50 realisations in each case, and assuming a normal distribution for the errors of the involved lines.
Following, we present the mapped ionic abundances.

%
\begin{table*}[th!]
\caption{Ionic and total abundance statistics for light elements. Number of spaxels, mean and standard deviation were calculated for the data between the 5 and 95 percentiles. \label{TabIon}} 
\centering                          
\begin{tabular}{lcccccccc}        
\hline\hline                 
 & N spaxels & 5\% & Q1 & median & Q3 & 95\% & mean & $\sigma$\\
\hline
$10^5\times$ O$^{++}$/H$^+$ ($\lambda$4959) &  57743 &    14.83 &    24.35 &    34.38 &    44.21 &    52.75 &    34.18 &    12.00 \\
Uncert.                                     &        &     1.46 &     2.24 &     2.86 &     3.55 &     4.74 &     2.95 &     1.00 \\
$10^5\times$ O$^{++}$/H$^+$ ($\lambda$5007) &  57743 &    14.77 &    24.32 &    34.40 &    44.28 &    52.81 &    34.20 &    12.05 \\
Uncert.                                     &        &     1.45 &     2.23 &     2.87 &     3.56 &     4.77 &     2.95 &     1.01 \\
$10^5\times$ O$^{+}$/H$^+$ ($\lambda$7320)  &  77315 &    10.28 &    17.30 &    34.37 &    44.32 &    58.69 &    32.99 &    16.23 \\
Uncert.                                     &        &     0.75 &     1.48 &     3.17 &     5.47 &    12.40 &     4.40 &     4.72 \\
Recom corr.                                 &  57628 &     8.83 &    13.75 &    26.69 &    38.98 &    49.68 &    27.11 &    13.94 \\
$10^5\times$ O$^{+}$/H$^+$ ($\lambda$7331)  &  77315 &     9.63 &    16.07 &    31.60 &    40.54 &    53.44 &    30.25 &    14.70 \\
Uncert.                                     &        &     0.72 &     1.41 &     2.96 &     5.07 &    11.37 &     4.08 &     4.33 \\
Recom corr.                                 &  57628 &     8.26 &    12.80 &    24.75 &    35.96 &    45.64 &    25.07 &    12.78 \\
$10^5\times$ O$^{0}$/H$^+$ ($\lambda$5577)  &   3277 &     2.32 &     5.71 &     8.71 &    11.21 &    17.11 &     8.89 &     4.59 \\
Uncert.                                     &        &     0.17 &     0.55 &     1.11 &     1.58 &     2.70 &     1.19 &     0.86 \\
$10^5\times$ O$^{0}$/H$^+$ ($\lambda$6300)  &  77748 &     0.94 &     2.80 &     9.99 &    16.89 &    29.13 &    11.21 &     9.23 \\
Uncert.                                     &        &     0.06 &     0.19 &     0.74 &     1.67 &     4.82 &     1.35 &     1.97 \\
$10^5\times$ O$^{++}$/H$^+$ ($\lambda$6364) &  77748 &     0.98 &     2.92 &    10.39 &    17.55 &    30.34 &    11.67 &     9.60 \\
Uncert.                                     &        &     0.06 &     0.19 &     0.76 &     1.73 &     5.01 &     1.40 &     2.03 \\
$10^5\times$ O/H$^+$ (TPP77)  &  57065 &    60.32 &    67.08 &    71.18 &    75.95 &    84.12 &    71.58 &     7.20 \\
Uncert.                       &        &     4.01 &     5.20 &     6.38 &     8.11 &    12.73 &     7.12 &     3.08 \\
$10^5\times$ O/H$^+$ (KB94)   &  57065 &    56.57 &    65.35 &    70.32 &    75.63 &    84.06 &    70.41 &     8.12 \\
Uncert.                       &        &     3.88 &     5.07 &     6.28 &     8.06 &    12.73 &     7.04 &     3.11 \\
$10^5\times$ O/H$^+$ (DIMS14) &  57065 &    55.73 &    64.92 &    70.16 &    75.60 &    84.06 &    70.16 &     8.36 \\
Uncert.                       &        &     3.85 &     5.04 &     6.26 &     8.06 &    12.72 &     7.02 &     3.12 \\\hline
$10^5\times$ N$^{+}$/H$^+$ ($\lambda$6548) &  77852 &     3.84 &     7.45 &    17.56 &    22.20 &    27.51 &    15.87 &     7.99 \\
Uncert.                 &        &    0.20 &     0.44 &     1.17 &     2.01 &     4.51 &     1.56 &     1.67 \\
$10^5\times$ N$^{+}$/H$^+$ ($\lambda$6584)  &  77849 &     3.91 &     7.59 &    17.87 &    22.59 &    27.99 &    16.15 &     8.13 \\
Uncert.                 &        &     0.21 &     0.45 &     1.19 &     2.05 &     4.56 &     1.59 &     1.72 \\
$10^5\times$ N$^0$/H$^+$ ($\lambda$5199) &  65020 &     0.39 &     0.80 &     2.30 &     3.80 &     6.74 &     2.64 &     2.08 \\
Uncert.                 &        &     0.03 &     0.06 &     0.19 &     0.40 &     1.13 &     0.33 &     0.42 \\
\hline
$10^6\times$ S$^{++}$/H$^+$ ($\lambda$6312)   &  57743 &     4.03 &     5.20 &     6.17 &     7.24 &     8.83 &     6.25 &     1.44 \\
Uncert.                 &        &     0.38 &     0.50 &     0.61 &     0.73 &     1.00 &     0.64 &     0.19 \\
$10^6\times$ S$^{++}$/H$^+$ ($\lambda$9069)  &  57742 &     4.02 &     5.18 &     6.15 &     7.23 &     8.81 &     6.24 &     1.44 \\
Uncert.                 &        &     0.23 &     0.31 &     0.37 &     0.46 &     0.61 &     0.39 &     0.12 \\
$10^6\times$ S$^{+}$/H$^+$ ($\lambda$6716)      &  77849 &     1.06 &     1.93 &     4.21 &     5.48 &     7.24 &     3.97 &     2.04 \\
Uncert.                 &        &     0.06 &     0.12 &     0.30 &     0.51 &     1.18 &     0.40 &     0.43 \\
$10^6\times$ S$^{+}$/H$^+$ ($\lambda$6731)      &  77848 &     1.07 &     1.93 &     4.22 &     5.49 &     7.22 &     3.97 &     2.04 \\
Uncert.                 &        &     0.06 &     0.11 &     0.28 &     0.49 &     1.15 &     0.39 &     0.42 \\
\hline   
$10^7\times$ Cl$^{+++}$/H$^+$ ($\lambda$8046)  &  11003 &     0.18 &     0.32 &     0.44 &     0.56 &     0.71 &     0.44 &     0.16 \\
Uncert.                 &        &     0.02 &     0.03 &     0.04 &     0.05 &     0.08 &     0.04 &     0.02 \\
$10^7\times$ Cl$^{++}$/H$^+$ ($\lambda$5518)  &  41861 &     1.00 &     1.27 &     1.48 &     1.69 &     1.98 &     1.49 &     0.30 \\
Uncert.                 &        &     0.09 &     0.12 &     0.15 &     0.18 &     0.25 &     0.16 &     0.05 \\
$10^7\times$ Cl$^{++}$/H$^+$ ($\lambda$5538) &  41850 &     0.90 &     1.20 &     1.41 &     1.63 &     1.91 &     1.41 &     0.31 \\
Uncert.                 &        &     0.09 &     0.13 &     0.15 &     0.19 &     0.26 &     0.16 &     0.05 \\
\hline
$10^6\times$ Ar$^{++}$/H$^+$ ($\lambda$7136) &  57743 &     1.81 &     2.31 &     2.60 &     2.86 &     3.15 &     2.56 &     0.41 \\
Uncert.                 &        &     0.10 &     0.14 &     0.17 &     0.22 &     0.30 &     0.18 &     0.06 \\
$10^6\times$ Ar$^{++}$/H$^+$ ($\lambda$7751)&  57719 &     1.81 &     2.31 &     2.62 &     2.89 &     3.20 &     2.58 &     0.42 \\
Uncert.                 &        &     0.10 &     0.14 &     0.18 &     0.22 &     0.30 &     0.19 &     0.06 \\
\hline
\end{tabular}
\end{table*}

   \begin{figure*}
   \centering
     \includegraphics[angle=0,width=0.32\textwidth, clip=, viewport=30 0 540 550,]{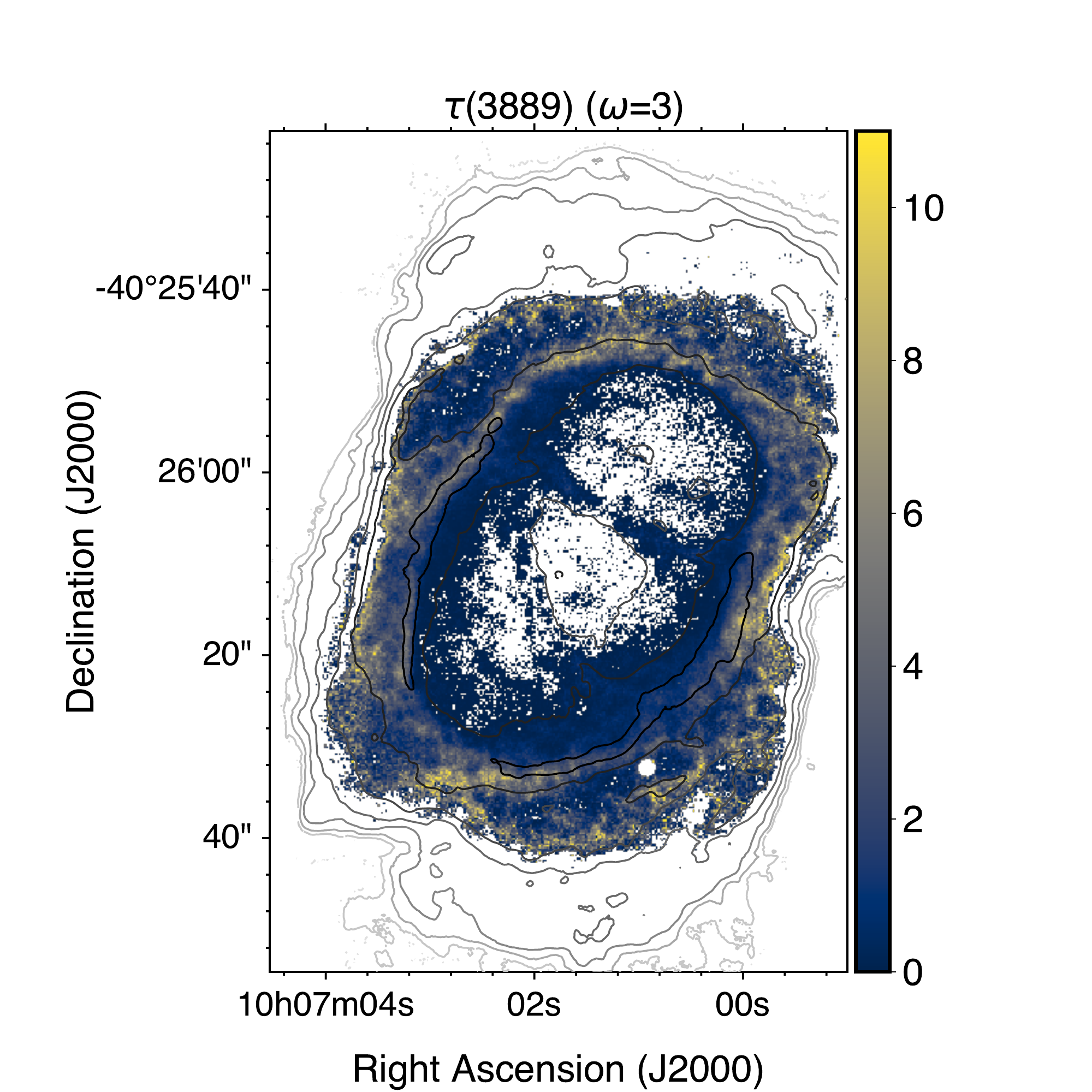}
   \includegraphics[angle=0,width=0.32\textwidth, clip=, viewport=30 0 540 550,]{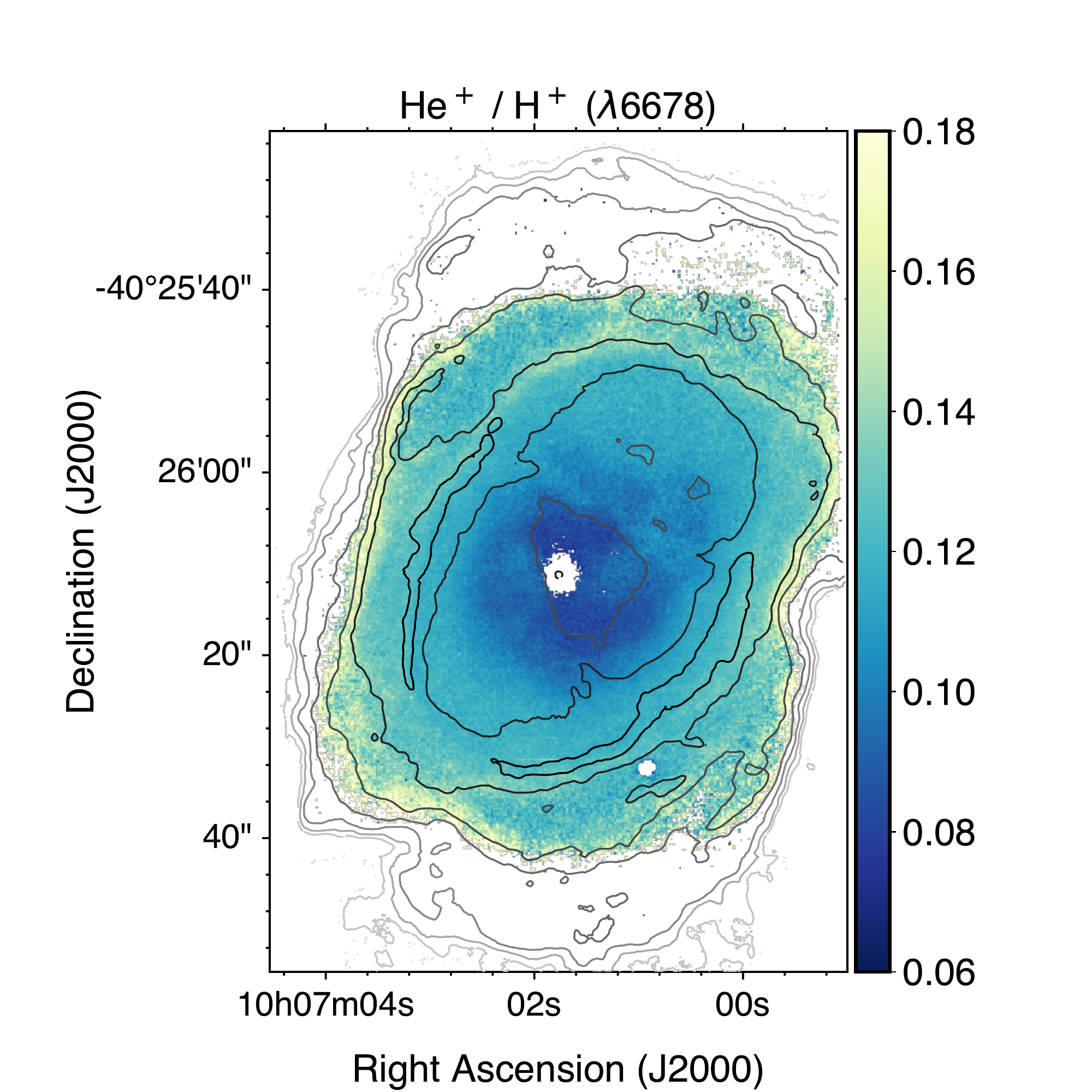}
   \includegraphics[angle=0,width=0.32\textwidth, clip=, viewport=30 0 540 550,]{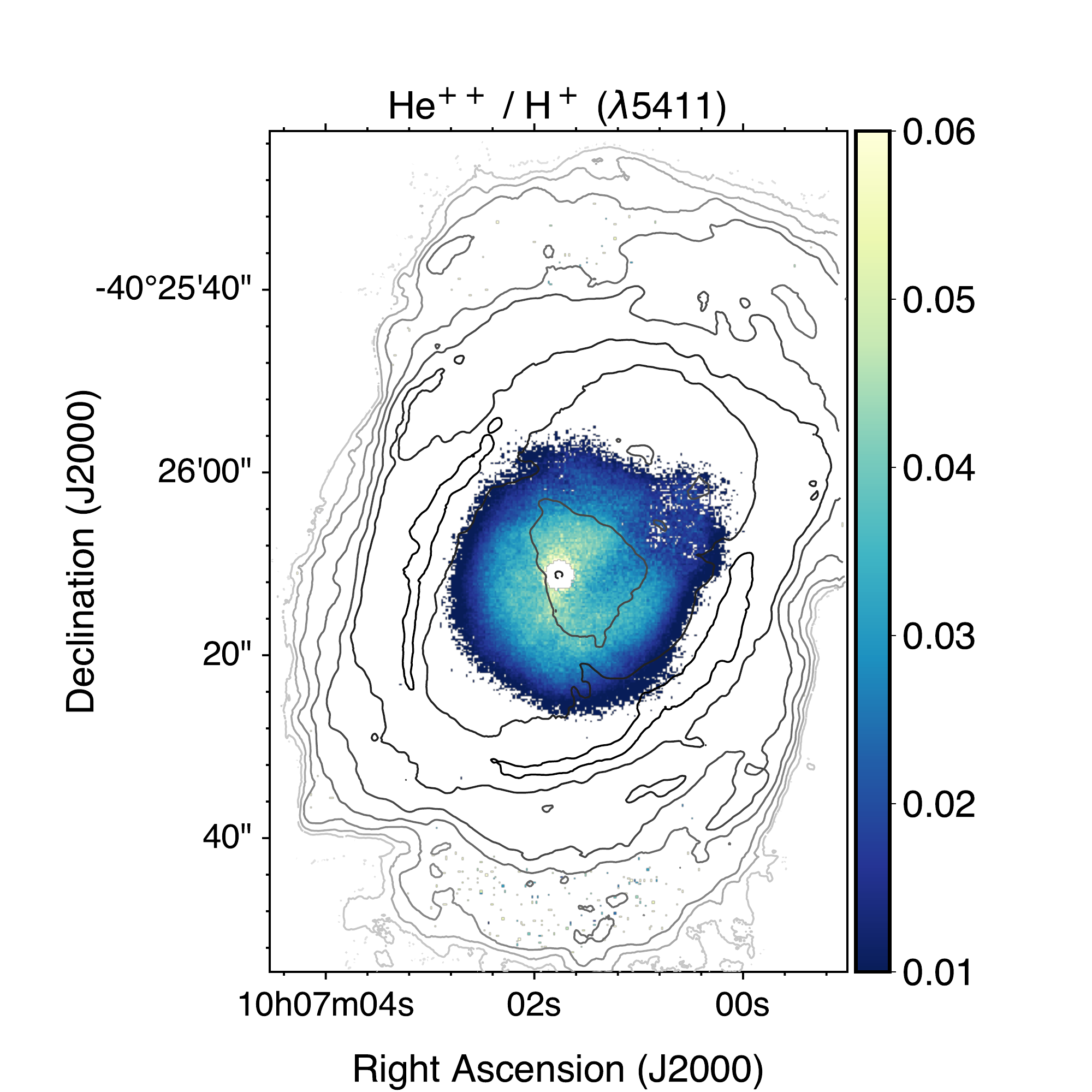}
   \caption{
   \emph{Left:} Derived map for $\tau$(3889) using the \hei\, $\lambda$7065 and $\lambda$6678 emission lines and a ratio of expansion velocity to thermal velocity for the nebula of 3.
\emph{Centre:} Map for singly ionised helium as determined using the extinction corrected flux map for the singlet line at $\lambda$6678 and the $N_e$ and $T_e$ simultaneously derived with the \sii\, and \nii\, lines.
\emph{Right:} Map for doubly ionised helium as determined using the extinction corrected flux map for the line at $\lambda$5411, the median $N_e$ as determined from the \cliii\, doublet, and the median $T_e$ as determined from the \siii\ lines. 
   }
   \label{mapionichelium}
    \end{figure*}

   \begin{figure*}
   \centering
   \includegraphics[angle=0,width=0.32\textwidth, clip=, viewport=30 0 540 550,]{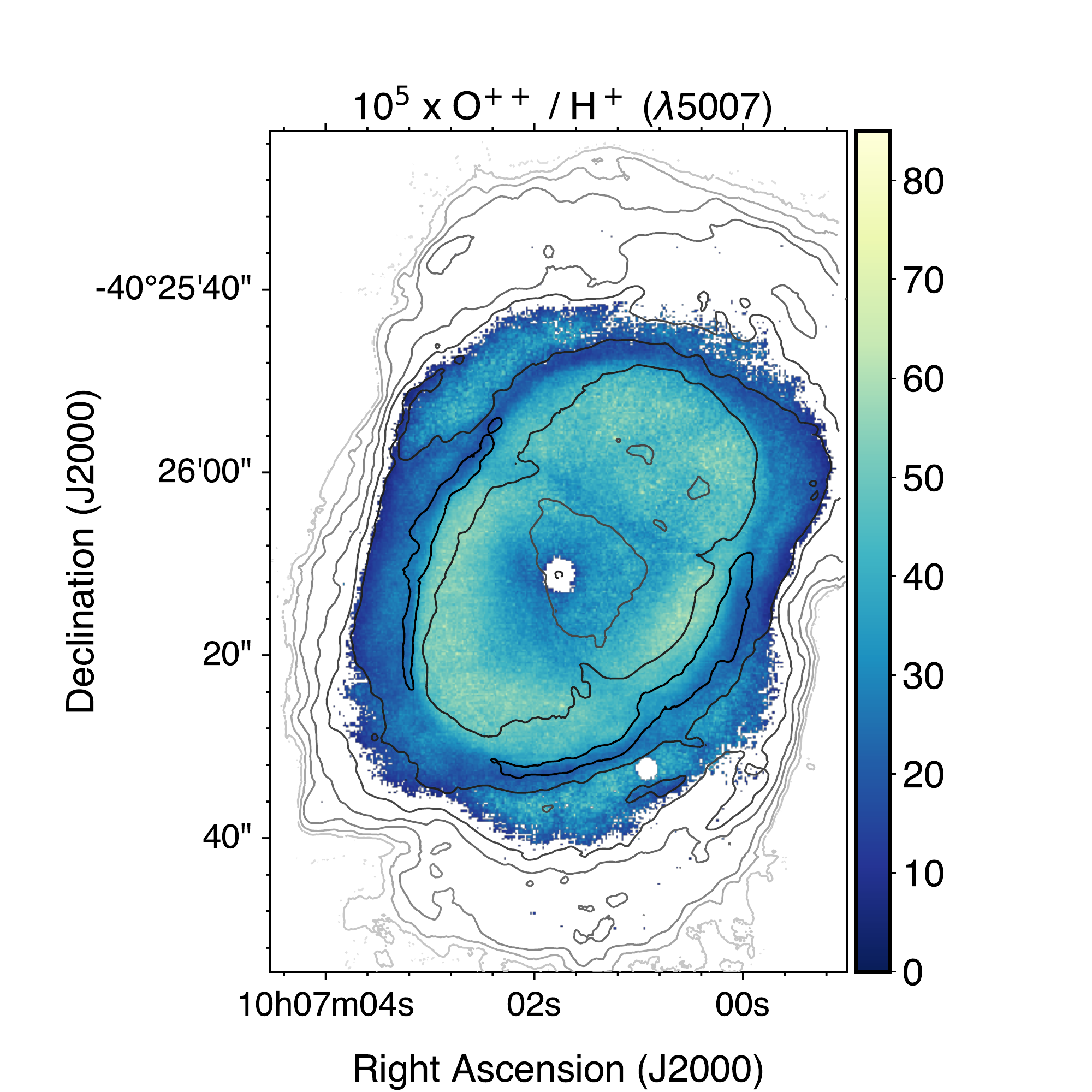}
   \includegraphics[angle=0,width=0.32\textwidth, clip=, viewport=30 0 540 550,]{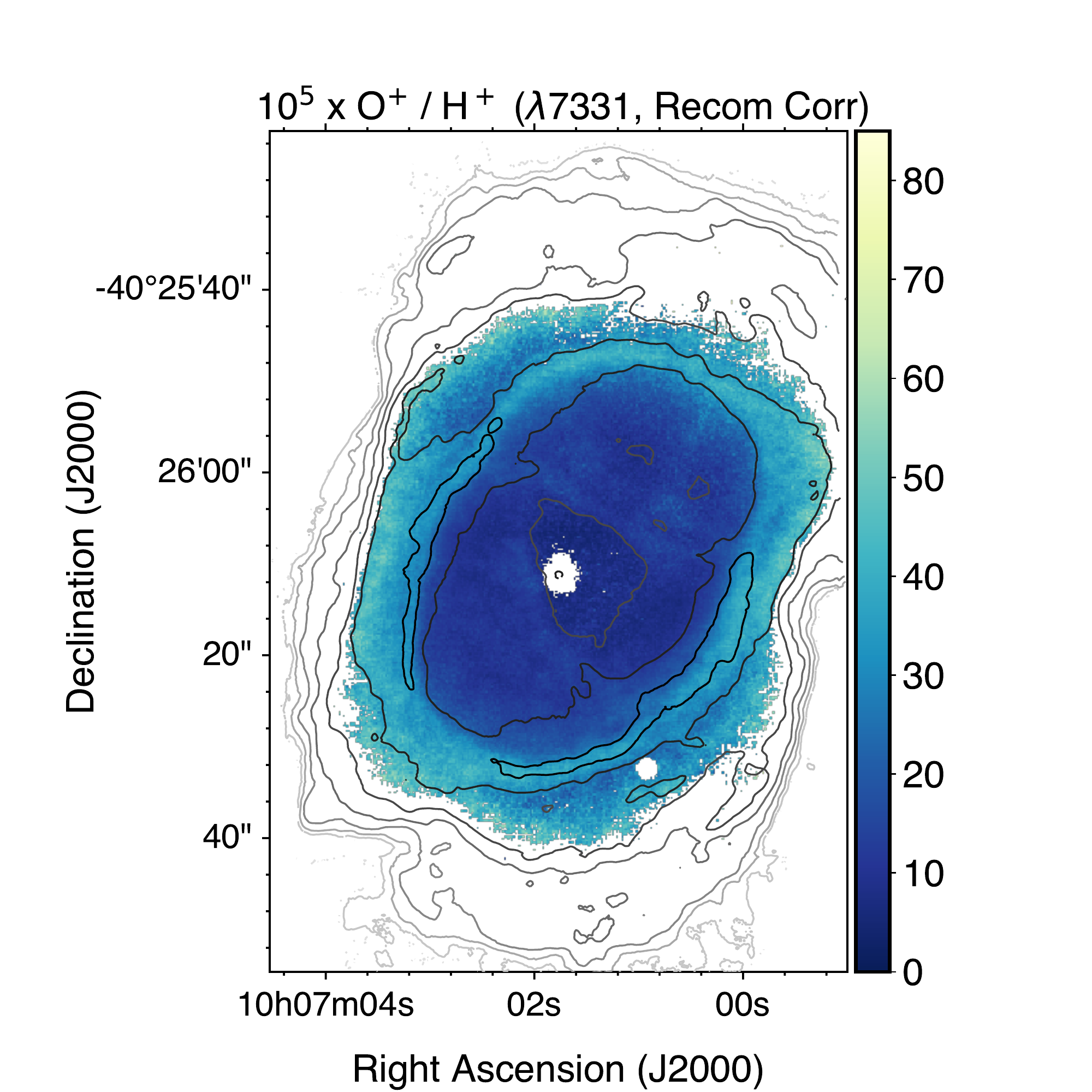}
   \includegraphics[angle=0,width=0.32\textwidth, clip=, viewport=30 0 540 550,]{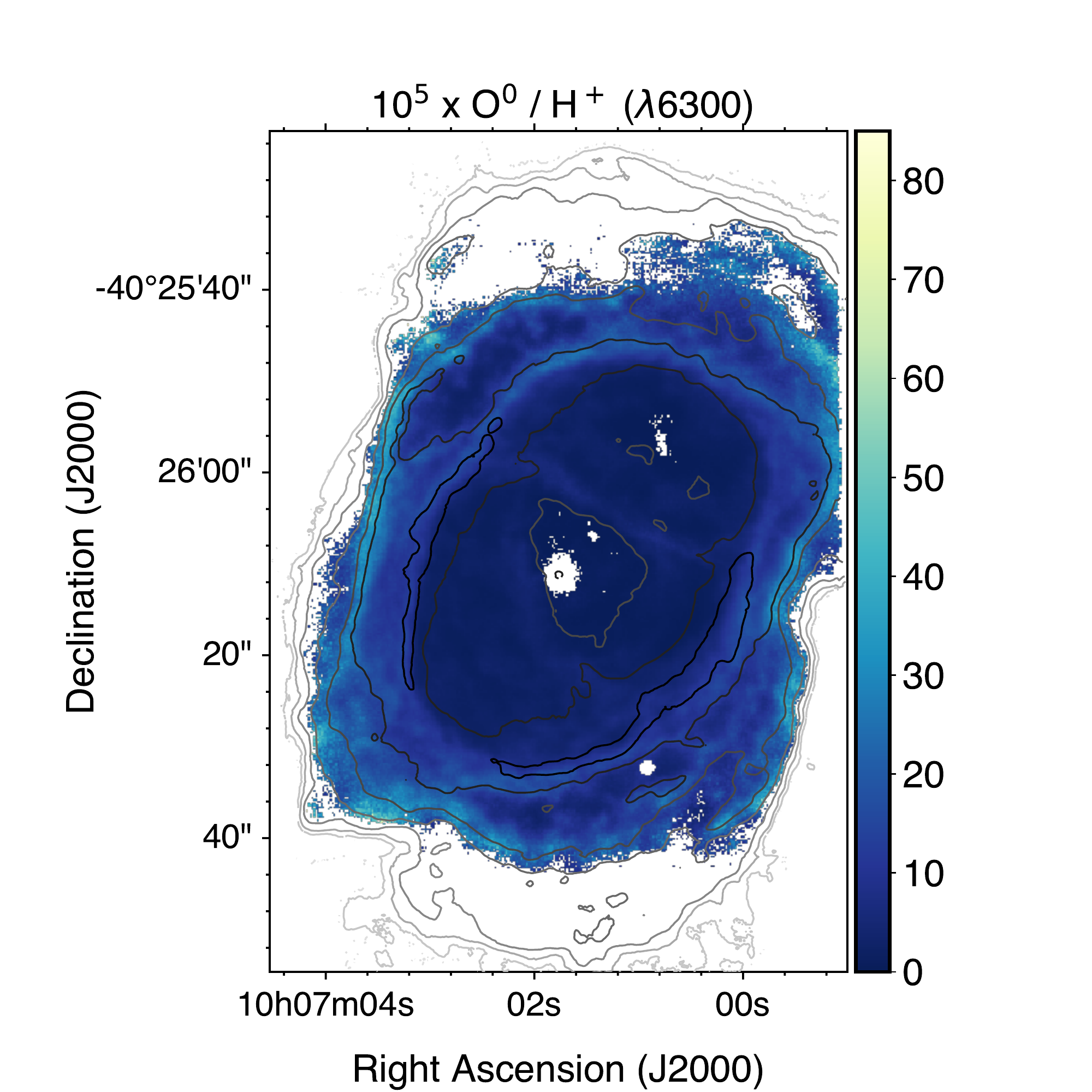}
   \caption{Maps for the O$^{++}$ (\emph{left}),  O$^{+}$,  (\emph{centre}), and O$^0$ (\emph{right}) ionic abundances, derived as described in the text.
   All the three maps display the same range in abundance in order to emphasise the relative contribution of each ion in the different parts of the nebula.}
   \label{mapOion}
    \end{figure*}

   \begin{figure*}
   \centering
   \includegraphics[angle=0,width=0.32\textwidth, clip=, viewport=30 0 540 550,]{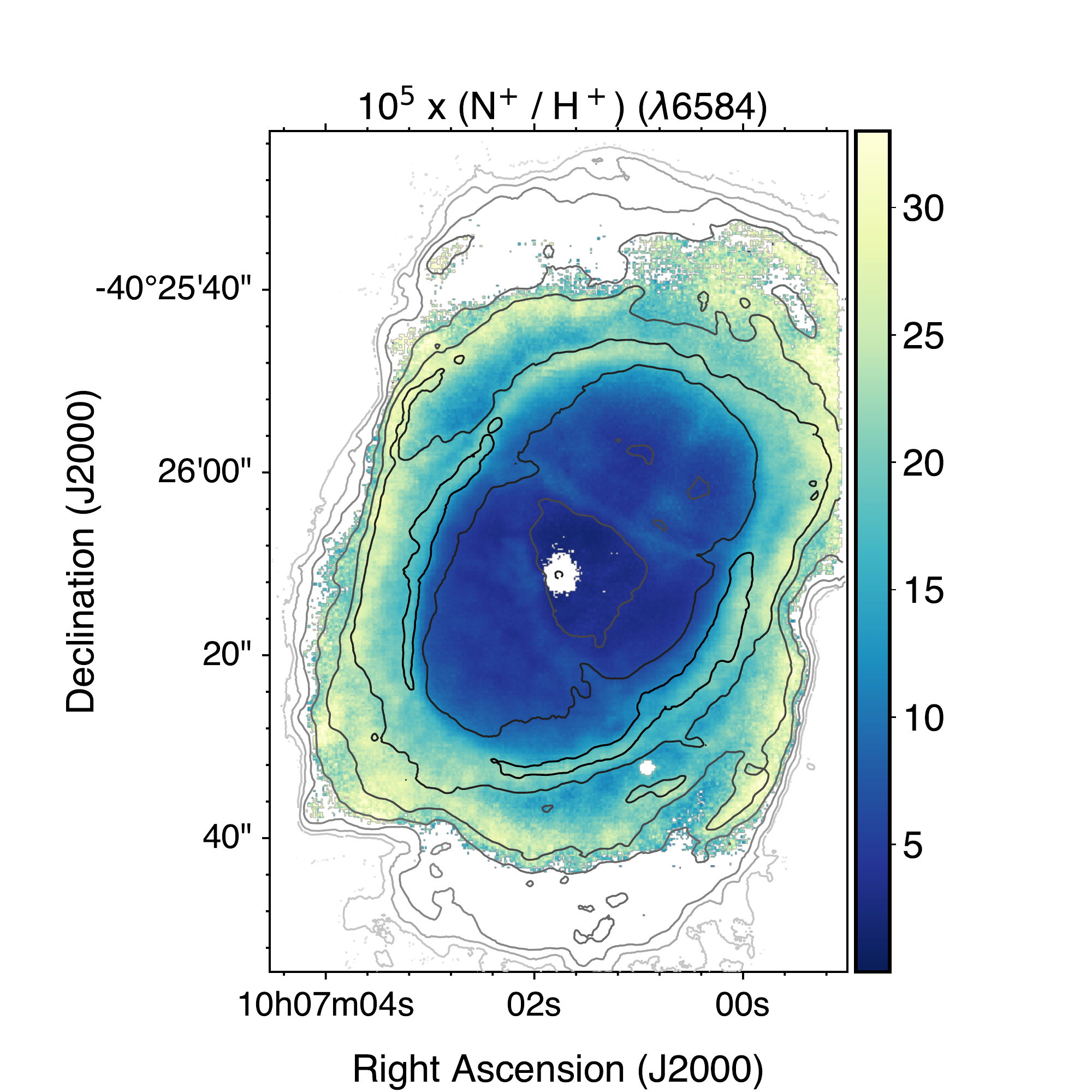}
   \includegraphics[angle=0,width=0.32\textwidth, clip=, viewport=30 0 540 550,]{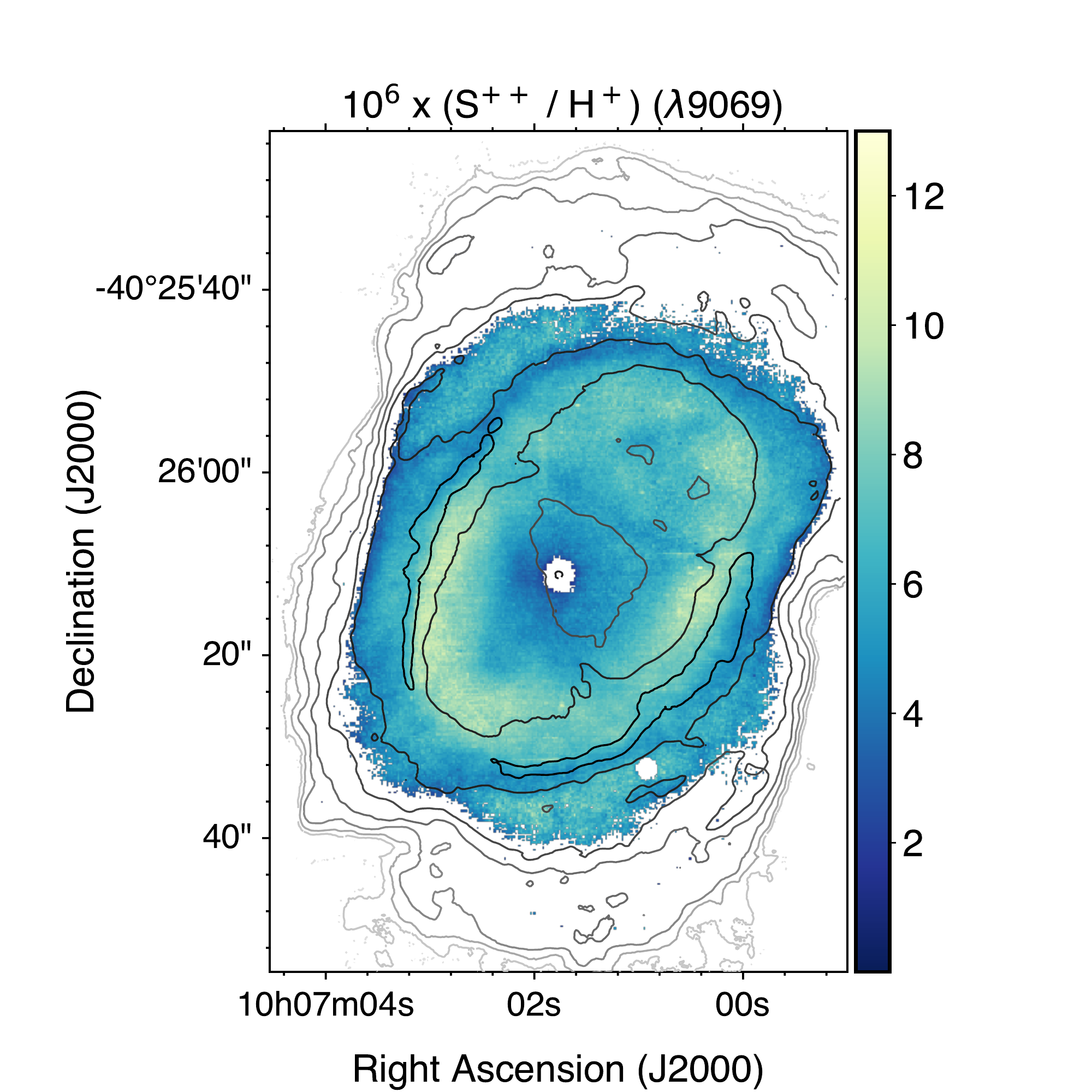}
   \includegraphics[angle=0,width=0.32\textwidth, clip=, viewport=30 0 540 550,]{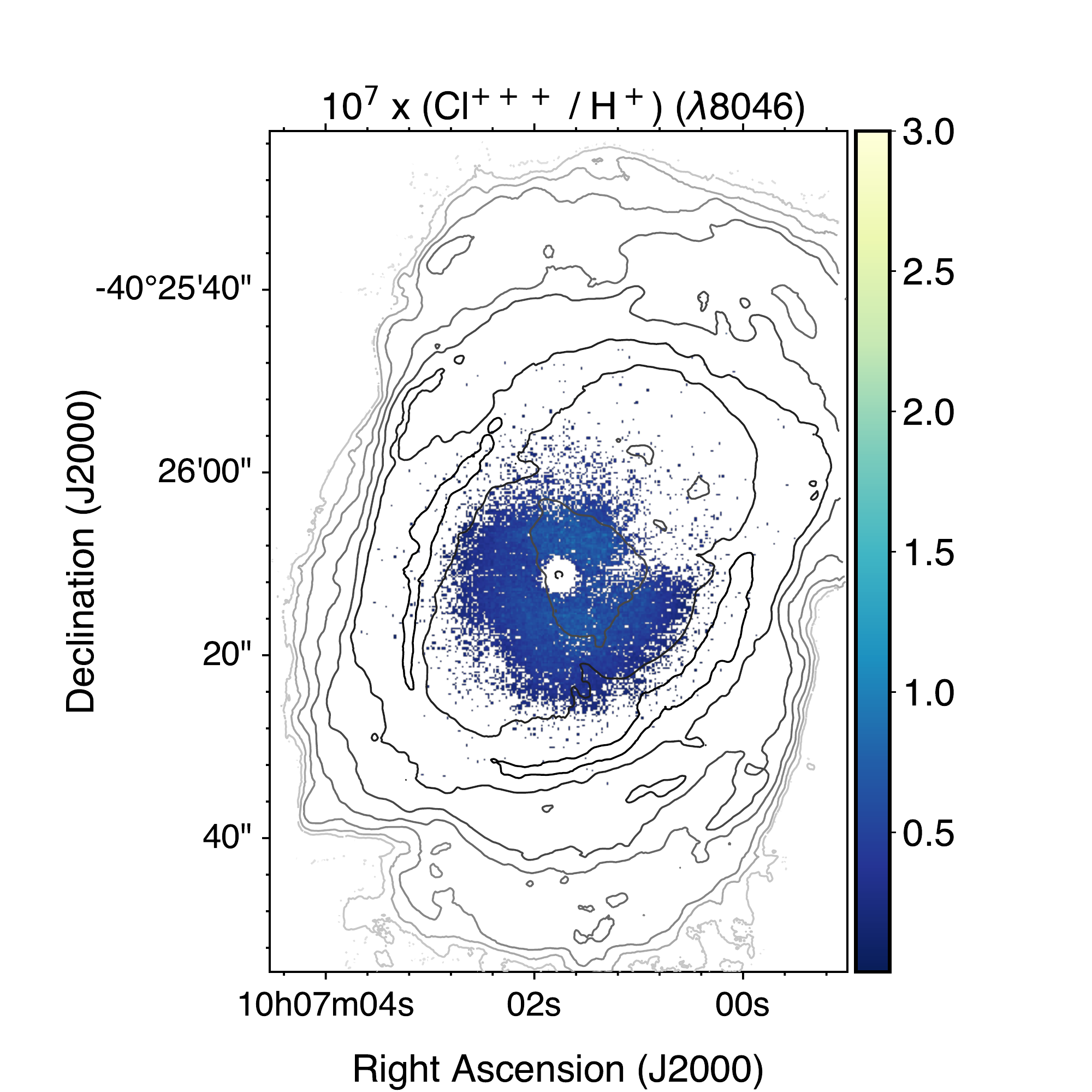}
   \includegraphics[angle=0,width=0.32\textwidth, clip=, viewport=30 0 540 550,]{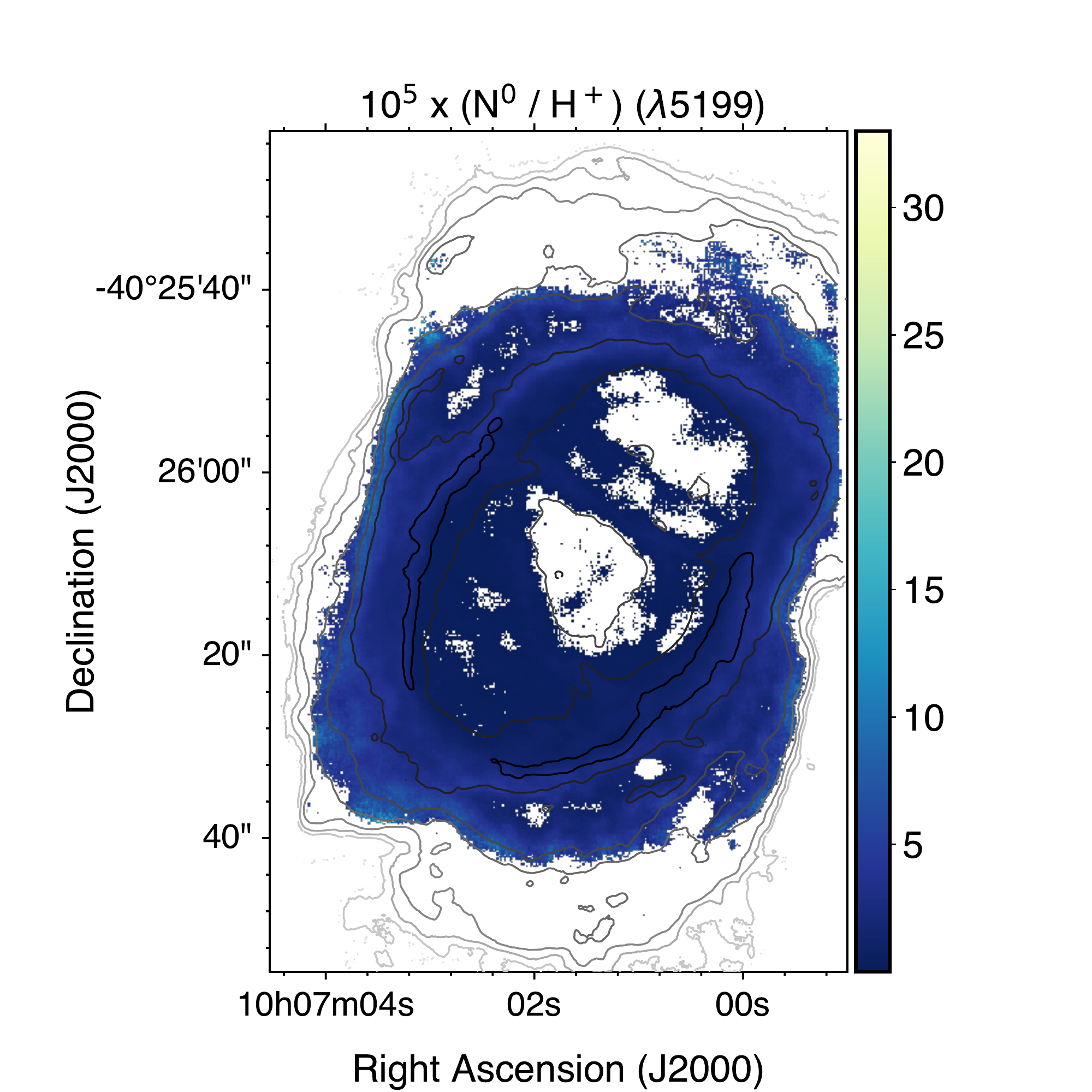}
   \includegraphics[angle=0,width=0.32\textwidth, clip=, viewport=30 0 540 550,]{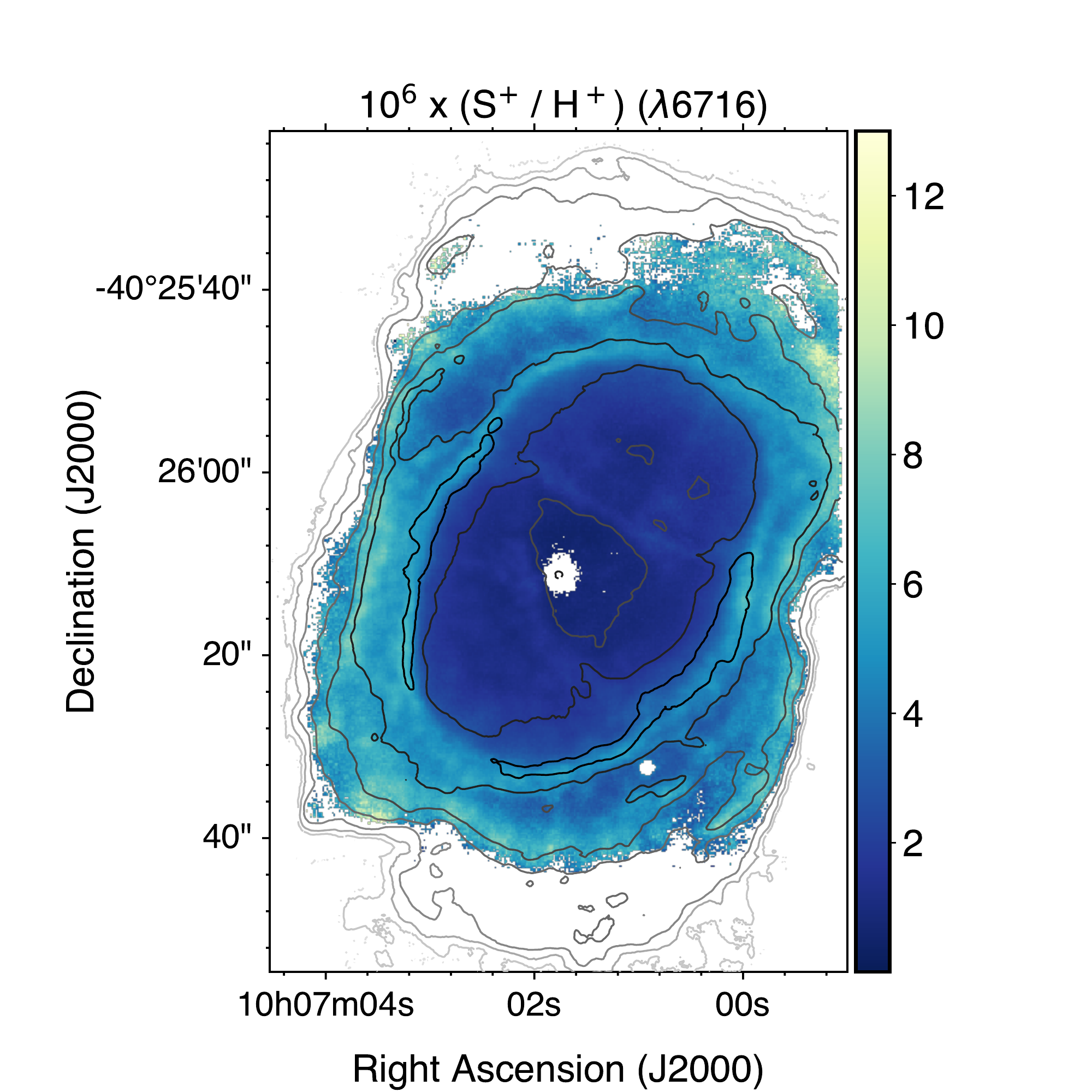}
   \includegraphics[angle=0,width=0.32\textwidth, clip=, viewport=30 0 540 550,]{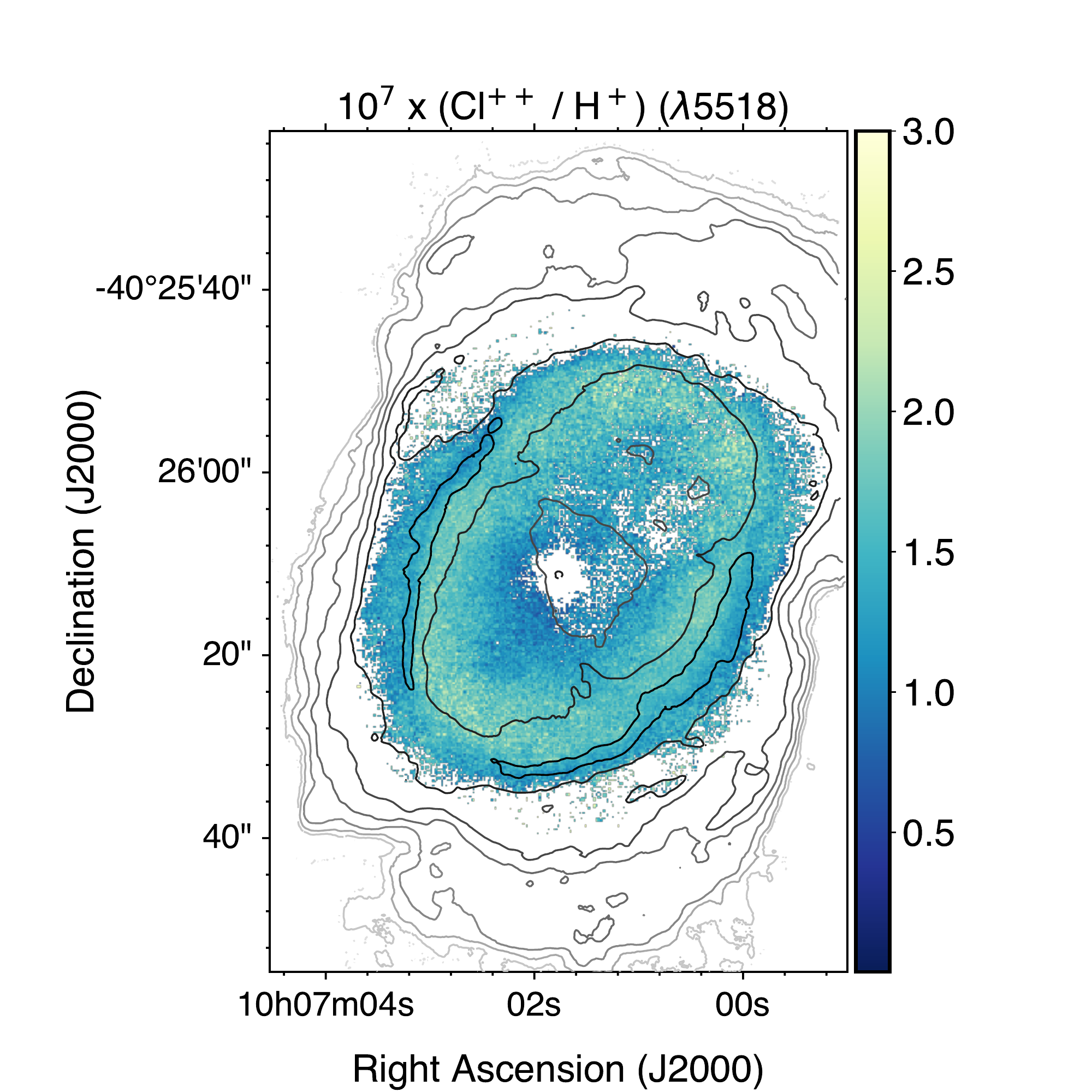}
   \caption{Maps for the ionic abundances of nitrogen (\emph{left}), sulfur (\emph{centre}), and chlorine (\emph{right}). There are two maps tracing different ions for each element, with the more ionised one presented in the upper row. For a given element, both maps display the same range in abundance in order to emphasize the relative contribution of each ion in the different parts of the nebula.
}
   \label{mapNSClion}
    \end{figure*}

\subsection{He$^+$}

There are several singlet and triplet recombination He$^+$ lines within the MUSE spectral coverage, that may or may not be affected by collisional and radiative transfer effects. In the particular case of helium, the 2$^3$S level is metastable and, under certain conditions, the optical depths in lower 2$^3$S–n$^3$P$^0$ lines
imply non-negligible effects on the emission line strengths
\citep{Osterbrock06}.
Radiative effects are not important for photons from the singlet cascade. For transitions in the triplet cascade, the strongest net effect in the optical is for the $\lambda$3889 line (not covered with MUSE) which is weakened by self-absorption, followed by the $\lambda$7065 line, strengthened by resonance fluorescence. The strong $\lambda$5876 line, also in the triplet cascade, is only mildly affected.
Here, we use the four red and brightest \hei\, lines (see \citet{Walsh18} and \citet{MonrealIbero13} for a discussion of the choice of lines). Specifically, these are two singlet lines, insensitive to radiative transfer effects ($\lambda$6678 and $\lambda$7281), and two triplet lines ($\lambda$5876 and $\lambda$7065).

The relative importance of radiative transfer
effects is quantified by the correction factor, $f_\tau(\lambda)$,
for each line which is a function of the optical depth at $\lambda$3889, $\tau$(3889).
This factor depends on the ratio between the radial and thermal velocity.
\citet{Meatheringham88} report a  $v_{exp}$(\oii)$\sim20$ km s$^{-1}$ and  $v_{exp}(\oiii)\sim15$ km s$^{-1}$, suggesting a ratio between radial and thermal velocity of $\omega\sim$2.
\citet{Robbins68} provide tabulated values for $f_\tau(\lambda)$ for $\omega\sim$0, 3, 5 and $T_e$ = 10000, 20000~K. We calculated $\tau$(3889) using the values for  $\omega\sim$3 and $T_e = 10000$ K, which are the closest to the values measured for \object{NGC\,3132}.
In the left panel of  Figure \ref{mapionichelium}, the map for the derived optical depth $\tau$(3889) is presented. This is relatively low in most of the nebula, with the exception of the rim, where it reaches values of $\sim$5.

The He$^+$ map is presented in the central panel of Figure \ref{mapionichelium}. This was created using the strongest \ion{He}{i} singlet line ($\lambda$6678), and the $T_e$ and $N_e$ maps derived from the \sii\, and \nii\, emission lines. Because of the low values of $\tau$(3889) in \object{NGC\,3132} and the mild dependence of the $\lambda$5876 line on radiative transfer effects, the He$^+$ map using this latter line was comparable. The He$^+$ map created from $\lambda$7281 was comparable too, although covering a smaller area, since this is a fainter line.

\subsection{He$^{++}$}

The best optical line to derive He$^{++}$ abundances would be the strong recombination line at $\lambda$4686. This is, however, only accessible with the MUSE extended mode, not used here. Still, the \heii\, line at $\lambda$5411 was detected over an important fraction of spaxels (see Fig. \ref{panel3colores}). Thus, we used this line instead.
The produced He$^{++}$ / H$^{+}$ map is presented in the right panel of Fig. \ref{mapionichelium}. Since He$^{++}$ should be confined to the regions with higher excitation, it was created using the $T_e$(\siii) map and the median value for the $N_e$(\cliii) map. Still, similar maps were produced using other combinations of $T_e$ and $N_e$ (e.g., $T_e$(\nii,\sii) and $N_e$(\nii,\sii)). The derived ionic abundances were equivalent within the uncertainties, as expected, given the weak dependence of recombination lines on $T_e$ and $N_e$ within the range of values in NGC\,3132. 

The He$^{++}$ abundance decreases with radius. The nebula is almost isotropically thick to high energy ($h\nu>54.4$ eV) photons, running out of them at a distance of $\sim$15$^{\prime\prime}$ from the central star. Only in the north-west direction, the nebula presents a blister-like structure where high energy photons can reach radial distances up to $\sim$20$^{\prime\prime}$. 

\subsection{O$^{0}$, O$^{+}$, O$^{++}$}

The derived maps for the three ions are presented in Fig. \ref{mapOion}.
There are several oxygen forbidden lines falling in the MUSE spectral range available this purpose. For the doubly ionised oxygen, we used both the \oiii$\lambda$4959 and  \oiii$\lambda$5007 lines together with the $T_e$(\siii) maps, as presented in Fig. \ref{mapNeTe}, and the median of the derived $N_e$(\cliii). The derived median abundance is compatible with those previously reported measured using integrated spectrum over specific areas \citep{Tsamis03,DelgadoInglada15}.

For single ionised oxygen, we adopted the $T_e$(\nii,\sii) and $N_e$(\nii,\sii) maps. We used both the \oii$\lambda$7320 and \oii$\lambda$7331 lines. This doublet may be affected by a certain O$^ {++}$ recombination contribution. We estimated this contribution using the correction scheme proposed by \citet{Liu00}, and the O$^{++}$/H map derived above and found that for the particular case of \object{NGC\,3132}, this was negligible ($<$1\%) at the rim and dust lanes, but could reach values up to $\sim$10\% in the low surface brightness part of the inner nebula. The range of covered O$^ {+}$/H  values is compatible with those previously reported, measured using integrated spectra over specific areas \citep{Tsamis03,DelgadoInglada15}.

Finally, neutral oxygen abundance was derived using the \oi\, lines observed by MUSE: $\lambda$6364, $\lambda$6300, and $\lambda$5577 (although for this line in a much smaller number of spaxels), together with  $N_e$(\nii,\sii) and $T_e$(\nii,\sii).

\subsection{N, S, Cl, and Ar  ionisation maps}

A blend of four \ion{N}{ii} recombination lines at $\sim$5650-5690~\AA, with the strongest line at $\lambda$5679.6 is covered by the MUSE spectral range. Still, these lines are extremely faint, and even if the $\lambda$5679.6 could be identified in some of the spaxels, the depth of the data was not enough to adequately de-blend all the lines for a significant fraction of the spectra on a spaxel-by-spaxel basis. Thus, all the maps for ionic nitrogen abundance presented here were derived from CELs. 
Specifically, we used the [\ion{N}{i}]$\lambda$5199, \nii$\lambda$6548, and \nii$\lambda$6584 lines. For both, N$^{+}$ and  N$^{0}$, we employed the $N_e$ and $T_e$ maps as derived from the \sii\, and \nii\, lines. 
A representative map for each ion is presented in the left column of Fig. \ref{mapNSClion}. 
%

   \begin{figure}
   \centering
   \includegraphics[angle=0,width=0.32\textwidth, clip=, viewport=30 0 540 550,]{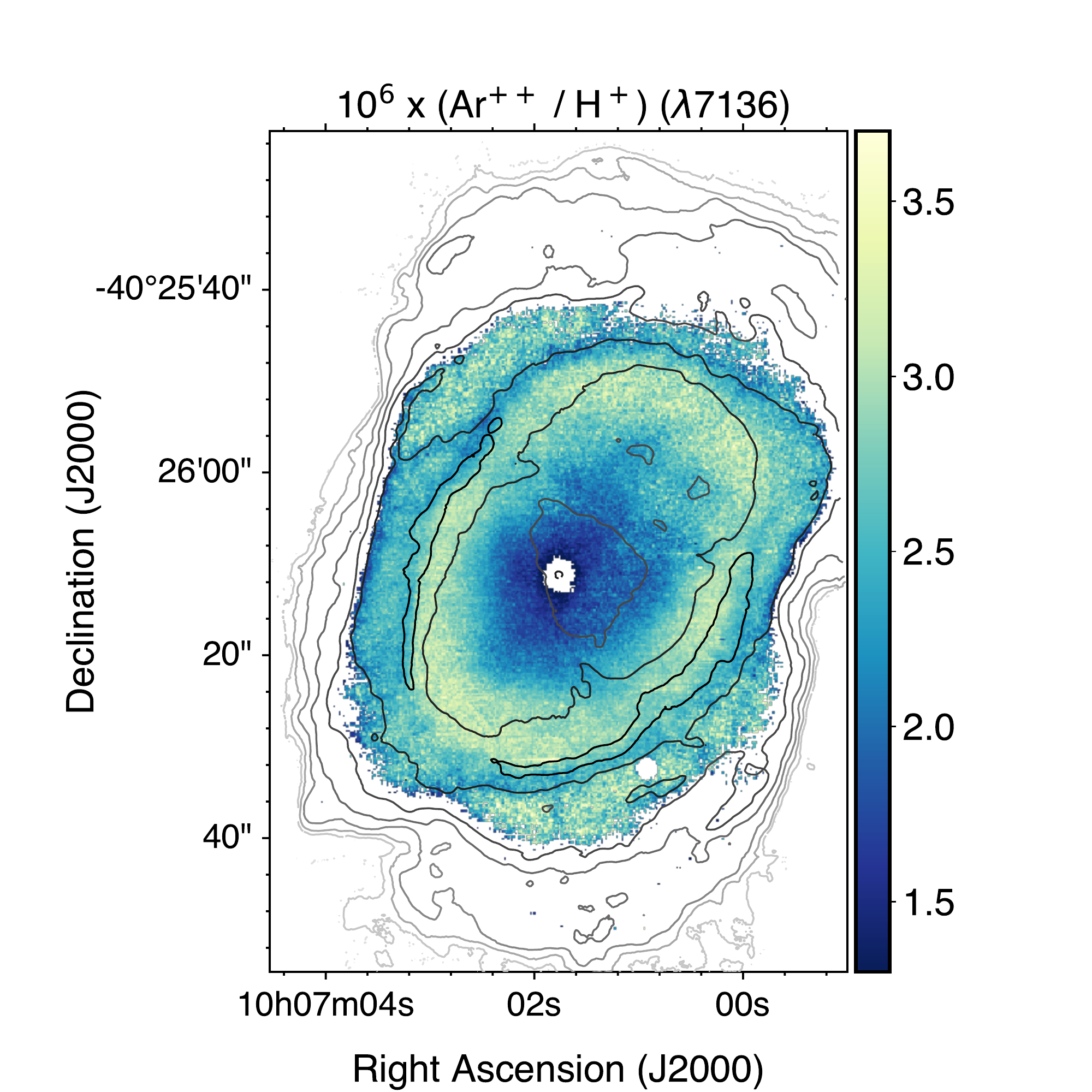}
   \caption{Map for the ionic abundances of Ar$^{+++}$.}
   \label{mapArion}
    \end{figure}

Likewise, abundances in several ionisation states from CEL could be measured for sulfur. In particular, the  S$^+$ abundance was estimated using the \sii$\lambda$6716 and \sii$\lambda$6731 lines, while that for S$^{++}$ was calculated using \siii$\lambda$9069 and \siii$\lambda$6312.
For S$^+$, we used the maps for $T_e$(\nii,\sii), as presented in Fig. \ref{mapNeTe}, while for  S$^{++}$, we used the $T_e$(\siii) map and the median of $N_e$(\cliii) map.
A representative map for each ion is presented in the central column of Fig. \ref{mapNSClion}. 

Also, two chlorine ionisation states are sampled with CELs falling in the MUSE spectral range. The Cl$^{++}$ abundance can be estimated from the \cliii$\lambda$5518 and \cliii$\lambda$5538 lines, while for Cl$^{+++}$ the [\ion{Cl}{iv}]$\lambda$8044, although faint, was detected in the inner parts of the nebula.
To determine the abundances for both of these ions, we simply used  the $T_e$(\siii) map and the median of $N_e$(\cliii) map, as for S$^{++}$.
A representative map for each ion is presented in the right column of Fig. \ref{mapNSClion}.

Finally, some argon CELs also fall in the MUSE spectral range. \ariv$\lambda$7237 and \ariv$\lambda$7263 were too faint to derive a reliable 
Ar$^{+++}$ abundance on a spaxel-by-spaxel basis. Thus, for this element we determined only the Ar$^{++}$ abundance. We used the \ariii$\lambda$7136 and \ariii$\lambda$7751 lines together with the $T_e$(\siii) map and the median of $N_e$(\cliii) map.
A representative map for this ion is presented in  Fig. \ref{mapArion}. 

In general, all the derived values (see Tab. \ref{TabIon} and Figs. \ref{mapNSClion} and \ref{mapArion}) are in accord with ionic abundances as measured in specific areas reported in the literature \citep{Tsamis03,DelgadoInglada15} but extend these determinations over the whole face of the nebula.

   \begin{figure}
   \centering
   \includegraphics[angle=0,width=0.32\textwidth, clip=, viewport=30 0 540 550,]{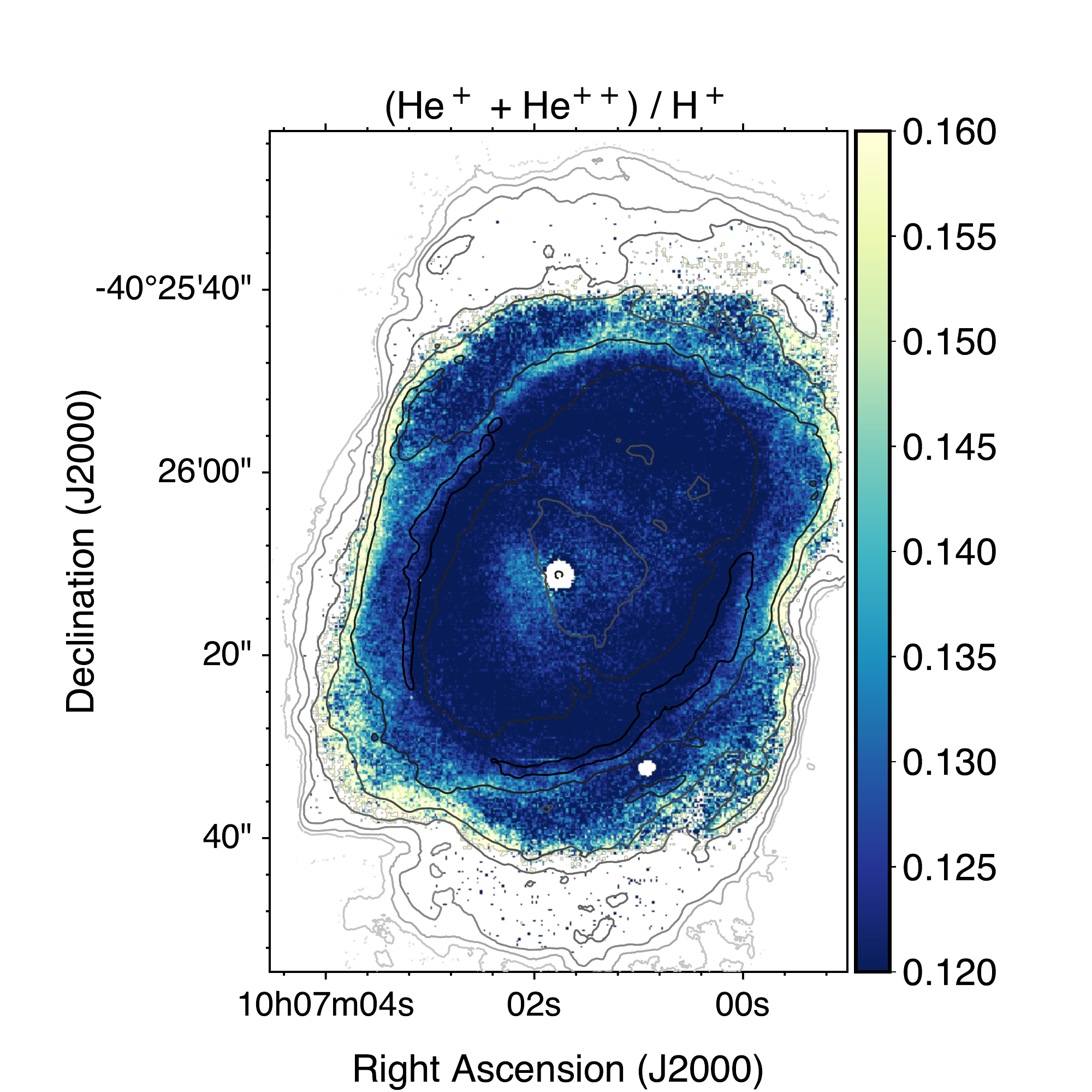}
   \caption{Map of the total helium abundance as the sum of the He$^{+}$/H$^{+}$ and the He$^{++}$/H$^{+}$ maps as presented in Fig. \ref{mapionichelium}.}
   \label{mapHeTot}
    \end{figure}

   \begin{figure*}
   \centering
   \includegraphics[angle=0,width=0.32\textwidth, clip=, viewport=30 0 540 550,]{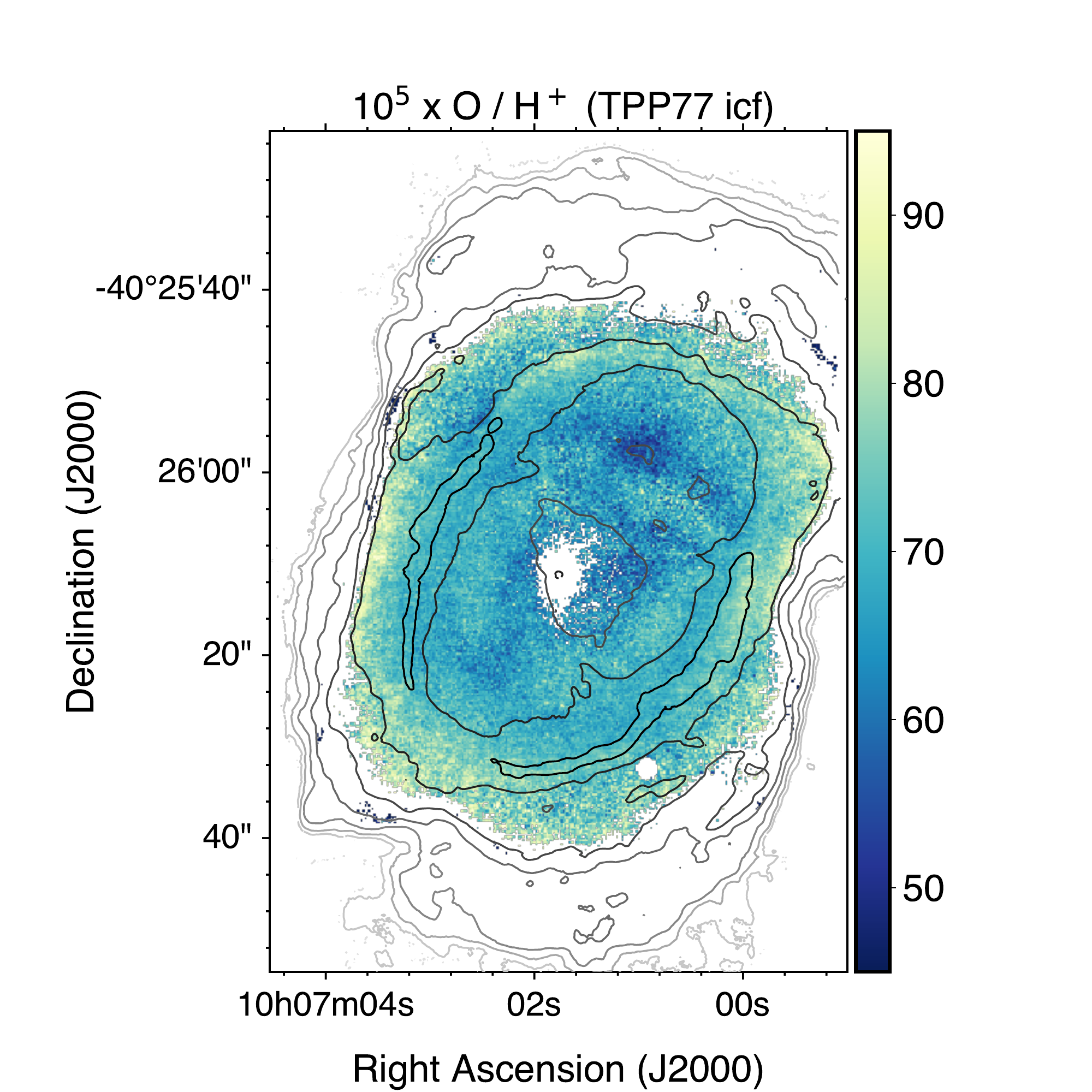}
   \includegraphics[angle=0,width=0.32\textwidth, clip=, viewport=30 0 540 550,]{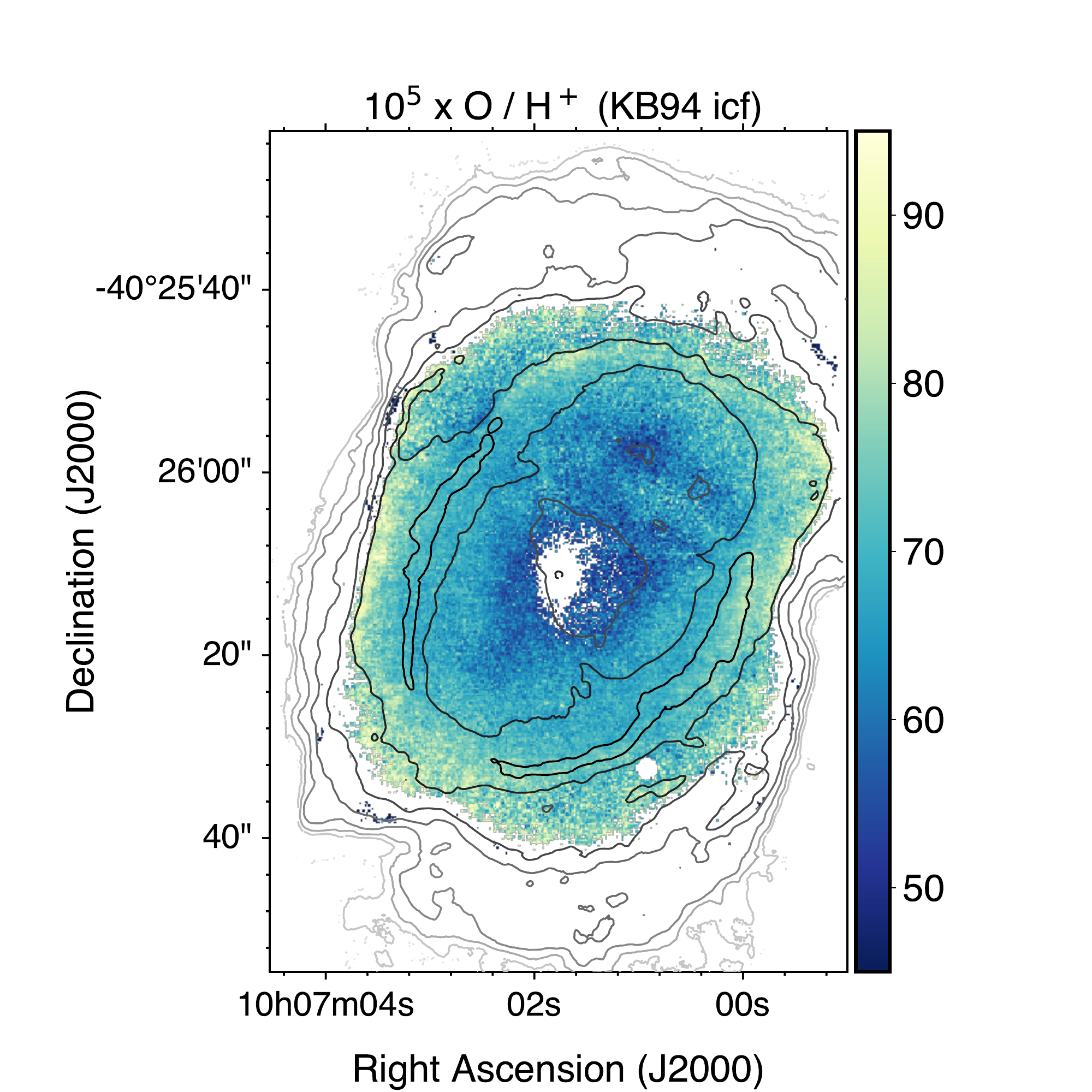}
   \includegraphics[angle=0,width=0.32\textwidth, clip=, viewport=30 0 540 550,]{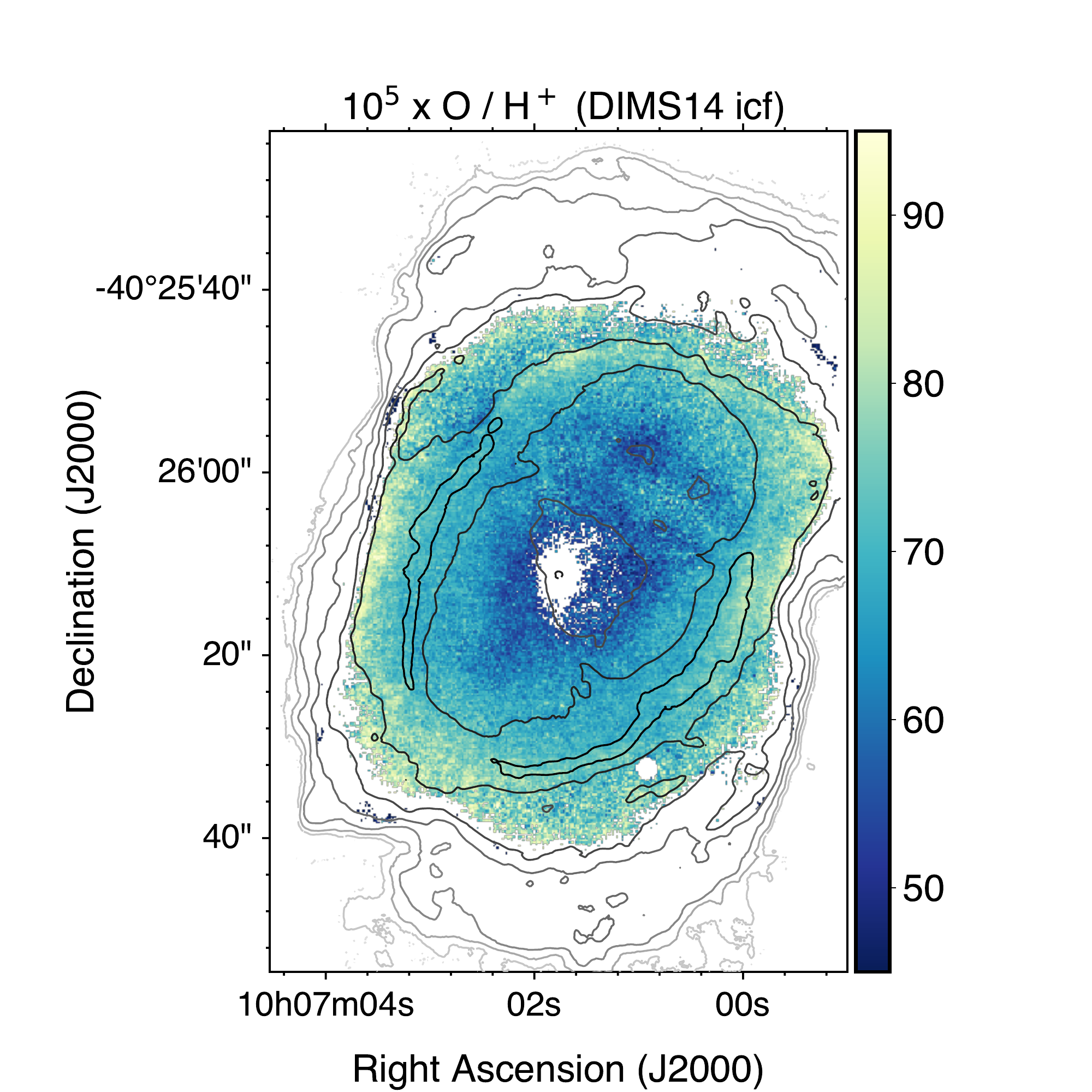}
   \caption{Maps of total O / H$^+$ using the ionisation correction factors to account for the presence of O$^{+++}$ as provided by \citet{TorresPeimbert77} (\emph{left}), \citet{Kingsburgh94} (\emph{centre}), and \citet{DelgadoInglada14} (\emph{right}).  All the three maps display the same range in abundance in order to better compare the different ionisation correction schemes.}
   \label{mapOtot}
    \end{figure*}

\section{Element abundance maps \label{secelement}}

\subsection{Helium}

As starting point, we assumed that the amount of He$^0$ is negligible in \object{NGC\,3132} (see also discussion about Fig. \ref{panel3colores} in Sect. \ref{seclinempas}), and we calculated the total amount of helium by simply summing the He$^+$ and He$^{++}$ maps derived in previous section.
The simple 1D photoionisation model presented in Sect. \ref{sec1Dmodel} predicts that 3\% of the detected helium is neutral, although of course provides no hint as to its spatial dependence.
Basic statistics are included in Table \ref{TabIon} while
the resulting map is presented in Figure \ref{mapHeTot}. The derived abundances are compatible with those previously reported using integrated spectra over specific areas \citep{Tsamis03,GarciaHernandez16}.

The map in Fig. \ref{mapHeTot} shows slightly higher values at the rim and some parts of the shell, but still these enhancements are always $<10$\% of the typical median helium abundance.
This is at a level comparable with our estimated uncertainties. Thus, it cannot be concluded that this implies an overabundance in helium at those locations.
Still, it is worth noting that the spaxels with higher He/H values are localised and present a rather coherent distribution. A similar situation was also identified in \object{NGC\,7009}, also observed with MUSE \citep{Walsh18}.

\subsection{Oxygen}

There are no [\ion{O}{iv}] lines in the optical spectral range. Thus, in absence of observations in the ultraviolet or infrared, in order to derive the total oxygen abundance, one must correct for the presence of O$^{+++}$. When the abundance of one or several ions of a given element is not available, it is customary the use of different ionisation correction factor (ICF) schemes.
These rely on the available abundances of ions of other elements with comparable ionisation potentials to those of the ionic species of the element of interest.
For the particular case of oxygen, typical schemes make use of certain combinations of the He$^{+}$ and  He$^{++}$ abundances. The first one was proposed by \citet{TorresPeimbert77}. Later on \citet{Kingsburgh94} and \citet{DelgadoInglada14} proposed additional schemes based on this same ratio.
All of them were determined for the nebula as a whole. 
Now that integral field spectroscopic observations are carried out in a routine manner and several 3D ionisation codes are available, \citet{Morisset17} suggested the possibility of using ICF schemes that change depending on the local conditions within the nebula.

   \begin{figure*}
   \centering
   \includegraphics[angle=0,width=0.32\textwidth, clip=, viewport=30 0 540 550,]{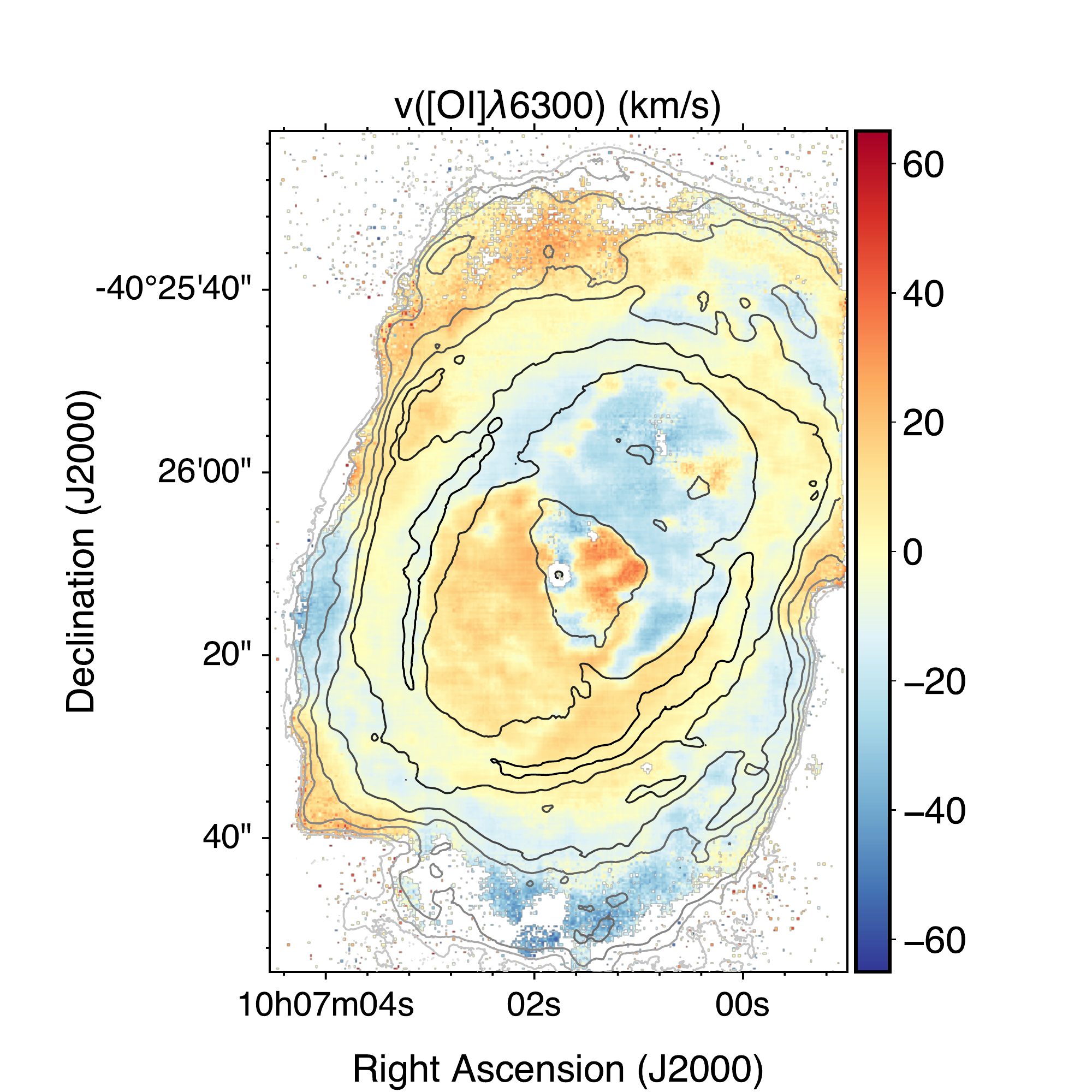}
   \includegraphics[angle=0,width=0.32\textwidth, clip=, viewport=30 0 540 550,]{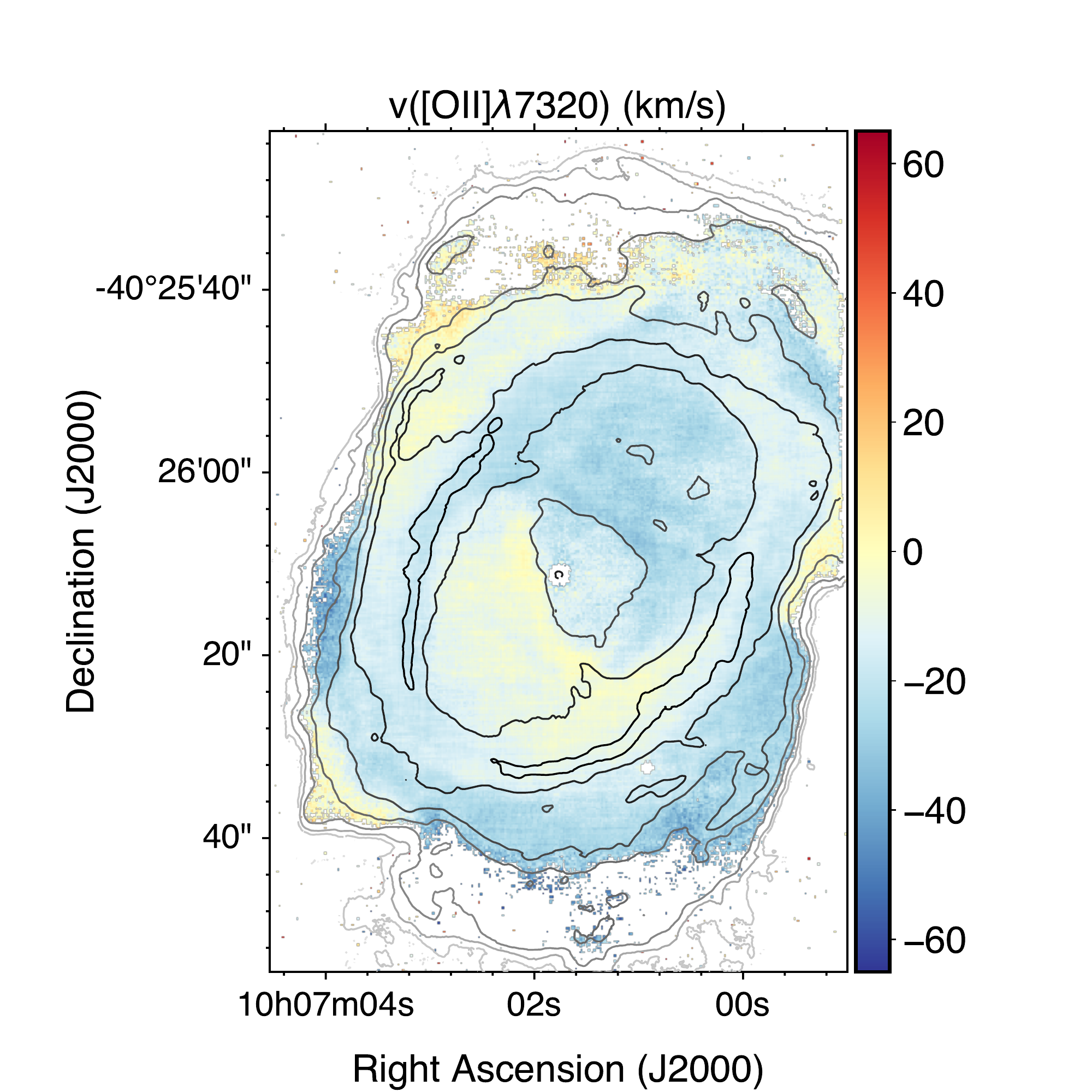}
   \includegraphics[angle=0,width=0.32\textwidth, clip=, viewport=30 0 540 550,]{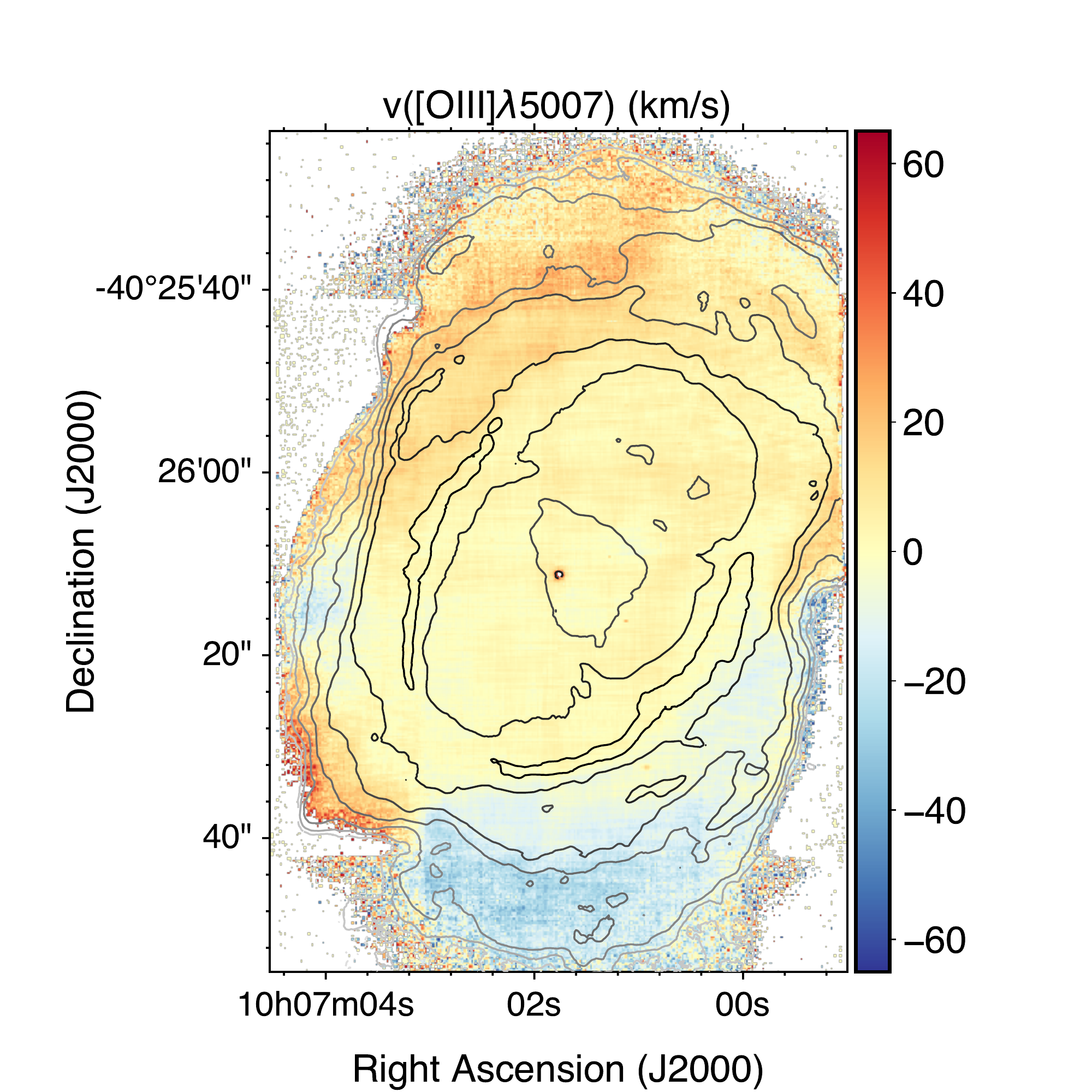}
   \includegraphics[angle=0,width=0.32\textwidth, clip=, viewport=30 0 540 550,]{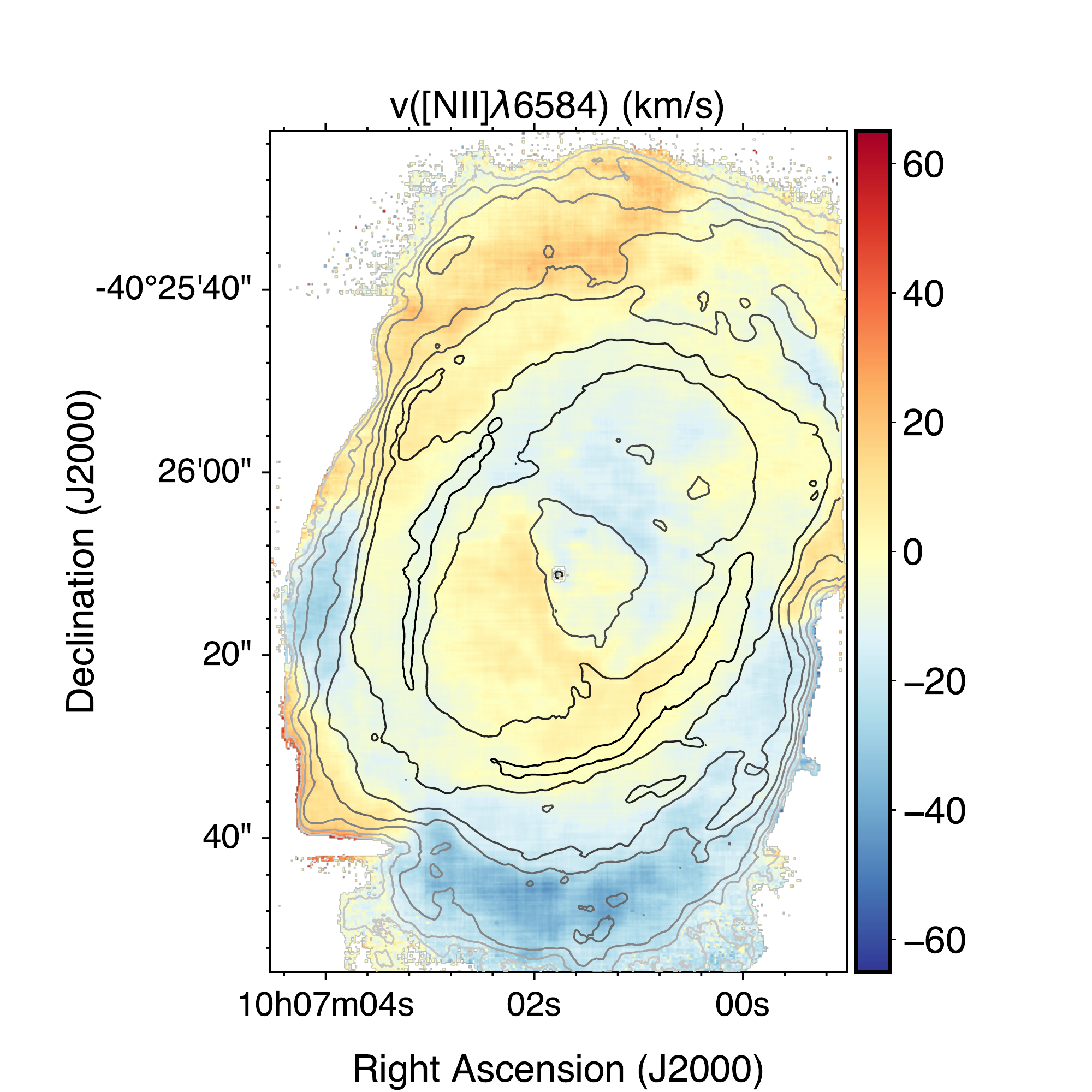}
   \includegraphics[angle=0,width=0.32\textwidth, clip=, viewport=30 0 540 550,]{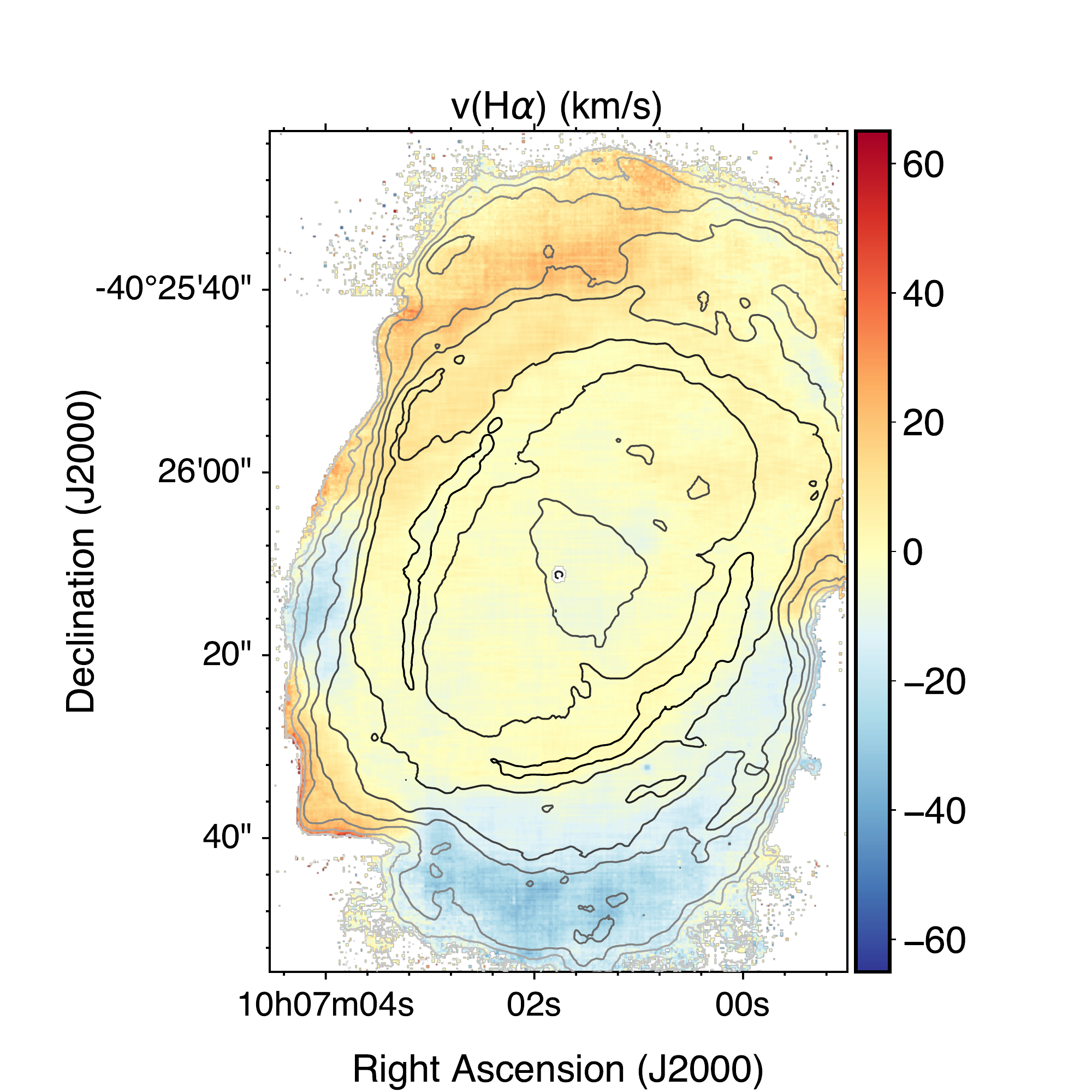}
   \includegraphics[angle=0,width=0.32\textwidth, clip=, viewport=30 0 540 550,]{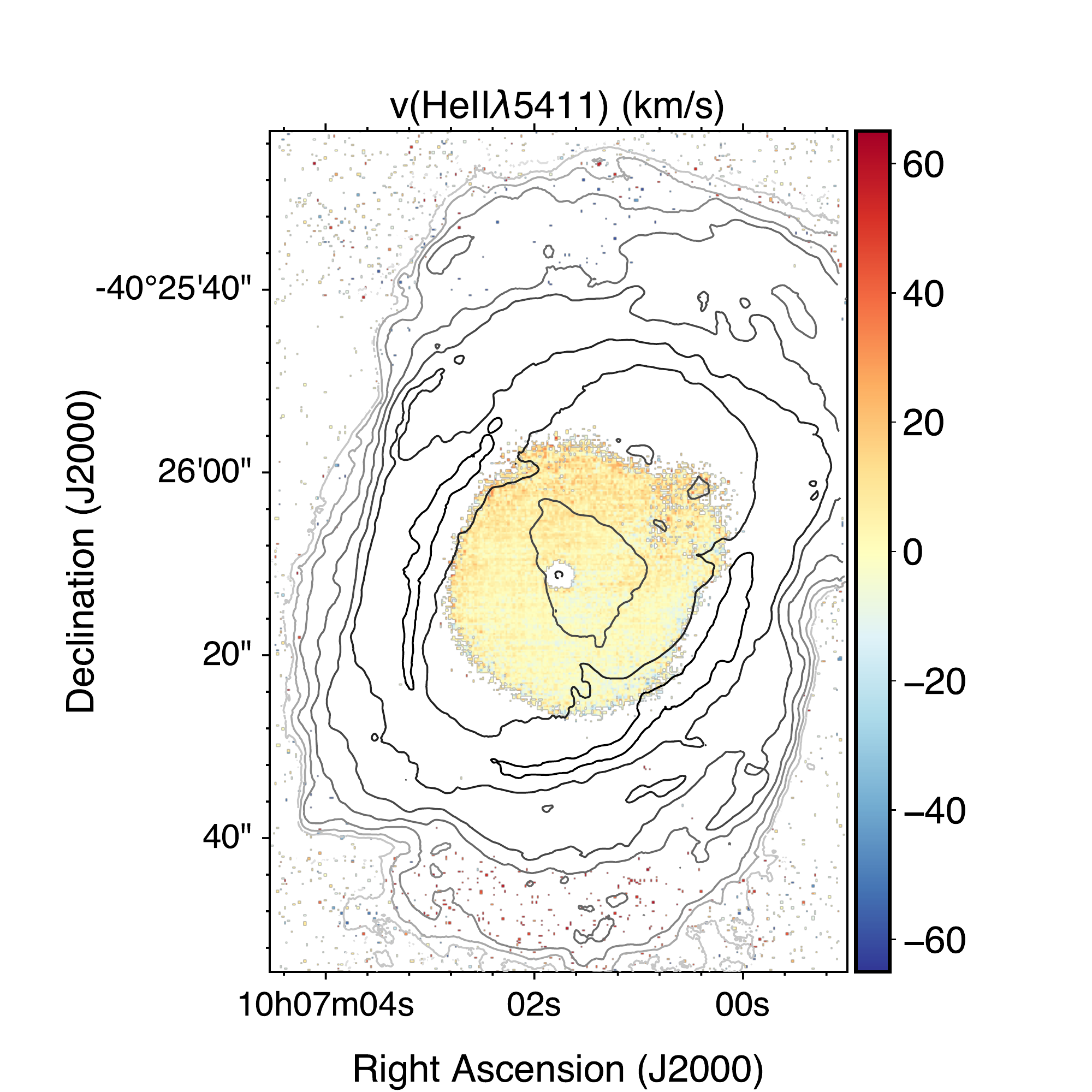}
   \caption{Velocity field maps for some relevant emission lines. In the upper row, lines corresponding to different ions of the same element (oxygen) are presented (from left to right: neutral, singly ionised, and doubly ionised). In the lower row, three additional maps for ions with ionisation potential increasing from left to right are presented. They are the bright \nii$\lambda$6584 (\emph{left}), \ha\, (\emph{centre}), and \heii$\lambda$5411 (\emph{right}) emission lines.}
   \label{mapvel}
    \end{figure*}

Still, the first step should always be exploring the consequences of using the same scheme all over the nebula. 
The simplest expectation is that, after correction, the nebula should present an (almost) homogeneous abundance. Departures from this homogeneity may suggest the path to be followed in the search for an optimal location-dependent ICF scheme.

Fig. \ref{mapOtot} presents the oxygen abundance maps, using all the three mentioned ICF schemes.
These are not affecting the outer part of the nebula, since \heii\, was not detected there. Thus, all the three maps are equivalent there.
The abundance maps are relatively flat in all the cases, being the flattest when using the scheme proposed by \citet{TorresPeimbert77}. Both the schemes by \citet{Kingsburgh94} and \citet{DelgadoInglada14} still show a small deficit and would need an additional correction of about 7-10\% to attain a similar level of homogeneity as with the former ICF scheme.
A similar exercise with \object{NGC\,7009} also showed better results (understood as closer to flatness in abundance) when using the scheme proposed by \citet{TorresPeimbert77}.
Proposing the optimal correction scheme for oxygen abundances in PNe when spatial information is available is well beyond the scope of this contribution. It would certainly need similar data for more examples of PNe, as well as detailed photoionisation modeling.
Still, the two cases studied so far suggest that a scheme not that different to that proposed by \citet{TorresPeimbert77} should suffice.

\section{Velocity maps \label{seckinema}}

A key piece of information to understand the evolutionary status of a planetary nebula is its expansion velocity, or
more correctly, the expansion velocity field in a 2D study. It can be used to estimate the kinematic age of the nebula, and put constraints on the post-AGB age. In  the case of time-dependent nebular expansion, this simple estimate of kinematic age will need refinement. However, rigorous measurement of this information is not straight-forward.
The expansion velocity could be estimated by e.g., disentangling the approaching and receding side of the nebula by means of  high-resolution spectroscopy. Yet, the derived velocities would depend on the used ion and the position within the nebula where the measurement is taken. For the specific case of \object{NGC\,3132}, \citet{Sahu86} measured  $v_{\rm{exp}}$(\oiii)$\lesssim10$ km s$^{-1}$ at five specific 15$^{\prime\prime}$-diameter apertures, while
\citet{Meatheringham88} measured $v_{\rm{exp}}$(\oii)$\sim20$ km s$^{-1}$ and $v_{\rm{exp}}$(\oiii)$\sim15$ km s$^{-1}$, at a non-reported location.
More recently, \citet{Hajian07} provided with position-velocity diagrams in \nii\, and \oiii\, along the minor axis and several cuts perpendicular to it (see their figures 17 and 19). The diagrams showed that lines were double peaked with differences between peaks always $\lesssim100$~km~s$^{-1}$ for \nii. In the northern half of the nebula, the bluest peak was largely dominant, and the red peak was barely detected, whilst this behaviour was reversed in southern half. For \oiii, differences between peaks were smaller ($\lesssim50$~km~s$^{-1}$) with a less clear separation between peaks.

Fig. \ref{mapvel} shows the derived MUSE velocity maps for a selection of the strongest emission lines. The upper row contains those for neutral to doubly ionised oxygen, while the lower row contains three ions with very different ionisation potential.
For the lines presented in this figure, MUSE resolution ranges from $\sim$1700 (\oiii$\lambda$5007) to  $\sim$2800 (\oii$\lambda$7320), corresponding to a FWHM ranging from $\sim$180 to $\sim$110~km~s$^{-1}$. This is clearly not enough to resolve the  double-peaked profiles previously reported.
Nonetheless, MUSE 2D mapping capability in many emission lines offer valuable information to understand the dynamical status of the nebula. The figure shows that kinematics in the nebula is complex. Yet, some patterns are recognisable:
i) in the area inside the rim, those ions with higher ionisation potential (\heii, \oiii) or extending in the whole extent of the nebula along a given line of sight (\ha) present a velocity field without any structure tracing the systemic velocity of the nebula ($v_{\rm{sys}}=5\pm1$~km s$^{-1}$), defined as the mean of the medians in the maps presented in Fig. \ref{mapvel};
ii) always inside the rim, those ions that, in a stratified view of the nebula, should occupy a more external shell-like structure (\oii, \oi, \nii) display a more structured velocity field with approaching velocities in the northern half of the nebula and receding velocities in the south;
iii) the contrast between the approaching and receding side is larger for ions with lower ionisation potential, following the sequence   \oii$\rightarrow$\nii$\rightarrow$\oi;
iv) when lines are detected beyond the rim, in the doily-like shell, the behaviour is reversed, with receding velocities in the northern side and approaching velocities in the southern one;
v) the map for the ion with the highest ionisation potential (\heii), which should be confined to the innermost layer of the nebula presents somewhat redder velocities in the northern half of the nebula, with differences between the northern and southern halves of $\sim$4~km~s$^{-1}$

An expanding shell-like picture seems difficult to reconcile with the whole set of these patterns. Instead, the derived velocity fields could fit much better in the diabolo-like picture proposed  by \citet{Monteiro00}. Additionally it can provide some constraints regarding the orientation and inclination of the diabolo structure.
In the map with the highest contrast ($v$(\oi)), the line emission should come from a comparatively thin layer along the line of sight. Thus, this map can be used to define the geometry of the ''walls'' of diabolo structure. 
The global velocity field, including both the inner and outer parts of the nebula, can be seen as two intertwined ring-like structures, each of which would delineate one of the sides of the diabolo. The position angle of the major axis of the nebula ($\sim-22^\circ$, see Sect. \ref{seclinempas}) should give the relative orientation of the two structures, while the ellipticity of each of the rings would give the inclination of the diabolo with respect to the plane of the sky. We measured a ratio between minor and major axes in both ellipses of about $b/a\sim0.7-0.8$, implying an inclination angle of $i\sim40-45^\circ$.
In this picture, the orientation with respect to the plane of the sky is quite in accord with that proposed by \citet{Monteiro00}. However, the position angle at which the diabolo should be oriented is somewhat different.
The other low-ionisation lines would trace a similar structure, but inner and closer to the ionising source.

On their side, higher ionisation species would occupy the innermost part of the nebula. The spectral resolution of the present data does not allow to discern whether they will do so as a wholly filled structure or a similar diabolo-like structure but with smaller velocity differences between the northern and southern sides of the diabolo.
In spite of the relatively low spectral resolution provided by MUSE, the analysis presented here nicely illustrates the power of 2D kinematic information in many emission lines to probe the intrinsic structure of the nebula.
Currently, work using IFS data to disentangle the 3D intrinsic structure of PNe by means of detailed kinematic analysis exist, but are relatively rare and generally rely on one or only a few lines \cite{Danehkar15,Danehkar16}.
The complementary approach of using Fabry-P\'erot instruments to derive 2D kinematic information at high spectral resolution (again typically for one or a few spectral features) has also already proved its potential \citep[e.g.][]{Pismis89,Arias01}.
Clearly, kinematic studies of this (and other) PNe would greatly benefit from the existence of instruments capable of simultaneously observing many lines at relatively high spectral resolution, such as the recently proposed BlueMUSE \citep{Richard19}.

   \begin{figure}[th!]
   \centering
   \includegraphics[angle=0,width=0.49\textwidth, clip=, viewport=0 0 928 280,]{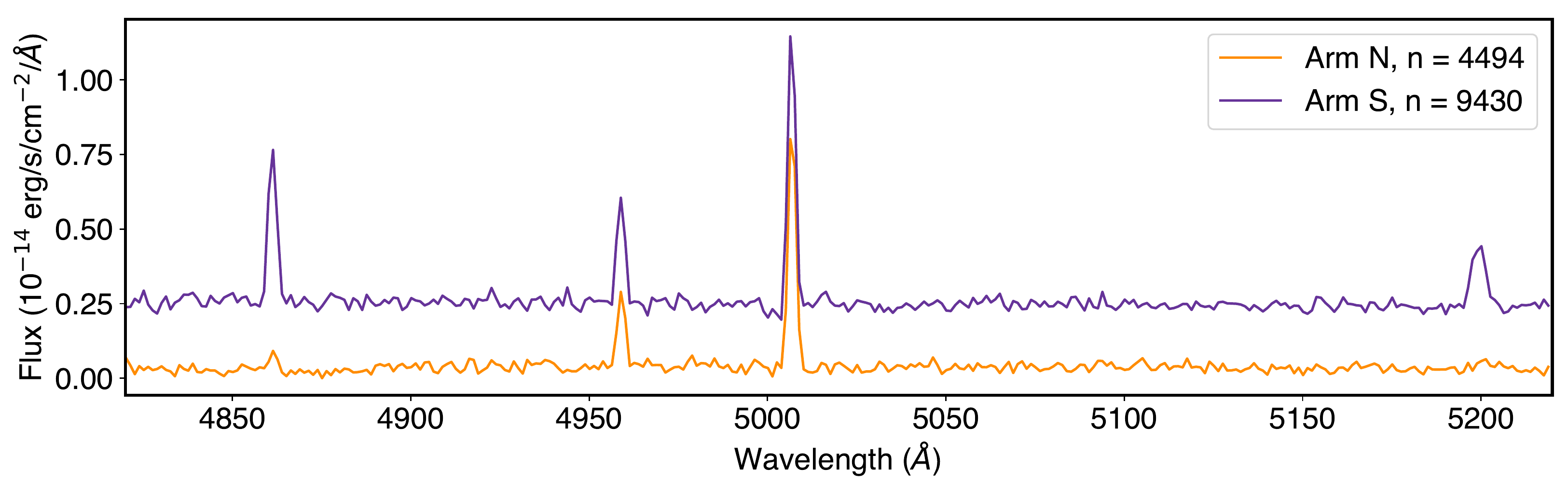}
   \includegraphics[angle=0,width=0.49\textwidth, clip=, viewport=0 0 928 280,]{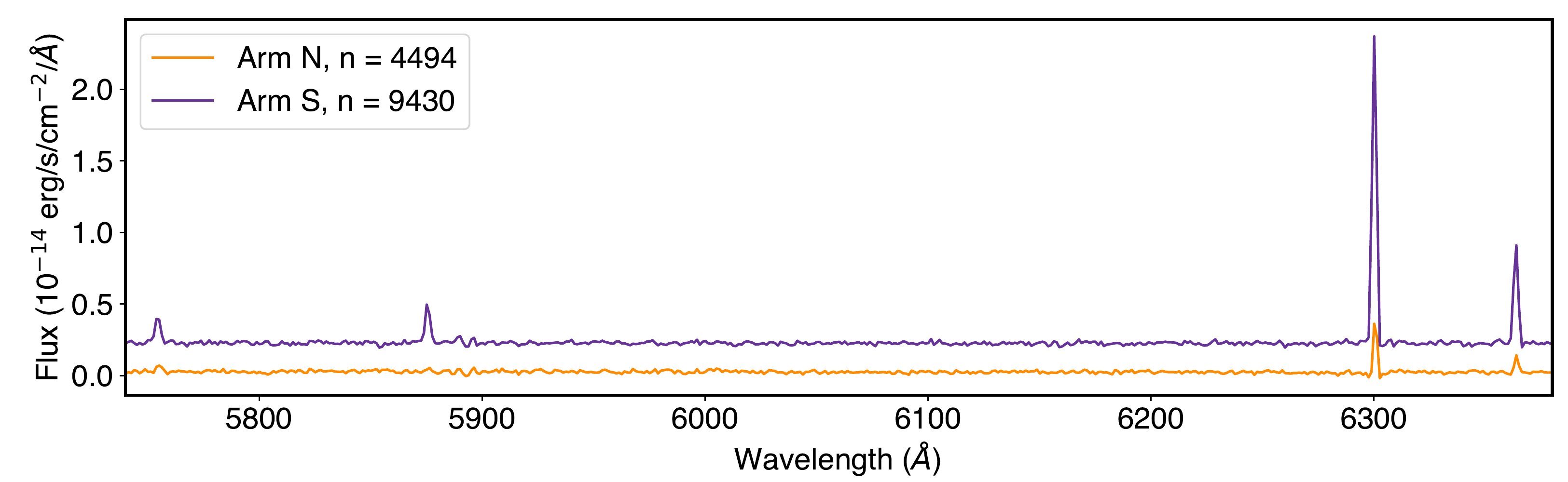}
   \includegraphics[angle=0,width=0.49\textwidth, clip=, viewport=0 0 928 280,]{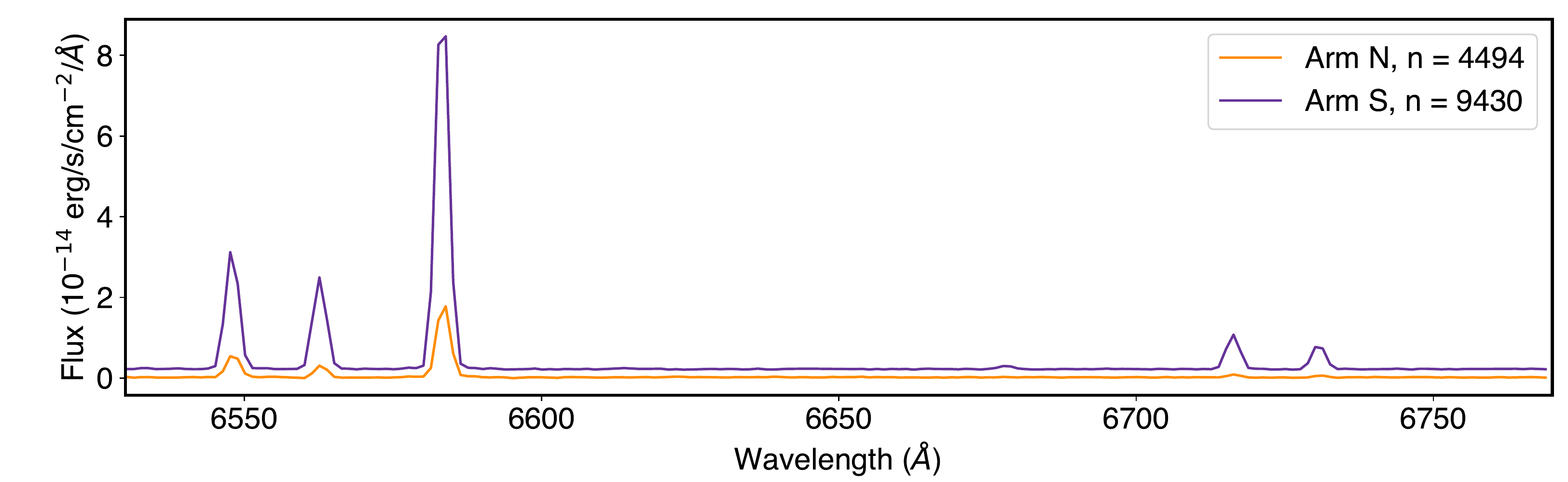}
   \includegraphics[angle=0,width=0.49\textwidth, clip=, viewport=0 0 928 280,]{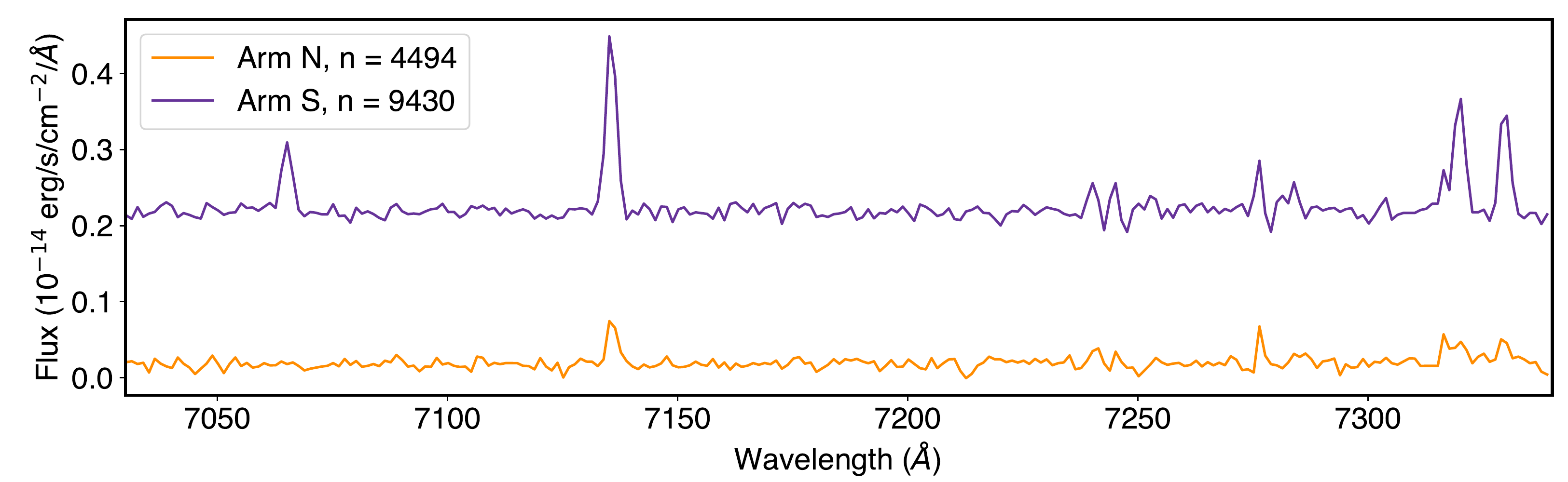}
   \caption{Selected spectral ranges showing the emission lines detected in the northern (\emph{orange}) and southern (\emph{violet}) LIS. Additionally, \siii$\lambda$9069 (not shown)  was the only emission line detected in the reddest part of the MUSE spectral range. To improve visibility, we applied an offset of 0.2 in the y axis to the spectrum of the southern arm. 
   \label{specarcs}}
    \end{figure}

\section{The low-ionisation structures \label{secarms}}

In Sect. \ref{seclinempas}, we reported the detection in \nii\, (and to a certain extent, also in \ha\, and \sii) of two extended arc-like structures at the northern and southern poles of the nebula. 
The relative strength of these lines suggests that they might be some kind of low-ionisation structures \citep[LISs,][]{Goncalves01,Goncalves04}.
LIS is a label encompassing a variety of small-scale structures including among others knots, filaments, and jets, that may or may not move at supersonic velocities with respect to the large-scale component in which they are located.
According to their velocities, LISs can be classified as FLIERs \citep{Balick93}, with radial velocities of 24-200 km~ s$^{-1}$, or SLOWERs \citep[slow moving low-ionisation emission regions, ][]{Perinotto00}, which share the expansion velocities of their host PNe.
Models can explain, to some extent, their morphology and kinematics, but questions such as what are the physical and chemical properties of the LISs, what are the specific physical processes for their origin or what is the dominant excitation mechanism remain open.

From the morphological point of view the two LISs detected in \object{NGC\,3132} are located roughly at $\sim$1$^{\prime}$ (0.25 pc) from the central star, well beyond the rim, and in a point-like symmetry.
They have an extremely low-surface brightness, between $\sim$200 (southern LIS) and $\sim$1000 (northern LIS) fainter than the brightest structure in the nebula (the rim) in the \nii\, emission lines.
The structures resemble the pair of jets reported for \object{IC 4846} \citep{Akras16}, \object{IC 4634} \citep{Hajian97} and \object{NGC 3918} \citep{Corradi96} or the jetlike pairs in NGC 6309, M3-1 \citep{Corradi96}, and He 2-5429 \citep{Guerrero99} but with much better developed arc-like structures (sizes of about $\sim30^{\prime\prime}-45^{\prime\prime}$, $\sim0.12-0.18$~pc).
Velocities in both LISs, as traced by  \nii$\lambda$6584  (not shown) are low, with those in the 
northern LIS being redder than those in the southern one. Still, within each LIS, velocities are
relatively homogeneous, without any obvious structure. In that sense, they could either be SLOWERs or fast (FLIERs) very close to the plane of the sky.

As stated before, only very few emission lines where detected in the arcs on a spaxel-by-spaxel bases. Thus, to characterise the physical and chemical properties of the arcs in a similar manner to the rest of the nebula, we created two high signal-to-noise spectra by co-adding the spaxels in the orange and violet boxes in Fig. \ref{apuntado}. All those spaxels with an integrated flux between 6\,800 and 6\,840~\AA\, $>$6$\times10^{-18}$~erg~s$^{-1}$~cm$^{-2}$ were masked to minimise the contribution of stellar continua from several faint stars in the the co-added spectrum. The resulting extracted spectra are displayed in Fig. \ref{specarcs}, where the total number of co-added spaxels is also indicated. The co-addition allows for the detection of some additional emission lines, useful to derive some basic physical and chemical properties for the LISs.
We measured their fluxes in the same manner as in the spaxel-by-spaxel analysis (see Sect. \ref{seclinempas}). These are listed in Table \ref{tabthearms}. Likewise, the derived physical and chemical properties are listed in Table \ref{tabthearmsprop}. Again, these were derived as in Sections \ref{secextin}, \ref{secionic}, and \ref{secelement}. Unsurprisingly, He$^{++}$ was not detected in the LISs. Thus, we do not list a total helium abundance. 
Regarding oxygen, the contribution of recombination to the O$^{++}$ was estimated following the correction scheme proposed by \citet{Liu00} and it was found negligible.
Also, we assumed no O$^{+++}$ in the LISs, which a reasonable assumption for a structure dominated by low-ionisation lines, and thus we list the total oxygen abundance as simply the sum of the O$^0$, O$^+$, and O$^{++}$ abundance (although
some contribution to O$^0$ may arise from a PDR).
In view of Table \ref{tabthearmsprop}, the two LISs in \object{NGC\,3132} display some remarkable properties.

\begin{table}[th!]
\small
\caption{Observed and de-reddened relative line fluxes with respect to 100$\times$\hb\, for the two LISs under discussion. \label{tabthearms}} 
\centering                          
\begin{tabular}{lcccccccc}        
\hline\hline                 
 &   \multicolumn{2}{c}{Arc N} & \multicolumn{2}{c}{Arc S} \\ 
Line & $F(\lambda)$ & $I(\lambda)$ & $F(\lambda)$  & $I(\lambda)$ \\
\hline 
c(\hb)              &  \multicolumn{2}{c}{0.56$\pm$0.23}  &    \multicolumn{2}{c}{0.52 $\pm$0.06}  \\
\hline
\hb &      100$\pm$     15 &      100$\pm$     14 &      100$\pm$      4 &      100$\pm$      3 \\
\oiii$\lambda$4959 &      381$\pm$     21 &      365$\pm$     18 &       66$\pm$      5 &       64$\pm$      4 \\
\oiii$\lambda$5007 &     1227$\pm$     34 &     1154$\pm$     32 &      166$\pm$      6 &      160$\pm$      6 \\
\niA$\lambda$5199 &       65$\pm$     19 &       56$\pm$     18 &       54$\pm$      4 &       50$\pm$      3 \\
\nii$\lambda$5755 &       79$\pm$     13 &       62$\pm$      8 &       36$\pm$      2 &       29$\pm$      2 \\
\hei$\lambda$5876 &       34$\pm$     16 &       27$\pm$     12 &       49$\pm$      3 &       38$\pm$      2 \\
\oi$\lambda$6300 &      404$\pm$     44 &      287$\pm$     30 &      361$\pm$      8 &      262$\pm$      6 \\
\oi$\lambda$6364 &      146$\pm$     17 &      100$\pm$     10 &      118$\pm$      3 &       85$\pm$      2 \\
\nii$\lambda$6548 &      839$\pm$     16 &      569$\pm$     12 &      542$\pm$      6 &      381$\pm$      4 \\
\ha     &      428$\pm$     24 &      287$\pm$     15 &      411$\pm$      6 &      289$\pm$      5 \\
\nii$\lambda$6584 &     2744$\pm$     34 &     1846$\pm$     20 &     1697$\pm$      7 &     1188$\pm$      5 \\
\hei$\lambda$6678 &       12$\pm$      7 &        7$\pm$      4 &       16$\pm$      1 &       11$\pm$      1 \\
\sii$\lambda$6716 &      102$\pm$      8 &       68$\pm$      5 &      156$\pm$      2 &      107$\pm$      2 \\
\sii$\lambda$6731 &       71$\pm$     10 &       46$\pm$      6 &      112$\pm$      2 &       76$\pm$      1 \\
\hei$\lambda$7065 &        4$\pm$      8 &        2$\pm$      5 &       17$\pm$      1 &       11$\pm$      1 \\
\ariii$\lambda$7136 &       90$\pm$      9 &       54$\pm$      4 &       44$\pm$      1 &       29$\pm$      1 \\
\oii$\lambda$7320 &       52$\pm$     20 &       29$\pm$     13 &       28$\pm$      3 &       18$\pm$      1 \\
\oii$\lambda$7331 &       43$\pm$     17 &       26$\pm$     13 &       26$\pm$      3 &       16$\pm$      2 \\
\siii$\lambda$9069  &       73$\pm$     19 &       35$\pm$      8 &       35$\pm$      3 &       17$\pm$      1 \\
\hline
\end{tabular}
\end{table}

\begin{table}[th!]
\caption{Derived properties for the two LISs. Estimated uncertainties for the listed line ratios are $\sim0.1$~dex. \label{tabthearmsprop}} 
\centering                          
\begin{tabular}{lcccccccc}        
\hline\hline                 
Property &  Arc N & Arc S \\ 
\hline 
c(\hb)              &  0.56$\pm$0.23  &    0.52 $\pm$0.06  \\
$v_{\lambda6584}$ (km s$^{-1}$)& $-0.2\pm0.6$ & $-9.7\pm0.2$ \\
$N_e$(\sii) (cm$^{-3}$) & $\lesssim$100 & $\lesssim$100\\
$T_e$(\nii) (K) & 14\,700$\pm$400 & 12\,f500$\pm$100\\
$\tau$($\lambda$3889) ($\omega$=3) & \ldots & 3.1$\pm$1.0\\
He$^+$/H$^+$ ($\lambda$5876)    &   \ldots &     0.29$\pm$   0.01 \\
He$^+$/H$^+$ ($\lambda$6678)   &   \ldots &     0.29$\pm$   0.01 \\
$10^5\times$ O$^{++}$/H$^+$ ($\lambda$5007) &   12.70$\pm$   1.24 &     2.79$\pm$   0.10 \\
$10^5\times$ O$^+$/H$^+$ ($\lambda$7320)   &   16.05$\pm$   3.46 &    20.99$\pm$   1.50\\
$10^5\times$ O$^+$/H$^+$ ($\lambda$7331) &    13.49$\pm$   3.29 &    19.96$\pm$   1.48 \\
$10^5\times$ O$^0$/H$^+$ ($\lambda$6300)&    15.01$\pm$   1.71 &    23.20$\pm$   0.93  \\
$10^5\times$ O/H$^+$     &    42.48$\pm$   5.52 &    46.47$\pm$   2.17 \\
$10^6\times$ S$^{++}$/H$^+$ ($\lambda$9069)&    2.60$\pm$   0.44 &     1.72$\pm$   0.10 \\
$10^6\times$ S$^{+}$/H$^+$ ($\lambda$6717)&    1.25$\pm$   0.11 &     2.72$\pm$   0.08  \\
$10^6\times$ S$^{+}$/H$^+$ ($\lambda$6731)&     1.16$\pm$   0.12 &     2.71$\pm$   0.08\\
$10^6\times$ Ar$^{++}$/H$^+$ ($\lambda$7136) &    2.01$\pm$   0.18 &     1.44$\pm$   0.04\\
$\log$ (\oiii$\lambda$5007/\ha) & $+0.60$ &   $-0.26$ \\
$\log$ (\nii$\lambda$6584/\ha) & $+0.81$ &  $+0.61$\\
$\log$ (\sii$\lambda\lambda$6716,6731/\ha) &  $-0.40$  & $-0.20$\\
$\log$ (\oi$\lambda$6300/\ha) & $+0.00$ &    $-0.04$\\
\hline
\end{tabular}
\end{table}

Firstly, extinction in both LISs is extremely high, actually higher than in any location at the main body of the PN. Both LISs are detected here for the first time as such. However \citet{Hora04} have already reported extended mid-IR extended emission  at the position of both LISs (and beyond) detected by means of imaging with IRAC on board of \emph{Spitzer}.
This emission was attributed to H$_2$ lines (mostly) and, in the case of the 8.0~$\mu$m band, an \ariii\, line. UIBs have been detected in the main body of the nebula \citep{Mata16} and the high extinction in the LISs implies the existence of a considerable amount of dust. Given the connection between dust and PAHs, it is plausible to expect a contribution of UIBs  at 3.3, 6.2 and 7.7 $\mu$m in the \emph{Spitzer} images. In order to evaluate the relative contribution of the molecular hydrogen and UIBs in these structures, mid-IR spectroscopy of the LISs (currently unavailable) would be needed.

Secondly, $N_e$ in both LISs is low, while $T_e$ is somewhat higher that in the main body of the nebula. Measurements of physical and chemical properties in LISs, and in particular in jets or jet-like structures are still scarce in the literature to infer what would be the expected behaviour. One of the few examples is provided by \citet{Akras16}, who also found smaller $N_e$ and higher $T_e$ in the jets of \object{IC\,4846}.

Regarding abundances, in comparison to the main body of the nebula, the only arc for which we could derive the helium abundance is highly enriched in helium and both arcs present significantly lower oxygen abundance than rest of the nebula (see Tab. \ref{tabthearmsprop}.

The listed line ratios give some clues regarding the ionisation mechanism.
All of them except $\log$(\nii/\ha) are within the range of line ratios for fast LISs predicted by \citet{Raga08} for shocked cloudlets. However \nii$\lambda$6584 is somewhat enhanced with respect to the predictions of these models. An investigation of the precise nature of this enhancement would require detailed modeling beyond the scope of this contribution. However, it is important to bear in mind that it would not necessarily imply an overabundance in nitrogen. Previous detailed modeling of LISs in other nebulae \citep[e.g., the knots on NGC\,7009,][]{Goncalves06} have already demonstrated how spatial variations in the ICF scheme can be enough to explain apparent nitrogen enhancements in LISs,
although the \nii/\ha\, ratio in the NGC\,3132 LISs is very large.

What would be the origin of the two arc-like structures identified here? \citet{Phillips83} suggested that precessing jets from a
binary system can launch the point-symmetric structures observed in some PNe. However, there are very few clear-cut examples of binaries actively shaping their surrounding planetary nebulae. One spectacular example is presented by \citet{Boffin12}, who established that the gigantic ($>2$ pc) jets in \object{Fleming~1}  \citep{Palmer96} can be explained by a precessing accretion disk around a companion.
\object{NGC\,3132} is a wide binary system, and thus this mechanism could be a good candidate to be explored.

\citet{Hora04} mentioned the difficulty of reconciling the existence of the extended mid-infrared emission in the northern and southern sides of the nebula (now known coincident with these LISs), with the orientation  proposed for the diabolo model by \citet{Monteiro00}.
However, if we adopt the geometry suggested by the kinematic analysis (see Sect. \ref{seckinema}), 
the direction along which the two LISs are aligned, is close enough to the direction of the major axis of the nebula (see Sect. \ref{seclinempas}) and, more interestingly, with the proposed orientation of the diabolo, as suggested by the kinematics. 
The redder velocities for the northern LIS and the bluer velocities for the southern one also fit in this scheme.
Thus, the picture delineated by the geometry proposed here is in reasonably good accord with the assumption of these LISs caused by precession, making this mechanism plausible.

Finally, it is interesting to note that the current estimation of PNe with some kind of LISs is $\sim$10\% \citep{Goncalves04}. More LISs, as found here up to $\sim$3 orders of magnitude fainter than the brightest parts of the nebula, could be identified through deep observations provided by MUSE. 
Future observations at similar depth may well reveal similar low surface brightness structures, making LISs a much more common phenomenon. 

\section{Integrated spectrum and 1D photoionisation model \label{sec1Dmodel}}

Although the main aim of this study is the 2D spectroscopy of NGC~3132, an IFU
can also provide a well-defined integrated spectrum of the nebula sampling the full 
range of conditions that single slit observations cannot achieve. Such an
integrated spectrum can be directly compared to the spectra of unresolved
PNe in nearby galaxies and also with photoionization models. 
We extracted a spectrum representative of the whole nebula by co-adding all the spectra in the spaxels with a flux in \nii$\lambda$6584 $>$10$^{-17}$ erg cm$^{-2}$ s$^{-1}$ after masking the spaxels corresponding to the central and field stars. A total of 121021 spaxels were considered, corresponding to an area of  $\sim$1.34~arcmin$^2$ or 0.0849 pc$^2$, for the adopted distance. Line fluxes for the strongest emission lines were measured using the same methodology as for the individual spaxels and the LIS spectra. Total \hb\, flux is 6.50$\pm$0.03$\times$10$^{-11}$ erg cm$^{-2}$ s$^{-1}$  ($-10.19$ in $\log_{10}$). This compares well with the $-10.45$ in $\log_{10}$ reported by \citet{Tsamis03} for a given aperture, much smaller than the area used for integration here. Likewise the measured \ha\, flux of 2.09$\pm$0.02$\times$10$^{-17}$ erg cm$^{-2}$ s$^{-1}$ ($-9.69$ in $\log_{10}$)) compares well with the $-9.57$ in $\log_{10}$  in a 6\, arcmin aperture reported by \citet{Frew08}. Measured and de-reddened fluxes are reported in Table \ref{tabintegspec}. Note that, not unsurprinsingly, the integrated c(\hb) is fully in accord with the median value reported in Sect. \ref{secextin}.

The simplest possible Cloudy model \citep[version 17.01]{Ferland17} was constructed 
to better understand the nebular conditions (central star temperature and luminosity,
ionization conditions and chemical abundances) and how these relate to the 2D 
structure. A spherical shell of constant density 1000 
cm$^{-3}$ with inner and outer radii 0.0013 and 0.24\,pc was adopted with an initial set 
of abundances adapted from \cite{Krabbe05} and \cite{Monteiro00} and the derived He
abundance of 0.128 (Tab. \ref{TabIon}). A black body central star of luminosity
250 L$_{\odot}$ was used based on the photoionization model by \cite{Monteiro00} 
scaled to the adopted distance of 860\,pc and the star temperature was varied to
match the strongest diagnostic lines. A satisfactory match was found with a 105000\,K
black body and the set of abundances (He: 0.124; N: 3.6e-4; O: 8.6e-4; Ne: 3.5e-4; 
S: 1.3e-5; Cl: 2.5e-7; Ar: 3.8e-6) with other abundances taken from \cite{Monteiro00}). 
The relative fluxes with respect to
\hb\, are listed in the 4th column of Tab. \ref{tabintegspec}. 

A confirmation of the black body stellar luminosity independent of
the Cloudy modelling can be provided by photometry of the Hubble 
Space Telescope (HST) imaging of the central hot star. Broad band 
filter images, in F438W ($B$), F555W ($V$) and F814W ($I$) were obtained 
in HST Proposal 11599 (PI R. A. Wade) with the UVIS channel of Wide 
Field Camera 3 in 2010.
The drizzle combined images, downloaded 
from the HST MAST archive, were analysed (associations $ib5708010$,
$ib5708020$, $ib5708030$, $ib5708040$, $ib5708050$ and $ib5708060$). 
The A0 primary (HD 87892) may be saturated even in the shortest 
exposures (0.48s)  but the hot companion star, offset by 
1.69$''$, is unaffected. Given the close proximity of the $\sim$ 
5.5 mag brighter A star, only a small photometric aperture could be 
used for the hot companion, so an appropriate aperture correction 
(based on HD 87892) was applied \citep[c.f.,][]{Gennaro2018}: the 
measured V magnitude on the Vega System is 16.00, with $(B-V)$ = $-$0.24 
by comparison to the F435W photometry.
For an extinction of $c$(\hb) = 0.15
from Tab. \ref{tabintegspec} (A$_v$ = 0.31), the intrinsic $V$ mag is 15.69,
fainter than the expected value of 14.85
for a 250 L$_{\odot}$  105\,000K black body at the assumed distance.

The match of the model to the integrated dereddened spectrum for the strong lines 
of He, N, O, S, Ar is satisfactory (within $\sim$15\%) but the largest discrepancies 
were for the weak lines of \textsc{[N\,i]} $\lambda$5199, and the auroral \nii$\lambda$5755 and 
\siii$\lambda$6312. The modelled electron temperatures from
\nii\, and \siii\, are lower than observed (respectively $\sim$500\,K and 1000\,K)
and the \sii\, density is modelled higher than observed (by $\sim$300 cm$^{-3}$),
indicating that the assumption of a constant density shell is inadequate
(as is clear from the $N_{\rm e}$ maps shown in Fig. \ref{mapNeTe})
and the morphology deduced from the velocity maps (Sect. \ref{seckinema}) and the 3D diabolo model
of \cite{Monteiro00}). There is clearly potential for a full 3D model of NGC~3132
taking account of all the structural information on this nebula.
The low central star luminosity together with moderate densities gives rise to 
the strong neutral lines from optically thick emission and the detection of 
H$_{2}$ \citep{Lupu06} and CO \citep{Sahai90, Phillips92, GuzmanRamirez18}. Given 
that the central star luminosity is low and temperature high, NGC\,3132 has
probably entered the PN cooling track with declining luminosity and the ionized 
nebula may be starting to recombine.

The N/O ratio from the Cloudy model is 0.42, and together with the He/H
value puts NGC~3132 in the category of typical Galactic Plane PN (Type II). 
The low value of ADF \citep{Tsamis04} appears to rule out any association 
between its central star binarity and high ADF, as established by e.g., 
\cite{Wesson18}, although NGC~3132 is a wide separation binary;
0.0062\,pc). The Cloudy model model predicts that the recombination 
contribution to \nii$\lambda$5755 of 1\% thus justifying
its neglect on the $T_{\rm e}$(\nii,\sii) determination (Sect. \ref{subsec_tenecol})
at least on a global scale. Also the total recombination contribution to
the \oii $\lambda$7320 and $\lambda$7331 lines is 3.1\% and 3.0\% respectively.

\begin{table}[th!]
\small
\caption{Observed and de-reddened relative line fluxes with respect to 100$\times$\hb\, for the integrated spectrum together with the predictions of the 1D Cloudy model. \label{tabintegspec}} 
\centering                          
\begin{tabular}{lcccccccc}        
\hline
\hline
Line & $F(\lambda)$ & $I(\lambda)$ & Model \\
\hline 
c(\hb)              &  \multicolumn{2}{c}{0.15$\pm$0.02} &  \\
\hline
\hb &   100.00$\pm$   0.82 &   100.00$\pm$   0.75 & 100.0 \\
\oiii$\lambda$4959 &   281.22$\pm$   1.19 &   278.21$\pm$   1.04 & 259.4\\
\oiii$\lambda$5007 &   845.19$\pm$   1.64 &   836.16$\pm$   1.30 & 774.0 \\
\niA$\lambda$5199 &     8.40$\pm$   4.78 &     8.17$\pm$   4.07 & 10.7\\
\heii$\lambda$5412 &     0.76$\pm$   1.12 &     0.73$\pm$   1.07 & 0.8 \\
\cliii$\lambda$5518 &     1.08$\pm$   1.27 &     1.03$\pm$   1.26 & 0.9\\
\cliii$\lambda$5538 &     0.85$\pm$   1.35 &     0.81$\pm$   1.33 & 0.8\\
\nii$\lambda$5755 &     8.13$\pm$   0.59 &     7.63$\pm$   0.59 & 6.3  \\
\hei$\lambda$5876 &    17.70$\pm$   0.63 &    16.49$\pm$   0.57 & 15.4\\
\oi$\lambda$6300 &    36.67$\pm$   1.31 &    33.47$\pm$   1.45 & 29.7 \\
\siii$\lambda$6312 &     2.46$\pm$   1.14 &     2.24$\pm$   0.97 & 1.6 \\
\oi$\lambda$6364 &    12.03$\pm$   0.99 &    10.96$\pm$   0.97 & 9.4\\
\nii$\lambda$6548 &   199.34$\pm$   1.47 &   179.64$\pm$   1.48 & 182.1 \\
\ha &   320.56$\pm$   1.60 &   288.33$\pm$   1.47 & 287.7\\
\nii$\lambda$6584 &   617.73$\pm$   1.99 &   556.27$\pm$   1.90 & 536.8\\
\hei$\lambda$6678 &     5.10$\pm$   0.95 &     4.58$\pm$   0.80 & 4.3\\
\sii$\lambda$6716 &    58.04$\pm$   1.58 &    51.79$\pm$   1.61 & 47.9 \\
\sii$\lambda$6731 &    56.67$\pm$   1.10 &    50.73$\pm$   1.05 & 52.6 \\\
\hei$\lambda$7065 &     4.51$\pm$   1.04 &     3.97$\pm$   0.97 & 4.5 \\
\ariii$\lambda$7136 &    31.08$\pm$   0.71 &    27.29$\pm$   0.66 & 25.2 \\
\oii$\lambda$7320 &     8.01$\pm$   2.02 &     6.96$\pm$   2.09 & 6.3\\
\oii$\lambda$7331 &     6.44$\pm$   1.33 &     5.60$\pm$   1.21 & 5.2 \\
\siii$\lambda$9069 &    47.30$\pm$   1.44 &    38.39$\pm$   1.37 & 36.8 \\
\hline
\end{tabular}
\end{table}

\section{Conclusions and perspectives \label{secconclu}}

We release here the MUSE data cube and a set of high-quality emission line maps for the planetary nebula \object{NGC 3132}.
The cube has a spatial size of $\sim$63$^{\prime\prime}\times$123$^{\prime\prime}$ (corresponding to $\sim$0.26$\times$0.51~pc$^2$ for the adopted distance) at a spatial sampling of 0\farcs2 pix$^{-1}$,  and a wavelength coverage of 4750-9300~\AA\, at a spectral sampling of 1.25~\AA\,pix~$^{-1}$.
Together with the variance cube, the file occupies a size of 5.4 GiB on disk.
These data reach an unprecedented depth covering a flux range of more than three orders of magnitude for the strongest lines (e.g., \nii$\lambda$6584). 
To exemplify the potential of the data, we carried out a detailed analysis of the physical and chemical properties of the nebula on a spaxel-by-spaxel basis. Our main conclusions are:

   \begin{enumerate}
      \item The data reveal an immense complexity in terms of morphology and ionization structure. In particular, in addition to the rim and a doily-like shell, both previously reported, two low-ionization arc-like structures have been identified towards the north and south of the nebula for the first time. 
      \item The extinction map, as derived from the hydrogen recombination lines, displays considerable structure, with large areas showing values up to 3-4 times higher than the median value, specially beyond the eastern side of the rim. A compilation of the results obtained for the few PNe with extinction maps available to date suggests that this characteristic may be common, directly demonstrating the survival of dust as an internal component of PNe.
      \item Maps for the electron temperature and density from several diagnostics were derived. Density maps are compatible with an inner high-ionisation plasma at relatively high density ($\sim$1000~cm$^{-3}$) while the low-ionisation plasma presents a structure in density with typical values of $\sim$300~cm$^{-3}$ and peaking at the rim with values $\sim$700~cm$^{-3}$.
      \item Maps for five electron temperature diagnostics are provided. Those derived from recombination lines are typically lower than those derived from CELs.  Median $T_e$ decreases according to the following sequence: \nii,\sii $\rightarrow$ \siii  $\rightarrow$ \oi $\rightarrow$ \hei $\rightarrow$ PJ. Likewise the range of temperatures covered by recombination lines is much larger than those from CELs.  It has been proposed that this behaviour, together with large values of $t^2$, can be attributed to the existence of high density clumps. The spatial variations of the differences in temperature and $t^2$ within the nebula suggest variations in the properties and distrubution of the clumps, should these exist.
      \item Several ionic and elemental abundances have been mapped. Regarding helium, we determined a median of He / H = 0.124, with somewhat higher values at the rim and outer shell. Still, with the estimated uncertainties, we cannot conclude that there are chemical inhomogeneities in the main body of the nebula.
      \item Regarding light elements, we derived maps for ionic abundances of oxygen, nitrogen, sulphur, argon and chlorine. The ranges of measured abundances are compatible with values available in the literature.
      \item The analysis of the derived velocity maps for a selection of the strongest emission lines, sampling a set of ions covering a variety of ionisation potentials, support a geometry for the nebula similar to the diabolo-like model proposed by \citet{Monteiro00} but with a different orientation for the diabolo, (major axis roughly at P.A.$\sim-22^\circ$). This analysis nicely illustrates the power of 2D kinematic information in many lines to understand the intrinsic (3D) structure of PNe.
      \item We identified two low-surface brightness LISs towards the northern and sourthern side of the nebula. The LISs display an arc-like morphology, extremely high extinction, and helium abundance, with the northern LIS having redder (more positive) velocities than the southern one. They are spatially coincident with some extended low-surface brightness mid-IR emission attributed mainly to H$_2$ \citep{Hora04}. Still, the high extinction within the structures suggests an important contribution of PAH emission in the mid-IR. We discuss the putative origin of these structures in the context of their point-like symmetry and the proposed orientation of the nebula. We conclude that they may be the consequence of precessing jets caused by the binary star system. 
      \item A simple 1D Cloudy model assuming a spherical shell of constant density is able to reproduced the strong lines in the integrated spectrum of the nebula with an accuracy of $\sim$15\%. However, the modelled $T_e$(\nii) and $T_e$(\siii) are lower than observed by $\sim$500~K and $\sim$1\,000~K, and density higher than observed by $\sim$300~cm$^{-3}$. This indicates that such simple assumptions are not adequate and support a full 3D model of the nebula taking into account all the structural information.
   \end{enumerate}
This ensemble of results nicely illustrate the potential of IFS, in general, and wide-field integral field spectrographs like MUSE in particular, to learn about PNe.
Specifically, this set of data, together with the results listed above, supply a large and detailed set of constraints regarding the physical, chemical and kinematic properties of the nebula, thus providing with an invaluable opportunity to revisit the 3D modelling of the nebula. This would be the most natural next step.
Moreover, with the exception of the two newly identified LISs, and the basic 1D photoionisation modeling, all the analysis presented here was done on a spaxel-by-spaxel basis. A more detailed analysis, in terms of e.g. emission line content, can be done by means of a carefully chosen spatial binning of the data (at the cost of some loss of spatial resolution).
Despite the suitability of MUSE to such a study, this work also illustrates some drawbacks: with its spectral range excluding bluer wavelengths and availability of total abundances for only two elements (helium and oxygen). Also, the kinematic description that can be extracted at its spectral resolution is still somehow limited, since it does not allow the de-blending of different spectral components, should these exist. 
Both, issues can be overcome with instruments similar to MUSE but with somewhat higher spectral resolution and access to the bluer part of the optical spectral range, like the recently proposed BlueMUSE \citep{Richard19}.

\begin{acknowledgements}
The authors are very grateful to David Carton and Jarle Brinchmann for their initial assessment of the quality of the MUSE Commissioning data for PNe. Their suggestions prompted us to reduce the data presented in this contribution.
Likewise, we thank Peter Weilbacher for useful discussions about the MUSE data. 
We also thank the referee for the valuable and supportive comments that have helped us to improve the first submitted version of this paper.
AMI acknowledges support from the Spanish MINECO through project
AYA2015-68217-P.
Based on observations collected at the European Organisation for Astronomical Research in the Southern Hemisphere under ESO programme 60.A-9100(A).
We are also grateful to the communities who developed the many Python packages used in this research, such MPDAF \citep{Piqueras17}, Astropy \citep{AstropyCollaboration13}, numpy \citep{Walt11}, scipy \citep{Jones01} and matplotlib \citep{Hunter07}.

\end{acknowledgements}

\bibliographystyle{aa}
\bibliography{mybib_aa}

\begin{thebibliography}{92}
\expandafter\ifx\csname natexlab\endcsname\relax\def\natexlab#1{#1}\fi

\bibitem[{{Akras} \& {Gon{\c{c}}alves}(2016)}]{Akras16}
{Akras}, S. \& {Gon{\c{c}}alves}, D.~R. 2016, \mnras, 455, 930

\bibitem[{{Ali} {et~al.}(2016){Ali}, {Dopita}, {Basurah}, {Amer}, {Alsulami},
  \& {Alruhaili}}]{Ali16}
{Ali}, A., {Dopita}, M.~A., {Basurah}, H.~M., {et~al.} 2016, \mnras, 462, 1393

\bibitem[{{Arias} {et~al.}(2001){Arias}, {Rosado}, {Salas}, \&
  {Cruz-Gonz{\'a}lez}}]{Arias01}
{Arias}, L., {Rosado}, M., {Salas}, L., \& {Cruz-Gonz{\'a}lez}, I. 2001, \aj,
  122, 3293

\bibitem[{{Astropy Collaboration} {et~al.}(2013){Astropy Collaboration},
  {Robitaille}, {Tollerud}, {Greenfield}, {Droettboom}, {Bray}, {Aldcroft},
  {Davis}, {Ginsburg}, {Price-Whelan}, {Kerzendorf}, {Conley}, {Crighton},
  {Barbary}, {Muna}, {Ferguson}, {Grollier}, {Parikh}, {Nair}, {Unther},
  {Deil}, {Woillez}, {Conseil}, {Kramer}, {Turner}, {Singer}, {Fox}, {Weaver},
  {Zabalza}, {Edwards}, {Azalee Bostroem}, {Burke}, {Casey}, {Crawford},
  {Dencheva}, {Ely}, {Jenness}, {Labrie}, {Lim}, {Pierfederici}, {Pontzen},
  {Ptak}, {Refsdal}, {Servillat}, \& {Streicher}}]{AstropyCollaboration13}
{Astropy Collaboration}, {Robitaille}, T.~P., {Tollerud}, E.~J., {et~al.} 2013,
  \aap, 558, A33

\bibitem[{{Bacon} {et~al.}(2010){Bacon}, {Accardo}, {Adjali}, {Anwand},
  {Bauer}, {Biswas}, {Blaizot}, {Boudon}, {Brau-Nogue}, {Brinchmann},
  {Caillier}, {Capoani}, {Carollo}, {Contini}, {Couderc}, {Daguis{\'e}},
  {Deiries}, {Delabre}, {Dreizler}, {Dubois}, {Dupieux}, {Dupuy}, {Emsellem},
  {Fechner}, {Fleischmann}, {Fran{\c c}ois}, {Gallou}, {Gharsa}, {Glindemann},
  {Gojak}, {Guiderdoni}, {Hansali}, {Hahn}, {Jarno}, {Kelz}, {Koehler},
  {Kosmalski}, {Laurent}, {Le Floch}, {Lilly}, {Lizon}, {Loupias}, {Manescau},
  {Monstein}, {Nicklas}, {Olaya}, {Pares}, {Pasquini}, {P{\'e}contal-Rousset},
  {Pell{\'o}}, {Petit}, {Popow}, {Reiss}, {Remillieux}, {Renault}, {Roth},
  {Rupprecht}, {Serre}, {Schaye}, {Soucail}, {Steinmetz}, {Streicher}, {Stuik},
  {Valentin}, {Vernet}, {Weilbacher}, {Wisotzki}, \& {Yerle}}]{Bacon10}
{Bacon}, R., {Accardo}, M., {Adjali}, L., {et~al.} 2010, in \procspie, Vol.
  7735, Ground-based and Airborne Instrumentation for Astronomy III, 773508

\bibitem[{{Balick} {et~al.}(1993){Balick}, {Rugers}, {Terzian}, \&
  {Chengalur}}]{Balick93}
{Balick}, B., {Rugers}, M., {Terzian}, Y., \& {Chengalur}, J.~N. 1993, \apj,
  411, 778

\bibitem[{{Benjamin} {et~al.}(1999){Benjamin}, {Skillman}, \&
  {Smits}}]{Benjamin99}
{Benjamin}, R.~A., {Skillman}, E.~D., \& {Smits}, D.~P. 1999, \apj, 514, 307

\bibitem[{{Boffin} {et~al.}(2012){Boffin}, {Miszalski}, {Rauch}, {Jones},
  {Corradi}, {Napiwotzki}, {Day-Jones}, \& {K{\"o}ppen}}]{Boffin12}
{Boffin}, H. M.~J., {Miszalski}, B., {Rauch}, T., {et~al.} 2012, Science, 338,
  773

\bibitem[{{Capitanio} {et~al.}(2017){Capitanio}, {Lallement}, {Vergely},
  {Elyajouri}, \& {Monreal-Ibero}}]{Capitanio17}
{Capitanio}, L., {Lallement}, R., {Vergely}, J.~L., {Elyajouri}, M., \&
  {Monreal-Ibero}, A. 2017, \aap, 606, A65

\bibitem[{{Cardelli} {et~al.}(1989){Cardelli}, {Clayton}, \&
  {Mathis}}]{Cardelli89}
{Cardelli}, J.~A., {Clayton}, G.~C., \& {Mathis}, J.~S. 1989, \apj, 345, 245

\bibitem[{{Ciardullo} {et~al.}(1999){Ciardullo}, {Bond}, {Sipior}, {Fullton},
  {Zhang}, \& {Schaefer}}]{Ciardullo99}
{Ciardullo}, R., {Bond}, H.~E., {Sipior}, M.~S., {et~al.} 1999, \aj, 118, 488

\bibitem[{{Corradi} {et~al.}(2015){Corradi}, {Garc{\'{\i}}a-Rojas}, {Jones}, \&
  {Rodr{\'{\i}}guez-Gil}}]{Corradi15}
{Corradi}, R.~L.~M., {Garc{\'{\i}}a-Rojas}, J., {Jones}, D., \&
  {Rodr{\'{\i}}guez-Gil}, P. 2015, \apj, 803, 99

\bibitem[{{Corradi} {et~al.}(1996){Corradi}, {Manso}, {Mampaso}, \&
  {Schwarz}}]{Corradi96}
{Corradi}, R.~L.~M., {Manso}, R., {Mampaso}, A., \& {Schwarz}, H.~E. 1996,
  \aap, 313, 913

\bibitem[{{Cuesta} {et~al.}(1993){Cuesta}, {Phillips}, \& {Mampaso}}]{Cuesta93}
{Cuesta}, L., {Phillips}, J.~P., \& {Mampaso}, A. 1993, \aap, 267, 199

\bibitem[{{Danehkar}(2015)}]{Danehkar15}
{Danehkar}, A. 2015, \apj, 815, 35

\bibitem[{{Danehkar} {et~al.}(2016){Danehkar}, {Parker}, \&
  {Steffen}}]{Danehkar16}
{Danehkar}, A., {Parker}, Q.~A., \& {Steffen}, W. 2016, \aj, 151, 38

\bibitem[{{Delgado-Inglada} {et~al.}(2014){Delgado-Inglada}, {Morisset}, \&
  {Stasi{\'n}ska}}]{DelgadoInglada14}
{Delgado-Inglada}, G., {Morisset}, C., \& {Stasi{\'n}ska}, G. 2014, \mnras,
  440, 536

\bibitem[{{Delgado-Inglada} {et~al.}(2015){Delgado-Inglada}, {Rodr{\'\i}guez},
  {Peimbert}, {Stasi{\'n}ska}, \& {Morisset}}]{DelgadoInglada15}
{Delgado-Inglada}, G., {Rodr{\'\i}guez}, M., {Peimbert}, M., {Stasi{\'n}ska},
  G., \& {Morisset}, C. 2015, \mnras, 449, 1797

\bibitem[{{Draine} \& {Li}(2007)}]{Draine07}
{Draine}, B.~T. \& {Li}, A. 2007, \apj, 657, 810

\bibitem[{{Evans}(1968)}]{Evans68}
{Evans}, D.~S. 1968, Monthly Notes of the Astronomical Society of South Africa,
  27, 129

\bibitem[{{Ferland} {et~al.}(2017){Ferland}, {Chatzikos}, {Guzm{\'a}n},
  {Lykins}, {van Hoof}, {Williams}, {Abel}, {Badnell}, {Keenan}, {Porter}, \&
  {Stancil}}]{Ferland17}
{Ferland}, G.~J., {Chatzikos}, M., {Guzm{\'a}n}, F., {et~al.} 2017, \rmxaa, 53,
  385

\bibitem[{{Frew}(2008)}]{Frew08}
{Frew}, D.~J. 2008, PhD thesis, Department of Physics, Macquarie University,
  NSW 2109, Australia

\bibitem[{{Garc{\'\i}a-D{\'\i}az} {et~al.}(2012){Garc{\'\i}a-D{\'\i}az},
  {L{\'o}pez}, {Steffen}, \& {Richer}}]{GarciaDiaz12}
{Garc{\'\i}a-D{\'\i}az}, M.~T., {L{\'o}pez}, J.~A., {Steffen}, W., \& {Richer},
  M.~G. 2012, \apj, 761, 172

\bibitem[{{Garc{\'{\i}}a-Hern{\'a}ndez}
  {et~al.}(2016){Garc{\'{\i}}a-Hern{\'a}ndez}, {Ventura}, {Delgado-Inglada},
  {Dell'Agli}, {Di Criscienzo}, \& {Yag{\"u}e}}]{GarciaHernandez16}
{Garc{\'{\i}}a-Hern{\'a}ndez}, D.~A., {Ventura}, P., {Delgado-Inglada}, G.,
  {et~al.} 2016, \mnras, 461, 542

\bibitem[{{Gathier} {et~al.}(1986){Gathier}, {Pottasch}, \& {Pel}}]{Gathier86}
{Gathier}, R., {Pottasch}, S.~R., \& {Pel}, J.~W. 1986, \aap, 157, 171

\bibitem[{{Gennaro}(2018)}]{Gennaro2018}
{Gennaro}, M. e.~a. 2018, {WFC3 Data Handbook, Version 4.0, (Baltimore:
  STScI)}, version 4.0 edn., Space Telescope Science Institute, STScI,
  Baltimore USA, provided as per the recommended reference

\bibitem[{{Gon{\c c}alves}(2004)}]{Gonsalves04}
{Gon{\c c}alves}, D.~R. 2004, in Astronomical Society of the Pacific Conference
  Series, Vol. 313, Asymmetrical Planetary Nebulae III: Winds, Structure and
  the Thunderbird, ed. M.~{Meixner}, J.~H. {Kastner}, B.~{Balick}, \&
  N.~{Soker}, 216

\bibitem[{{Gon{\c{c}}alves}(2004)}]{Goncalves04}
{Gon{\c{c}}alves}, D.~R. 2004, in Astronomical Society of the Pacific
  Conference Series, Vol. 313, Asymmetrical Planetary Nebulae III: Winds,
  Structure and the Thunderbird, ed. M.~{Meixner}, J.~H. {Kastner},
  B.~{Balick}, \& N.~{Soker}, 216

\bibitem[{{Gon{\c{c}}alves} {et~al.}(2001){Gon{\c{c}}alves}, {Corradi}, \&
  {Mampaso}}]{Goncalves01}
{Gon{\c{c}}alves}, D.~R., {Corradi}, R. L.~M., \& {Mampaso}, A. 2001, \apj,
  547, 302

\bibitem[{{Gon{\c{c}}alves} {et~al.}(2006){Gon{\c{c}}alves}, {Ercolano},
  {Carnero}, {Mampaso}, \& {Corradi}}]{Goncalves06}
{Gon{\c{c}}alves}, D.~R., {Ercolano}, B., {Carnero}, A., {Mampaso}, A., \&
  {Corradi}, R.~L.~M. 2006, \mnras, 365, 1039

\bibitem[{{Guerrero} {et~al.}(1999){Guerrero}, {V{\'a}zquez}, \&
  {L{\'o}pez}}]{Guerrero99}
{Guerrero}, M.~A., {V{\'a}zquez}, R., \& {L{\'o}pez}, J.~A. 1999, \aj, 117, 967

\bibitem[{{Guzman-Ramirez} {et~al.}(2018){Guzman-Ramirez},
  {G{\'o}mez-Ru{\'{\i}}z}, {Boffin}, {Jones}, {Wesson}, {Zijlstra}, {Smith}, \&
  {Nyman}}]{GuzmanRamirez18}
{Guzman-Ramirez}, L., {G{\'o}mez-Ru{\'{\i}}z}, A.~I., {Boffin}, H.~M.~J.,
  {et~al.} 2018, \aap, 618, A91

\bibitem[{{Hajian} {et~al.}(1997){Hajian}, {Balick}, {Terzian}, \&
  {Perinotto}}]{Hajian97}
{Hajian}, A.~R., {Balick}, B., {Terzian}, Y., \& {Perinotto}, M. 1997, \apj,
  487, 304

\bibitem[{{Hajian} {et~al.}(2007){Hajian}, {Movit}, {Trofimov}, {Balick},
  {Terzian}, {Knuth}, {Granquist-Fraser}, {Huyser}, {Jalobeanu}, \&
  {McIntosh}}]{Hajian07}
{Hajian}, A.~R., {Movit}, S.~M., {Trofimov}, D., {et~al.} 2007, \apjs, 169, 289

\bibitem[{{Hora} {et~al.}(2004){Hora}, {Latter}, {Allen}, {Marengo}, {Deutsch},
  \& {Pipher}}]{Hora04}
{Hora}, J.~L., {Latter}, W.~B., {Allen}, L.~E., {et~al.} 2004, \apjs, 154, 296

\bibitem[{Hunter(2007)}]{Hunter07}
Hunter, J.~D. 2007, Computing In Science \& Engineering, 9, 90

\bibitem[{Jones {et~al.}(2001)Jones, Oliphant, Peterson, {et~al.}}]{Jones01}
Jones, E., Oliphant, T., Peterson, P., {et~al.} 2001, {SciPy}: Open source
  scientific tools for {Python}

\bibitem[{{Juguet} {et~al.}(1988){Juguet}, {Louise}, {Macron}, \&
  {Pascoli}}]{Juguet88}
{Juguet}, J.~L., {Louise}, R., {Macron}, A., \& {Pascoli}, G. 1988, \aap, 205,
  267

\bibitem[{{Kimeswenger} \& {Barr{\'{\i}}a}(2018)}]{Kimeswenger18}
{Kimeswenger}, S. \& {Barr{\'{\i}}a}, D. 2018, \aap, 616, L2

\bibitem[{{Kingsburgh} \& {Barlow}(1994)}]{Kingsburgh94}
{Kingsburgh}, R.~L. \& {Barlow}, M.~J. 1994, \mnras, 271, 257

\bibitem[{{Kohoutek} \& {Laustsen}(1977)}]{Kohoutek77}
{Kohoutek}, L. \& {Laustsen}, S. 1977, \aap, 61, 761

\bibitem[{{Krabbe} \& {Copetti}(2005)}]{Krabbe05}
{Krabbe}, A.~C. \& {Copetti}, M.~V.~F. 2005, \aap, 443, 981

\bibitem[{{Kwitter} {et~al.}(2014){Kwitter}, {M{\'e}ndez}, {Pe{\~n}a},
  {Stanghellini}, {Corradi}, {De Marco}, {Fang}, {Henry}, {Karakas}, {Liu},
  {L{\'o}pez}, {Manchado}, \& {Parker}}]{Kwitter14}
{Kwitter}, K.~B., {M{\'e}ndez}, R.~H., {Pe{\~n}a}, M., {et~al.} 2014, \rmxaa,
  50, 203

\bibitem[{{Kwok}(1994)}]{Kwok94}
{Kwok}, S. 1994, \pasp, 106, 344

\bibitem[{{Lago} \& {Costa}(2016)}]{Lago16}
{Lago}, P.~J.~A. \& {Costa}, R.~D.~D. 2016, \rmxaa, 52, 329

\bibitem[{{Lame} \& {Pogge}(1996)}]{Lame96}
{Lame}, N.~J. \& {Pogge}, R.~W. 1996, \aj, 111, 2320

\bibitem[{{Leal-Ferreira} {et~al.}(2011){Leal-Ferreira}, {Gon{\c{c}}alves},
  {Monteiro}, \& {Richards}}]{LealFerreira11}
{Leal-Ferreira}, M.~L., {Gon{\c{c}}alves}, D.~R., {Monteiro}, H., \&
  {Richards}, J.~W. 2011, \mnras, 411, 1395

\bibitem[{{Liu} {et~al.}(2000){Liu}, {Storey}, {Barlow}, {Danziger}, {Cohen},
  \& {Bryce}}]{Liu00}
{Liu}, X.-W., {Storey}, P.~J., {Barlow}, M.~J., {et~al.} 2000, \mnras, 312, 585

\bibitem[{{Lupu} {et~al.}(2006){Lupu}, {France}, \& {McCandliss}}]{Lupu06}
{Lupu}, R.~E., {France}, K., \& {McCandliss}, S.~R. 2006, \apj, 644, 981

\bibitem[{{Luridiana} {et~al.}(2015){Luridiana}, {Morisset}, \&
  {Shaw}}]{Luridiana15}
{Luridiana}, V., {Morisset}, C., \& {Shaw}, R.~A. 2015, \aap, 573, A42

\bibitem[{{Mata} {et~al.}(2016){Mata}, {Ramos-Larios}, {Guerrero},
  {Nigoche-Netro}, {Toal{\'a}}, {Fang}, {Rubio}, {Kemp}, {Navarro}, \&
  {Corral}}]{Mata16}
{Mata}, H., {Ramos-Larios}, G., {Guerrero}, M.~A., {et~al.} 2016, \mnras, 459,
  841

\bibitem[{{Meaburn} \& {Walsh}(1981)}]{Meaburn81}
{Meaburn}, J. \& {Walsh}, J.~R. 1981, \apss, 78, 473

\bibitem[{{Meatheringham} {et~al.}(1988){Meatheringham}, {Wood}, \&
  {Faulkner}}]{Meatheringham88}
{Meatheringham}, S.~J., {Wood}, P.~R., \& {Faulkner}, D.~J. 1988, \apj, 334,
  862

\bibitem[{{Monreal-Ibero} {et~al.}(2005){Monreal-Ibero}, {Roth},
  {Sch{\"o}nberner}, {Steffen}, \& {B{\"o}hm}}]{MonrealIbero05}
{Monreal-Ibero}, A., {Roth}, M.~M., {Sch{\"o}nberner}, D., {Steffen}, M., \&
  {B{\"o}hm}, P. 2005, \apjl, 628, L139

\bibitem[{{Monreal-Ibero} {et~al.}(2006){Monreal-Ibero}, {Roth},
  {Sch{\"o}nberner}, {Steffen}, \& {B{\"o}hm}}]{MonrealIbero06}
{Monreal-Ibero}, A., {Roth}, M.~M., {Sch{\"o}nberner}, D., {Steffen}, M., \&
  {B{\"o}hm}, P. 2006, \nar, 50, 426

\bibitem[{{Monreal-Ibero} {et~al.}(2013){Monreal-Ibero}, {Walsh},
  {Westmoquette}, \& {V{\'{\i}}lchez}}]{MonrealIbero13}
{Monreal-Ibero}, A., {Walsh}, J.~R., {Westmoquette}, M.~S., \&
  {V{\'{\i}}lchez}, J.~M. 2013, \aap, 553, A57

\bibitem[{{Monteiro} {et~al.}(2013){Monteiro}, {Gon{\c c}alves},
  {Leal-Ferreira}, \& {Corradi}}]{Monteiro13}
{Monteiro}, H., {Gon{\c c}alves}, D.~R., {Leal-Ferreira}, M.~L., \& {Corradi},
  R.~L.~M. 2013, \aap, 560, A102

\bibitem[{{Monteiro} {et~al.}(2000){Monteiro}, {Morisset}, {Gruenwald}, \&
  {Viegas}}]{Monteiro00}
{Monteiro}, H., {Morisset}, C., {Gruenwald}, R., \& {Viegas}, S.~M. 2000, \apj,
  537, 853

\bibitem[{{Morisset}(2017)}]{Morisset17}
{Morisset}, C. 2017, in IAU Symposium, Vol. 323, Planetary Nebulae:
  Multi-Wavelength Probes of Stellar and Galactic Evolution, ed. X.~{Liu},
  L.~{Stanghellini}, \& A.~{Karakas}, 43--50

\bibitem[{{Osterbrock} \& {Ferland}(2006)}]{Osterbrock06}
{Osterbrock}, D.~E. \& {Ferland}, G.~J. 2006, {Astrophysics of gaseous nebulae
  and active galactic nuclei}, ed. D.~E. {Osterbrock} \& G.~J. {Ferland}

\bibitem[{{Palmer} {et~al.}(1996){Palmer}, {Lopez}, {Meaburn}, \&
  {Lloyd}}]{Palmer96}
{Palmer}, J.~W., {Lopez}, J.~A., {Meaburn}, J., \& {Lloyd}, H.~M. 1996, \aap,
  307, 225

\bibitem[{{Peimbert}(1967)}]{Peimbert67}
{Peimbert}, M. 1967, \apj, 150, 825

\bibitem[{{Perinotto}(2000)}]{Perinotto00}
{Perinotto}, M. 2000, \apss, 274, 205

\bibitem[{{Phillips} \& {Cuesta}(2000)}]{Phillips00}
{Phillips}, J.~P. \& {Cuesta}, L. 2000, \aj, 119, 335

\bibitem[{{Phillips} \& {Reay}(1983)}]{Phillips83}
{Phillips}, J.~P. \& {Reay}, N.~K. 1983, \aap, 117, 33

\bibitem[{{Phillips} {et~al.}(1992){Phillips}, {Williams}, {Mampaso}, \&
  {Ukita}}]{Phillips92}
{Phillips}, J.~P., {Williams}, P.~G., {Mampaso}, A., \& {Ukita}, N. 1992,
  \apss, 188, 171

\bibitem[{{Piqueras} {et~al.}(2017){Piqueras}, {Conseil}, {Shepherd}, {Bacon},
  {Leclercq}, \& {Richard}}]{Piqueras17}
{Piqueras}, L., {Conseil}, S., {Shepherd}, M., {et~al.} 2017, to be published
  in ADASS XXVI, ArXiv e-prints [\eprint[arXiv]{1710.03554}]

\bibitem[{{Pismis}(1989)}]{Pismis89}
{Pismis}, P. 1989, \mnras, 237, 611

\bibitem[{{Porter} {et~al.}(2013){Porter}, {Ferland}, {Storey}, \&
  {Detisch}}]{Porter13}
{Porter}, R.~L., {Ferland}, G.~J., {Storey}, P.~J., \& {Detisch}, M.~J. 2013,
  \mnras, 433, L89

\bibitem[{{Raga} {et~al.}(2008){Raga}, {Riera}, {Mellema}, {Esquivel}, \&
  {Vel{\'a}zquez}}]{Raga08}
{Raga}, A.~C., {Riera}, A., {Mellema}, G., {Esquivel}, A., \& {Vel{\'a}zquez},
  P.~F. 2008, \aap, 489, 1141

\bibitem[{{Richard} {et~al.}(2019){Richard}, {Bacon}, {Blaizot}, {Boissier},
  {Boselli}, {NicolasBouch{\'e}}, {Brinchmann}, {Castro}, {Ciesla}, \&
  {Crowther}}]{Richard19}
{Richard}, J., {Bacon}, R., {Blaizot}, J., {et~al.} 2019, arXiv e-prints,
  arXiv:1906.01657

\bibitem[{{Robbins}(1968)}]{Robbins68}
{Robbins}, R.~R. 1968, \apj, 151, 511

\bibitem[{{Sahai} {et~al.}(1990){Sahai}, {Wootten}, \& {Clegg}}]{Sahai90}
{Sahai}, R., {Wootten}, A., \& {Clegg}, R.~E.~S. 1990, \aap, 234, L1

\bibitem[{{Sahu} \& {Desai}(1986)}]{Sahu86}
{Sahu}, K.~C. \& {Desai}, J.~N. 1986, \aap, 161, 357

\bibitem[{{Sandin} {et~al.}(2008){Sandin}, {Sch{\"o}nberner}, {Roth},
  {Steffen}, {B{\"o}hm}, \& {Monreal-Ibero}}]{Sandin08}
{Sandin}, C., {Sch{\"o}nberner}, D., {Roth}, M.~M., {et~al.} 2008, \aap, 486,
  545

\bibitem[{{Sch{\"o}nberner} {et~al.}(2018){Sch{\"o}nberner}, {Balick}, \&
  {Jacob}}]{Schoenberner18}
{Sch{\"o}nberner}, D., {Balick}, B., \& {Jacob}, R. 2018, \aap, 609, A126

\bibitem[{{Storey}(1984)}]{Storey84}
{Storey}, J.~W.~V. 1984, \mnras, 206, 521

\bibitem[{{St{\"o}rzer} \& {Hollenbach}(2000)}]{Stoerzer00}
{St{\"o}rzer}, H. \& {Hollenbach}, D. 2000, \apj, 539, 751

\bibitem[{{Torres-Peimbert} \& {Peimbert}(1977)}]{TorresPeimbert77}
{Torres-Peimbert}, S. \& {Peimbert}, M. 1977, \rmxaa, 2, 181

\bibitem[{{Tsamis} {et~al.}(2003){Tsamis}, {Barlow}, {Liu}, {Danziger}, \&
  {Storey}}]{Tsamis03}
{Tsamis}, Y.~G., {Barlow}, M.~J., {Liu}, X., {Danziger}, I.~J., \& {Storey},
  P.~J. 2003, \mnras, 345, 186

\bibitem[{{Tsamis} {et~al.}(2004){Tsamis}, {Barlow}, {Liu}, {Storey}, \&
  {Danziger}}]{Tsamis04}
{Tsamis}, Y.~G., {Barlow}, M.~J., {Liu}, X.-W., {Storey}, P.~J., \& {Danziger},
  I.~J. 2004, \mnras, 353, 953

\bibitem[{{Tsamis} {et~al.}(2008){Tsamis}, {Walsh}, {P{\'e}quignot}, {Barlow},
  {Danziger}, \& {Liu}}]{Tsamis08}
{Tsamis}, Y.~G., {Walsh}, J.~R., {P{\'e}quignot}, D., {et~al.} 2008, \mnras,
  386, 22

\bibitem[{{Walsh} {et~al.}(2016){Walsh}, {Monreal-Ibero}, {Barlow}, {Ueta},
  {Wesson}, \& {Zijlstra}}]{Walsh16}
{Walsh}, J.~R., {Monreal-Ibero}, A., {Barlow}, M.~J., {et~al.} 2016, \aap, 588,
  A106

\bibitem[{{Walsh} {et~al.}(2018){Walsh}, {Monreal-Ibero}, {Barlow}, {Ueta},
  {Wesson}, {Zijlstra}, {Kimeswenger}, {Leal-Ferreira}, \& {Otsuka}}]{Walsh18}
{Walsh}, J.~R., {Monreal-Ibero}, A., {Barlow}, M.~J., {et~al.} 2018, \aap, 620,
  A169

\bibitem[{Walt {et~al.}(2011)Walt, Colbert, \& Varoquaux}]{Walt11}
Walt, S. v.~d., Colbert, S.~C., \& Varoquaux, G. 2011, Computing in Science and
  Engg., 13, 22

\bibitem[{{Weilbacher} {et~al.}(2015{\natexlab{a}}){Weilbacher},
  {Monreal-Ibero}, {Kollatschny}, {Ginsburg}, {McLeod}, {Kamann}, {Sandin},
  {Palsa}, {Wisotzki}, {Bacon}, {Selman}, {Brinchmann}, {Caruana}, {Kelz},
  {Martinsson}, {P{\'e}contal-Rousset}, {Richard}, \& {Wendt}}]{Weilbacher15}
{Weilbacher}, P.~M., {Monreal-Ibero}, A., {Kollatschny}, W., {et~al.}
  2015{\natexlab{a}}, \aap, 582, A114

\bibitem[{{Weilbacher} {et~al.}(2015{\natexlab{b}}){Weilbacher},
  {Monreal-Ibero}, {Mc Leod}, {Ginsburg}, {Kollatschny}, {Sandin}, {Wendt},
  {Wisotzki}, \& {Bacon}}]{Weilbacher15b}
{Weilbacher}, P.~M., {Monreal-Ibero}, A., {Mc Leod}, A.~F., {et~al.}
  2015{\natexlab{b}}, The Messenger, 162, 37

\bibitem[{{Weilbacher} {et~al.}(2014){Weilbacher}, {Streicher}, {Urrutia},
  {P{\'e}contal-Rousset}, {Jarno}, \& {Bacon}}]{Weilbacher14}
{Weilbacher}, P.~M., {Streicher}, O., {Urrutia}, T., {et~al.} 2014, in
  Astronomical Society of the Pacific Conference Series, Vol. 485, Astronomical
  Data Analysis Software and Systems XXIII, ed. N.~{Manset} \& P.~{Forshay},
  451

\bibitem[{{Wesson} {et~al.}(2018){Wesson}, {Jones}, {Garc{\'{\i}}a-Rojas},
  {Boffin}, \& {Corradi}}]{Wesson18}
{Wesson}, R., {Jones}, D., {Garc{\'{\i}}a-Rojas}, J., {Boffin}, H.~M.~J., \&
  {Corradi}, R.~L.~M. 2018, \mnras, 480, 4589

\bibitem[{{Woodward} {et~al.}(1992){Woodward}, {Pipher}, {Forrest}, {Moneti},
  \& {Shure}}]{Woodward92}
{Woodward}, C.~E., {Pipher}, J.~L., {Forrest}, W.~J., {Moneti}, A., \& {Shure},
  M.~A. 1992, \apj, 385, 567

\bibitem[{{Zhang} {et~al.}(2005){Zhang}, {Liu}, {Liu}, \& {Rubin}}]{Zhang05}
{Zhang}, Y., {Liu}, X.-W., {Liu}, Y., \& {Rubin}, R.~H. 2005, \mnras, 358, 457

\bibitem[{{Zhang} {et~al.}(2004){Zhang}, {Liu}, {Wesson}, {Storey}, {Liu}, \&
  {Danziger}}]{Zhang04}
{Zhang}, Y., {Liu}, X.~W., {Wesson}, R., {et~al.} 2004, \mnras, 351, 935

\end{thebibliography}

\appendix

\end{document}